\documentclass[usenatbib]{mn2e}
\usepackage{multirow}
\usepackage{rotating}
\usepackage{colortbl}
\usepackage{color}
\usepackage[fleqn]{amsmath}
\usepackage{amssymb}
\usepackage{amsfonts}
\usepackage{verbatim}
\usepackage{comment}
\usepackage{scalefnt}
\usepackage{lscape}

\usepackage[bookmarks=False,bookmarksnumbered,colorlinks=true, citecolor=blue, linkcolor=cyan]{hyperref}

\newcommand{\apjl}{ApJ}
\newcommand{\apjs}{ApJS}
\newcommand{\ao}{Appl.~Opt.}
\newcommand{\aap}{A\&A}

\newcommand{\beq}{\begin{equation}}
\newcommand{\eeq}{\end{equation}}

\definecolor{grey}{rgb}{0.5,0.6,0.7}
\def \simlt { \lower .75ex \hbox{$\sim$} \llap{\raise .27ex \hbox{$<$}} }
\definecolor{purple}{rgb}{0.65,0.15,0.9}
\definecolor{darkorange}{rgb}{0.8,0.3,0}
\definecolor{olive}{rgb}{0.4,0.6,0.25}
\definecolor{darkgreen}{rgb}{0,0.7,0}
\definecolor{darkred}{rgb}{0.5,0,0}

\title[Bringing faint active galactic nuclei (AGNs) to light]{Bringing faint active galactic nuclei (AGNs) to light: a view from large-scale cosmological simulations}
\author[Schirra \& Habouzit]{Adrian P.\ Schirra$^{1}$\thanks{E-mail: a.schirra@stud.uni-heidelberg.de},  M\'{e}lanie Habouzit$^{2, 1}$, Ralf S.\ Klessen$^{1,3}$, Francesca Fornasini$^{4,5}$, 
\newauthor Dylan Nelson$^{6,1}$, Annalisa Pillepich$^{2}$, Daniel Angl\'es-Alc\'azar$^{7,8}$, Romeel Dav\'e$^{9}$, 
\newauthor Francesca Civano$^{4}$\\
 $^{1}$ Universit\"{a}t Heidelberg, Zentrum f\"{u}r Astronomie, Institut f\"{u}r Theoretische Astrophysik, Albert-Ueberle-Str. 2, 69120 Heidelberg, Germany\\
 $^{2}$ Max-Planck-Institut f\"ur Astronomie, K\"onigstuhl 17, D-69117 Heidelberg, Germany\\
 $^{3}$ Universit\"{a}t Heidelberg, Interdisziplin\"{a}res Zentrum f\"{u}r Wissenschaftliches Rechnen, Im Neuenheimer Feld 205, 69120 Heidelberg, Germany\\
 $^{4}$ Harvard-Smithsonian Center for Astrophysics, 60 Garden Street, Cambridge, MA 02138, USA\\
 $^{5}$ Stonehill College, 320 Washington Street, North Easton, MA 02357, USA\\
 $^{6}$ Max-Planck-Institut f\"ur Astrophysik, 
 Karl-Schwarzschild-Strasse 1, D-85740 Garching bei M\"unchen, Germany\\
 $^{7}$ Department of Physics, University of Connecticut, 196 Auditorium Road, U-3046, Storrs, CT 06269-3046, USA\\
 $^{8}$ Center for Computational Astrophysics, Flatiron Institute, New York, NY 10010, USA\\
 $^{9}$ Institute for Astronomy, Royal Observatory, University of Edinburgh, Edinburgh EH9 3HJ, UK
}

\begin{document}
\date{2020}
\maketitle
\begin{abstract}

The sensitivity of X-ray facilities and our ability to detect fainter active galactic nuclei (AGN) will increase with the upcoming Athena mission and the AXIS and Lynx concept missions, thus improving our understanding of supermassive black holes (BHs) in a luminosity regime that can be dominated by X-ray binaries.
We analyze the population of faint AGN ($L_{\rm x,\, 2-10 \, keV}\leqslant 10^{42}\, \rm erg/s$) in the Illustris, TNG100, EAGLE, and SIMBA cosmological simulations, and find that the properties of their host galaxies vary from one simulation to another. In Illustris and EAGLE, faint AGN are powered by low-mass BHs located in low-mass star-forming galaxies. In TNG100 and SIMBA, they are mostly associated with more massive BHs in quenched massive galaxies. 
We model the X-ray binary populations (XRB) of the simulated galaxies, and find that AGN often dominate the galaxy AGN+XRB hard X-ray luminosity at $z>2$, while XRBs dominate in some simulations at $z<2$. Whether the AGN or XRB emission dominates in star-forming and quenched galaxies depends on the simulations.
These differences in simulations can be used to discriminate between galaxy formation models with future high-resolution X-ray observations. 
We compare the luminosity of simulated faint AGN host galaxies to observations of stacked galaxies from \textit{Chandra}. 
Our comparison indicates that the simulations post-processed with our X-ray modeling tend to overestimate the AGN+XRB X-ray luminosity; luminosity that can be strongly affected by AGN obscuration.
Some simulations reveal clear AGN trends as a function of stellar mass (e.g., galaxy luminosity drop in massive galaxies), which are not apparent in the observations.  
\end{abstract}


\section{Introduction}

Some galaxies 
show a much broader energy distribution than some others. They are powered by active galactic nuclei (AGN), i.e. by supermassive black holes (BHs) accreting efficiently at their centers. Evidence shows that BHs could be present in the center of most massive galaxies around us \citep[][and references therein]{2009ApJ...698..198G,2013ARA&A..51..511K,2020ARA&A..58..257G}.
The population of AGN is constrained by the luminosity function of the AGN, e.g., in X-ray \citep{2015MNRAS.451.1892A,2015MNRAS.453.1946G}. 
However, these constraints from observations do not cover the faint regime defined by hard X-ray (2-10 keV) AGN luminosity of $L_{\rm AGN}\leqslant 10^{42}\,\rm erg/s$.
The properties of the faint AGN population are especially hard to determine because their luminosities are in a regime potentially dominated by X-ray binaries (XRBs). In this paper, we aim at constraining the population of faint AGN, and the relative contribution of the X-ray emission from the AGN and from the XRB population in galaxies of different masses in large-scale cosmological simulations.

X-ray binaries are formed by a compact object, stellar-mass black hole or neutron star, accreting from a normal star. 
In a binary system, the mass is transferred from the star to the compact object via Roche Lobe overflow or stellar wind mass transfer. 
The number and luminosity of XRBs are thought to primarily depend on the host galaxy properties, such as mass and specific star formation rate
\citep[sSFR,][]{2016ApJ...825....7L,2018ApJ...865...43F,2019ApJS..243....3L}.
The galaxy-wide emission from the binary population comes from both high-mass XRBs (HMXBs) and low-mass XRBs (LMXBs).   
The stellar companion in LMXBs is less massive than the compact object, and LMXB lifetime
is longer than 1 Gyr \citep{2010ApJ...724..559L}.
HMXBs are composed of a neutron star or stellar-mass black hole and a more massive stellar companion  
than in LMXBs. As such,  
their lifetime is shorter than 100 Myr. 
The galaxy-wide number of LMXBs 
increases with the 
galaxy stellar mass.  

A scaling of the X-ray emission with the stellar mass of the galaxies and their star formation rate (SFR) was found in several analyses of nearby galaxies \citep[$D\leqslant 50 \, \rm Mpc$, e.g.,][]{2003ChJAS...3..257G,2004AAS...20515501C,2004MNRAS.349..146G,2010ApJ...724..559L, 2011ApJ...729...12B,2012MNRAS.419.2095M, 2012MNRAS.426.1870M,2012A&A...546A..36Z}. 
The most recent scaling relation was derived in \citet{2019ApJS..243....3L}. In this work, galaxies with X-ray measurements from the \textit{Chandra} data are decomposed into spatially resolved maps of SFR and stellar mass, and the XRB population statistics is extracted for sSFR bins. The parameters of the scaling relation functional form are fitted in these sSFR bins (i.e., fits over the pixels of all the galaxies together, and galaxy by galaxy). The result scaling relation is then tested by fitting each galaxy of the sample.
These studies revealed that the total XRB emission of quiescent galaxies is dominated by LMXBs,
for which the emission is proportional to the galaxy mass since LMXBs are long lived. However, the total XRB emission of star-forming galaxies is dominated by young short lived luminous HMXBs 
\citep{Fabbiano:2005pj}, and the higher the galaxy SFR the higher the total XRB emission. The X-ray emission from LMXBs and HMXBs is comparable for sSFR $\approx 10^{-10}\, \rm yr^{-1}$ \citep{2019ApJS..243....3L}.

The X-ray emission of the XRB population could also depend on the metallicity and stellar ages of the galaxies \citep{2013ApJ...776L..31F,2016ApJ...825....7L,2017ApJ...840...39M}.
Studies focusing on more distant and diverse galaxies have found deviations from the empirical relations established in the local Universe. This is the case for samples of low-metallicity galaxies or for samples with a wider range of stellar masses \citep[e.g.,][]{2013HEAD...1330104B,2016HEAD...1540203B,2015A&A...579A..44D,2016cxo..prop.4835B,2016ApJ...817...95T,2010ApJ...721.1523K,2014ApJ...789...52L}.
Finally, an evolution with redshift was found in X-ray stacking analyses \citep[e.g.,][]{2007ApJ...657..681L,2016ApJ...825....7L,2013ApJ...762...45B,2014MNRAS.440L..26K,2017MNRAS.465.3390A}.
The emission from LMXBs per unit stellar mass is thought to be higher at higher redshifts. 
The donor stars 
of the LMXBs have higher masses at higher redshifts and the LMXBs are more luminous \citep{2016ApJ...825....7L}. Moreover, elliptical galaxies with many globular clusters show an excess of LMXBs. Globular clusters can produce LMXBs efficiently through stellar dynamical interactions  \citep{2018ApJ...869...52C,2019ApJS..243....3L}, i.e. by either capture of a stellar object by the second object, or hardening of soft binaries \citep[hard binaries are binaries with bound energy larger than the kinetic energy of the intruding star and stellar encounters involving hard binaries make them harder, see][]{2018ApJ...869...52C}. 
The HMXB emission per unit SFR is higher at higher redshifts. 
This is connected to the lower metallicities of the stellar populations, as observationally shown in \citet{2019ApJ...885...65F} and \citet{2020MNRAS.495..771F}.
The star formation at low metallicities is expected to result in a larger number of massive compact objects and Roche-Lobe overflow binaries because the mass loss of massive stars through stellar winds is less effective \citep{2006MNRAS.370.2079D,2010ApJ...725.1984L,2013ApJ...776L..31F,2016ApJ...825....7L}.

There is evidence for the presence of X-ray AGN in all types of galaxies, including dwarf galaxies \citep[$M_{\star}\leqslant 10^{9.5}\, \rm M_{\odot}$,][]{2016ApJ...817...20M,2018ApJ...863....1C}, where the galaxy-wide X-ray emission is higher than expected from the XRB population. In \citet{2020ApJ...898L..30M}, the authors identify AGN signatures in 37 local dwarf galaxies in the range $\log_{10} M_{\star}/\rm M_{\odot}=8.8-9.5$ using
integral field unit (IFU) spectroscopy
. These AGN are faint, sometimes with off-center X-ray emission
, and have bolometric luminosity in the range $\log_{10}L_{\rm bol}/\rm (erg/s)=38.9-41.4$. BH mass estimates for these AGN could lie in the range $\log_{10}M_{\rm BH}/\rm M_{\odot}=5.5-8$. The presence of AGN in dwarf galaxies could also shed new light on the role of AGN feedback in these galaxies, as studied in simulations \citep{2019MNRAS.484.2047K,2021MNRAS.503.3568K}, and observed with the detections of fast AGN-driven outflows \citep{2019ApJ...884...54M,2020ApJ...905..166L}.

Signatures of faint AGN have been identified in more massive galaxies than dwarf galaxies, particularly with stacking analysis of \textit{Chandra} COSMOS Legacy galaxies \citep{2017xru..conf...88G,2018ApJ...865...43F}. 
The BHs powering these faint AGN can have low accretion rates for different reasons: for example
BH accretion rates regulated by BH feedback, BH accretion rates regulated by the feedback in the environment of the BHs, BHs in quenched galaxies depleted in gas, off-center BHs that do not benefit from the galaxy potential well gas reservoir, and the accretion rate likely depends on BH mass and galaxy properties. Constraining the regime of faint AGN will help to understand the co-evolution and interplay between BHs and their host galaxies. In this study, we will 
investigate in which galaxies faint AGN reside 
in 
several cosmological simulations, and 
whether the low accretion rates onto the simulated faint AGN can have different causes (e.g., impact of AGN feedback, absence of cold gas), such as outlined above.

The 
galaxy-wide emission of the XRB population could depend on the SFR of the host galaxies, and in a similar way 
correlations between AGN activity and the SFR of their galaxies could exist. Even for the regime of the bright AGN (e.g., $L_{\rm AGN}\geqslant 10^{42}\, \rm erg/s$), the correlations between AGN activity and the properties of galaxies are still not clear, especially in terms of stellar masses, SFR and sSFR \citep[e.g.,][]{2019MNRAS.484.4360A}. 
For bright AGN of $L_{\rm AGN}\geqslant 10^{44}\, \rm erg/s$ in massive galaxies, there is evidence for a correlation between AGN luminosity and SFR \citep{2008ApJ...684..853L,2011MNRAS.416...13B,2012ApJ...749L..25M,2012A&A...545A..45R}.
There are very few observational constraints in the faint AGN regime, and therefore, in order to theoretically investigate possible dependences/correlations of our results with galaxy properties, we will split the simulated galaxies in starburst, star-forming, or quiescent galaxies.

We are also particularly interested in how we can compare the predictions from cosmological simulations to current and future observations to disentangle simulation sub-grid models and improve them. In this work, we compare the total X-ray luminosity of the simulated galaxies hosting faint AGN to observational constraints of stacked galaxies \citep{2018ApJ...865...43F}.
The high sensitivity of the upcoming X-ray mission Athena \citep{2013arXiv1306.2307N} and the NASA concept missions Lynx \citep{2018SPIE10699E..0NG} and AXIS \citep{2019BAAS...51g.107M} 
will allow us to investigate the properties of AGN up to high redshift, and will provide us with new constraints on the faint AGN population. Our work paves the way for future investigations on faint AGN, a goal aligned with the upcoming X-ray missions.
As in all studies on the AGN population, obscuration is critical and a major open question. The attenuation of radiation from intervening gas and dust can affect the contribution of both the AGN and the X-ray binaries to the observed galaxy X-ray luminosity. We will build several models for the obscuration of the faint AGN to account for this.

In this paper, we use the four Illustris, TNG100, EAGLE, and SIMBA cosmological hydrodynamical simulations of $\geqslant 100\,$ comoving $\rm Mpc$ (cMpc) box side length. 
For all these simulations, we model both the AGN luminosity and the XRB population of the simulated galaxies using several empirical scaling relations. 
We describe the physical models of the simulations, as well as our post-processing modeling of AGN and XRB luminosity in Section 2.
In Section 3, we analyze the AGN population of the simulations and in which galaxies they live. 
In order to derive the total X-ray luminosity of the simulated galaxies, we first analyze in Section 4 the properties of their AGN as a function of their host galaxy's mass and SFR. 
In Sections 5 and 6, we derive the relative contribution of the AGN and the XRB population to the galaxy total hard X-ray luminosity and we compare these results to recent findings in observations of stacked galaxies from the \textit{Chandra} COSMOS Legacy survey \citep{2018ApJ...865...43F}.

\section{Cosmological simulations and methods}
In this section, we describe the Illustris, TNG100, EAGLE, and SIMBA cosmological simulations, 
and our method to compute AGN luminosity, and the X-ray luminosity of the XRB population of the simulated galaxies. 
We also describe our method to divide the simulated galaxy in three samples with  different sSFR.

\subsection{Cosmological simulations}
In the following, we briefly summarize the BH subgrid models, i.e. seeding, accretion and feedback, of the Illustris, TNG100, EAGLE and SIMBA large-scale cosmological simulations. A more detailed comparison of the modeling of these simulations is presented in \citet{2021MNRAS.503.1940H}. Several other aspects differentiate these simulations, for example the presence of magnetic fields in the TNG100 simulation \citep{2018MNRAS.473.4077P}, the single mode AGN feedback of EAGLE \citep{2015MNRAS.446..521S}, the two mode accretion model of SIMBA \citep{2019MNRAS.486.2827D}. The detailed models can be found in the references given below.

\subsubsection{Illustris}
The Illustris hydrodynamical simulation \citep{2014Natur.509..177V,2014MNRAS.444.1518V,2014MNRAS.445..175G,2015MNRAS.452..575S} simultaneously follows the evolution of dark matter (DM) and baryonic matter, in a volume of $(106.5\, \rm cMpc)^3$. 
The Illustris simulation data is publicly available \citep{2015A&C....13...12N}.
The simulation produces a mix of galaxy morphologies in broad agreement with observations \citep{2014Natur.509..177V}.
The equations of gravity and hydrodynamics are evolved with the moving-mesh code AREPO \citep{2010MNRAS.401..791S}.
The galaxy formation model includes gas cooling, star formation, supernova (SN) feedback, and the physics of BHs.
BHs are seeded in halos with halo mass $\geqslant 7.1 \times 10^{10} \, $M$_{\odot}$ with seed mass $M_{\rm seed}=1.4 \times 10^5\, \rm M_{\odot}$. BHs can grow in mass with gas accretion and BH mergers. 
The BH accretion is modeled with the Bondi-Hoyle-Lyttleton formalism, as:
\begin{equation}
\dot{M}_{\rm BH}=\min\biggl[\alpha \dot{M}_{\rm Bondi}, \dot{M}_{\rm Edd}\biggr],
\label{eq:BH_accretion_rate}
\end{equation}
with
\begin{equation}
\dot{M}_{\rm Bondi}=\frac{4\pi G^2 M_{\rm BH}^2 \rho}{(c_{\rm s}^2 + v_{\rm BH}^2)^{3/2}},
\end{equation}
with $\rho$ and $c_{\rm s}$ the density and the sound speed of the surrounding gas, and $v_{\rm BH}$ the velocity of the BH relative to the gas, $G$ is the gravitational constant. The boost factor $\alpha = 100$ accounts for the unresolved gas around the AGN, which tend to underestimate the accretion rates.  
The boost factor is set to produce a BH population that agrees with the $M_{\rm BH}-M_{\star}$ diagram at $z=0$.
A re-positioning scheme for BH sink particles is used which ties them to the local minimum gravitational potential. 
$M_{\rm Edd}$ is the Eddington accretion rate of the BH:
\begin{equation}
\dot{M}_{\rm Edd}=\frac{4\pi G   m_{\rm p}}{\sigma_{T} \, c\,  \epsilon_{\rm r}} M_{\rm BH},
\end{equation}
with $m_{p}$ the proton mass, $\sigma_{\rm T}$ the Thomson cross section, $c$ the speed of light and $\epsilon_{\rm r}$ the radiative efficiency which is set to 0.2 in the simulation. 
The Eddington ratio is defined as $f_{\rm Edd}=\dot{M}_{\rm BH}/\dot{M}_{\rm Edd}$. 
The feedback from the AGN depends on the BH accretion rate and can operate in {\it high-accretion mode} ($f_{\rm Edd}>0.05$) or {\it low-accretion mode} ($f_{\rm Edd}<0.05$). 
The low-accretion mode model injects highly bursty thermal energy into large 'bubbles' ($\sim 50$ kpc) which are displaced away from the central galaxy. 
In the high-accretion mode, thermal energy is continuously injected into the surrounding gas. Moreover, radiative AGN feedback that modifies the ionisation state of the surrounding gas is included as well \citep{2013MNRAS.436.3031V}.

\subsubsection{TNG100}
The TNG100 simulation builds directly on the Illustris framework and has a volume of $(110.7\, \rm cMpc)^{3}$ \citep{2018MNRAS.477.1206N,2018MNRAS.475..648P,2018MNRAS.475..624N,2018MNRAS.480.5113M,2018MNRAS.475..676S,2019ComAC...6....2N}. The simulation includes magnetic fields. 
The BH seed mass is $M_{\rm seed}=1.1 \times 10^6\, \rm M_{\odot}$. The BH seed mass was increased in comparison with the Illustris seed mass by nearly one order of magnitude \citep{2018MNRAS.473.4077P}. 
The BH accretion rate is described by the Bondi-Hoyle-Lyttleton accretion rate limited by the Eddington accretion rate \citep{2017MNRAS.465.3291W}. However, in the TNG model, the additional boost factor $\alpha$ from Eq.~\ref{eq:BH_accretion_rate} is removed.
The effective sound speed $c_{s}^2=c_{\rm s,therm}^2+(B^2/4\pi\rho)$ includes the thermal and the magnetic signal propagation. 
The transition between the AGN feedback modes in the TNG model depends on the BH accretion rate and BH mass. 
A BH is assumed to be in the high accretion state as long as \citep{2017MNRAS.465.3291W}:
\begin{equation}
f_{\rm Edd} \geq \min\biggl[ 2 \times 10^{-3} \biggl(\frac{M_{\rm BH}}{10^8 {\rm M_{\odot}}}\biggr)^{2}, 0.1\biggr].
\label{Eq:chi_TNG}
\end{equation}
The low accretion feedback is modeled as kinetic outflows from the BHs \citep{2017MNRAS.465.3291W}, while the high accretion mode is modeled as injection of thermal energy in the BH surroundings \citep{2017MNRAS.465.3291W}.

\subsubsection{EAGLE}
The EAGLE simulation \citep{2015MNRAS.446..521S,2015MNRAS.450.1937C} has a volume of $(100\, \rm cMpc)^{3}$, and was run with the code {\sc ANARCHY}. This modified version of GADGET3 \citep{2005MNRAS.364.1105S} is based on the Smoothed Particle Hydrodynamics (SPH) method.
The simulation includes subgrid models for radiative cooling, star formation, stellar mass loss, metal enrichment and energy feedback from star formation \citep{2015MNRAS.446..521S}. 
The BH seeding mass is $M_{\rm seed}=1.48\times 10^{5}\, \rm M_{\odot}$ and BHs are seeded in halos with a mass of at least $M_{\rm h}=1.48\times 10^{10}\, \rm M_{\odot}$.
The accretion rate is determined by the Bondi-Hoyle-Lyttleton model \citep{2015MNRAS.454.1038R,2016MNRAS.462..190R}:
\begin{eqnarray}
\dot{M}_{\rm acc}=\min (
\dot{M}_{\rm Bondi}\times \min \left((c_{\rm s}/V_{\rm \Phi})^{3}/C_{\rm visc},1\right),
\dot{M}_{\rm Edd}),
\end{eqnarray}
with $\dot{M}_{\rm Bondi}$ and $\dot{M}_{\rm Edd}$ defined as in Eq. (2) and (3) and $\dot{M}_{\rm BH}=(1-\epsilon_r)\dot{M}_{\rm acc}$. The radiative efficiency is $\epsilon_r = 0.1$.
Here, $c_{\rm s}$ represents the sound speed of the surrounding gas, $V_{\Phi}$ is the rotation speed of the gas around the BH and $C_{\rm visc}$ is a free parameter. It is related to the viscosity of the accretion disk. The correction factor $\min \left((c_{\rm s}/V_{\rm \Phi})^{3}/C_{\rm visc},1\right)$ multiplied with the Bondi rate accounts for a lower accretion rate for gas with angular momentum. In that case, the accretion is not spherically symmetric and proceeds through an accretion disk \citep{2015MNRAS.446..521S}.
The simulation uses a single-mode AGN feedback model.
A fraction of the accreted gas onto the BHs is stochastically injected in the surroundings as thermal energy \citep{2015MNRAS.446..521S}.

\subsubsection{SIMBA}
SIMBA \citep{2019MNRAS.486.2827D} has a volume of $(147\, \rm cMpc)^{3}$ and builds on its predecessor Mufasa \citep{2016MNRAS.462.3265D}, using the code GIZMO
\citep{2015MNRAS.450...53H,2017arXiv171201294H} in its ``Meshless Finite Mass'' hydrodynamics mode. 
The BH seeding mass is $M_{\rm seed}=1.4\times 10^{4}\, \rm M_{\odot}$ and BHs are seeded in galaxies with $M_{\star}\sim 10^{9.5}\, \rm M_{\odot}$. This is motivated by FIRE simulations showing that stellar feedback strongly suppresses BH growth in lower mass galaxies \citep{2017MNRAS.472L.109A,2020arXiv200712185C}, in agreement with other models \citep[e.g.,][]{2017MNRAS.468.3935H}. Because of the seeding, we will not draw any conclusion on BH properties in galaxies with $M_{\star}<10^{9.5}\, \rm M_{\odot}$ for the SIMBA simulation.
The BH accretion rate is given by:
\begin{eqnarray}
\begin{aligned}
\dot{M}_{\rm BH}=(1-\epsilon_{\rm r})\times \\
\left[ \min(\dot{M}_{\rm Bondi},\dot{M}_{\rm Edd}) 
+ \min(\dot{M}_{\rm Torque},3\, \dot{M}_{\rm Edd})
\right],
\end{aligned}
\end{eqnarray}
with $\dot{M}_{\rm Bondi}$ and $\dot{M}_{\rm Edd}$ defined as in Eq. (2) and (3).
The radiative efficiency is $\epsilon_{\rm r}=0.1$. $\dot{M}_{\rm Torque}$ describes the gas inflow rate  driven by gravitational instabilities from the scale of the galaxy to the accretion disk of the BH \citep{2017MNRAS.464.2840A,2011MNRAS.415.1027H}. This accretion is only evaluated for the cold gas ($T<10^{5}\, \rm K$). For the hot gas ($T>10^{5}\, \rm K$), the Bondi-Hoyle-Lyttleton model is applied \citep{2019MNRAS.486.2827D}.
AGN feedback in SIMBA is modeled as an injection of kinetic energy following a two-mode approach, with high mass loading outflows in the radiative ``quasar'' mode and lower mass loading but faster outflows at low Eddington ratios in the jet mode \citep{2019MNRAS.486.2827D}. 
BHs begin to transition into jet mode for $f_{\rm Edd} < 0.2$, reaching full speed jets at  $f_{\rm Edd} \lesssim 0.02$.  Also, X-ray feedback is included for galaxies with $M_{\rm gas}<0.2 M_{\star}$ and full speed jets, following the implementation of \citet{2012ApJ...754..125C}.  See \citet{2019MNRAS.487.5764T,2021MNRAS.503.3492T} and \citet{2021MNRAS.503.1940H} for previous studies of BHs in SIMBA.

\subsection{Computation of the AGN luminosity and obscuration of the simulated BHs}
\subsubsection{AGN luminosity}
We compute in post-processing the luminosity of the BHs, following the model of \cite{Churazov2005}, i.e. explicitly distinguishing radiatively efficient and radiatively inefficient AGN. 
The bolometric luminosity of radiatively efficient BHs ($f_{\rm Edd}>0.1$) is given by:
\begin{eqnarray}
L_{\rm bol}=\frac{\epsilon_{\rm r}}{1-\epsilon_{\rm r}} \dot{M}_{\rm BH} c^{2}.
\label{eq:lum}
\end{eqnarray}
BHs with small Eddington ratios ($f_{\rm Edd}\leqslant 0.1$) are considered radiatively inefficient and their bolometric luminosities are given by \citep{2021MNRAS.503.1940H}:
\begin{eqnarray}
L_{\rm bol}=0.1 L_{\rm Edd} (10 f_{\rm Edd})^{2}=(10 f_{\rm Edd}) \epsilon_{\rm r} \dot{M}_{\rm BH} c^{2}.
\end{eqnarray}
We compute the hard (2-10 keV) X-ray luminosities of the AGN with the bolometric correction of \cite{Hop_bol_2007}:
\begin{eqnarray}
\log_{10} L_{\rm 2-10\, keV,\odot}=\log_{10} L_{\rm bol,\odot} - \log_{10} \rm BC,
\end{eqnarray}
with
\begin{eqnarray}
{\rm BC} = 10.83 \left(\frac{L_{\rm bol,\odot}}{10^{10}\, \rm L_{\odot}} \right)^{0.28} + 6.08 \left(\frac{L_{\rm bol,\odot}}{10^{10}\, \rm L_{\odot}} \right)^{-0.020}.
\end{eqnarray}
\noindent In the following, we present the luminosity of the AGN in erg/s.

\begin{figure}
\centering
\includegraphics[scale=0.53]{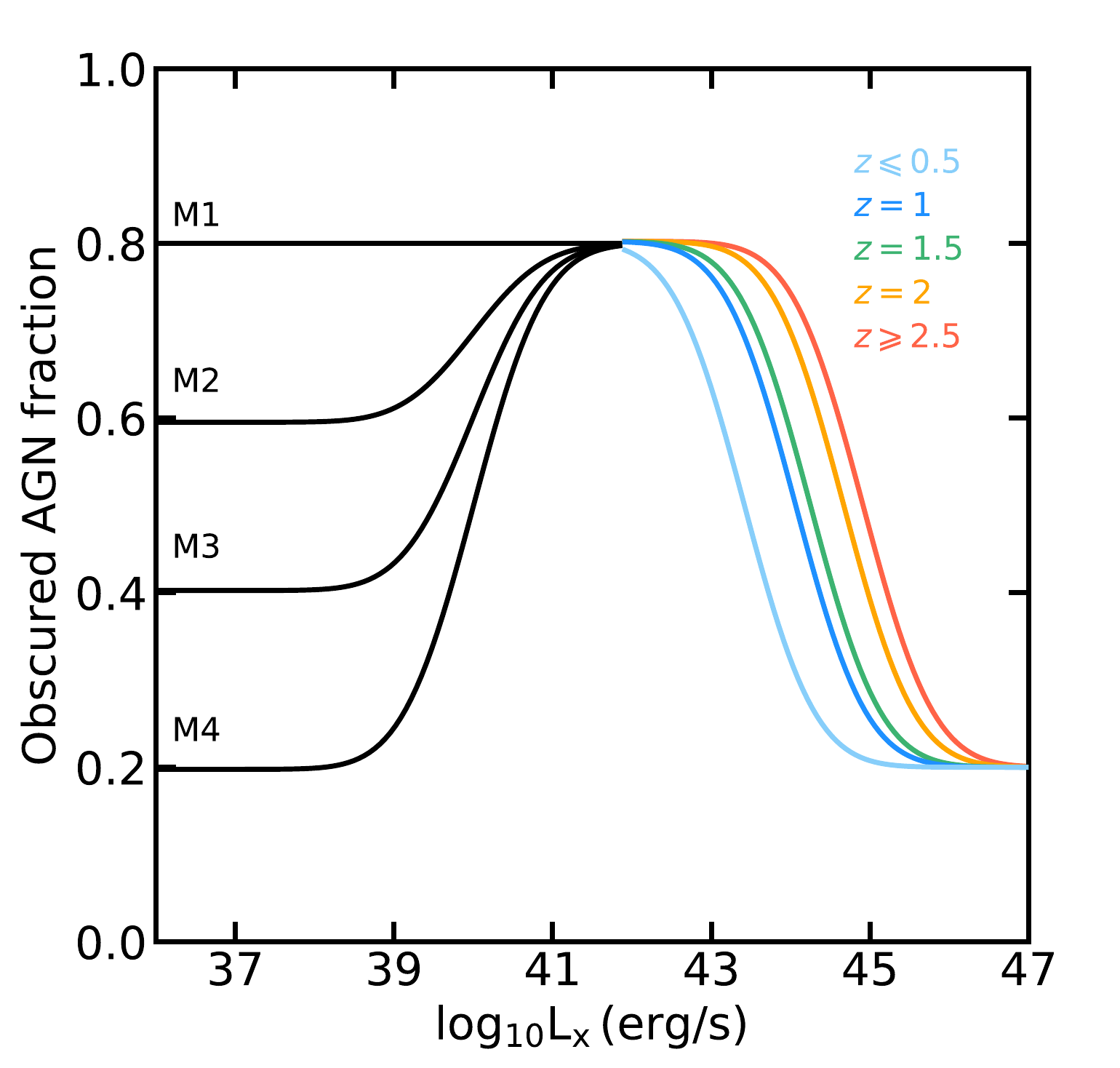}
\caption{Our theoretical models for the fraction of obscured AGN. For relatively bright AGN with $L_{\rm AGN}$ $\geqslant$ $10^{41}\rm erg/s$, the obscured AGN fractions depends on $L_{\rm AGN}$ and redshift \citep{2019MNRAS.484.4413H}; the models are based on observational constraints \citep{2014MNRAS.437.3550M,2014ApJ...786..104U}. 
Lacking observational constraints for fainter AGN ($L_{\rm AGN}$ $\leqslant$ $10^{41}\rm erg/s$), we build four different redshift-indepedent models to cover a broad range of possibilities (M1, M2, M3, M4).}
\label{fig:obscuration}
\end{figure}

\subsubsection{AGN obscuration}
AGN can be obscured by gas and/or dust, either physically close to the AGN and at larger scales within the host galaxies. We generally refer to two types of obscured AGN, either Compton-thin AGN with column densities of $N_{\rm H}\geqslant 10^{22}\, \rm cm^{-2}$, or more heavily obscured Compton-thick AGN with $N_{\rm H}\geqslant 10^{24}\, \rm cm^{-2}$. While the hard X-ray band suffers less from obscuration than softer bands, the observed luminosity of the AGN could still be attenuated.
The obscuration might depend on the intrinsic AGN luminosity and the redshift: there is more gas present at higher redshifts and therefore, the obscuration could be stronger \citep{2014MNRAS.437.3550M,2014A&A...562A..67G,2014MNRAS.445.3557V,2016MNRAS.463..348V}.

We build several models for the fraction of obscured AGN and apply them to the simulations. Our models depend on redshift and hard X-ray (2-10 keV) AGN luminosity for AGN with $L_{\rm AGN}\geqslant 10^{41} \,\rm erg/s$ \citep[][for more details]{2019MNRAS.484.4413H}, and only on AGN luminosity for fainter AGN. Obscuration could in principle also depend on galaxy SFR, a parameter that we do not consider here.

For the AGN with $L_{\rm x}\geqslant 10^{41}\rm \, erg/s$, we build the models based on the observational constraints of \citet{2014MNRAS.437.3550M} and \citet{2014ApJ...786..104U}: the colored lines of Fig.~\ref{fig:obscuration} are a fit to the observations. 
These constraints assume that the obscured AGN mostly include Compton-thin AGN, but that the presence of Compton-thick AGN can not be ruled out. Here we assume that our models based on these observations include all types of obscured AGN, and we apply them on the full AGN simulated samples.
Since we are lacking observational constraints on the population of faint AGN, and even more so on the fraction of those that are obscured, we build four models which cover a broad range of possible fractions of obscured AGN (Fig.~\ref{fig:obscuration}).
In our model M1, a large fraction ($80\%$) of the faint AGN with $L_{\rm AGN}\leqslant 10^{41}\, \rm erg/s$ are obscured, and this fraction does not depend on AGN luminosity. For M2, M3, and M4, the fraction does depend on $L_{\rm AGN}$ and can progressively go down to $60\%$ (M2), $40\%$ (M3), or $20\%$ (M4) of AGN being obscured.  
Our choice of low fractions of obscured AGN among the faint AGN with $L_{\rm AGN}\leqslant 10^{40} \,\rm erg/s$ is motivated by the fact that in simulations if these BHs do not have high accretion rates, their close surrounding is depleted of gas. Consequently, if the gas reservoir is small there may be little room for the gas/dust to obscure the emission from the AGN.
Since this is highly uncertain, we also have our model M1 with a high fraction of obscured AGN to bracket the range of possibilities.
In the following we apply our obscuration models only when specified in the text and figures.

\begin{table*}
    \centering
    \caption{Parametrizations from the literature of the hard X-ray luminosity of the XRB population in galaxies, as a function of their total stellar mass $M_{\star}\, (\rm M_{\odot})$, SFR, and redshift $z$. In this paper, we mostly use the scaling relation from \citet{2019ApJS..243....3L}.
    The M$_{\star}$, SFR an $z$ ranges describe the sample of galaxies used to derive the empirical relations.}
    \scalebox{0.8}{
    \begin{tabular}{lllll}
    \hline
    References & Scaling relations & $M_{\star}$ ranges & SFR ranges&Redshift ranges\\
    \hline
    \hline  
    \\
         L10 \citet{2010ApJ...724..559L} &  $L_{\rm XRB} = 10^{28.96} M_{\star} + 10^{39.21} \rm SFR$ & $M_{\star}=10^{10-11}\, \rm M_{\odot}$ &SFR$=10^{1-2} $M$_{\odot}$/yr& $z\sim 0$\\
    \\
    \hline 
    \\
         L16 \citet{2016ApJ...825....7L}&  $L_{\rm XRB} = 10^{29.37} (1+\rm z)^{2.0} M_{\star} + 10^{39.28} (1+ \rm z)^{1.3} \rm SFR$ & $M_{\star}=10^{9-12}\, \rm M_{\odot}$ &SFR$=10^{-2-3} $M$_{\odot}$/yr& $z=0-4$ \\
    \\
    \hline 
    \\
         L19 \citet{2019ApJS..243....3L}& $L_{\rm XRB} = 10^{29.15} (1+\rm z)^{2.0} M_{\star} + 10^{39.73} (1+ \rm z)^{1.3} \rm SFR$ & $M_{\star}=10^{9-11.5}\, \rm M_{\odot}$ &SFR$=10^{-2-1} $M$_{\odot}$/yr& $z\sim 0$\\
    \\
    \hline 
    \\
         A17 \citet{2017MNRAS.465.3390A}& $L_{\rm XRB} = 10^{28.81} (1+z)^{3.9} M_{\star} + 10^{39.50} (1+z)^{0.7} \rm SFR^{0.86}$ & $M_{\star}=10^{8.5-11.5}\, \rm M_{\odot}$ &SFR$=10^{-1-3} $M$_{\odot}$/yr& $z=0.1-4$\\
    \\
    \hline 
    \\
         F18 \citet{2018ApJ...865...43F}& $L_{\rm XRB} = 10^{29.98} (1+z)^{0.62} M_{\star} + 10^{39.78} (1+z)^{0.2} \rm SFR^{0.84}$ & $M_{\star}=10^{9.5-11.5}\, \rm M_{\odot}$ &SFR$=10^{-1-3} $M$_{\odot}$/yr& $z=0.1-5$\\
    \\     
    \hline  
  
    \end{tabular}
    }
    \label{tab:XRB_relations}
\end{table*}

\subsection{Computation of the XRB population luminosity}

We parametrize the X-ray emission ($2-10\, \rm keV$ band) from the XRB population of the simulations (which is not modeled in the simulations), including both LMXBs and HMXBs.
To do so, we employ XRB emission scaling relations that have been derived from observational samples. These relations provide the luminosity of the XRB population for a given galaxy as a function of its SFR, its stellar mass, and for some of them its redshift:
\begin{eqnarray}
L_{\rm XRB}=\alpha_{\rm LMXB} \left(1+z \right)^{\gamma} M_{\star}+\beta_{\rm HMXB} \left(1+z \right)^{\delta} \rm SFR.
\label{eq:xrb_relation}
\end{eqnarray}
For example, the parameters $\log_{10}\alpha_{\rm LMXB}=29.37\pm 0.15\, \rm erg/s/M_{\odot}$, $\log_{10}\beta_{\rm HMXB}=39.28\pm 0.05\, \rm erg/s/(M_{\odot}/yr)$, $\gamma=2.03\pm 0.60$, $\delta=1.31\pm0.13$ have been described, e.g., in \citet{2016ApJ...825....7L,2019ApJS..243....3L}.
The relations described in \citet{2017MNRAS.465.3390A} and \citet{2018ApJ...865...43F} use a power law of $\rm SFR^{0.86}$ and $\rm SFR^{0.84}$, respectively.
While it has been shown that the metallicity of the galaxies can play a role in the amplitude of the XRB emission, especially in some specific regimes of stellar mass and redshift, we do not include any metallicity dependence in our analysis below.

\begin{figure}
\centering
\includegraphics[scale=0.87]{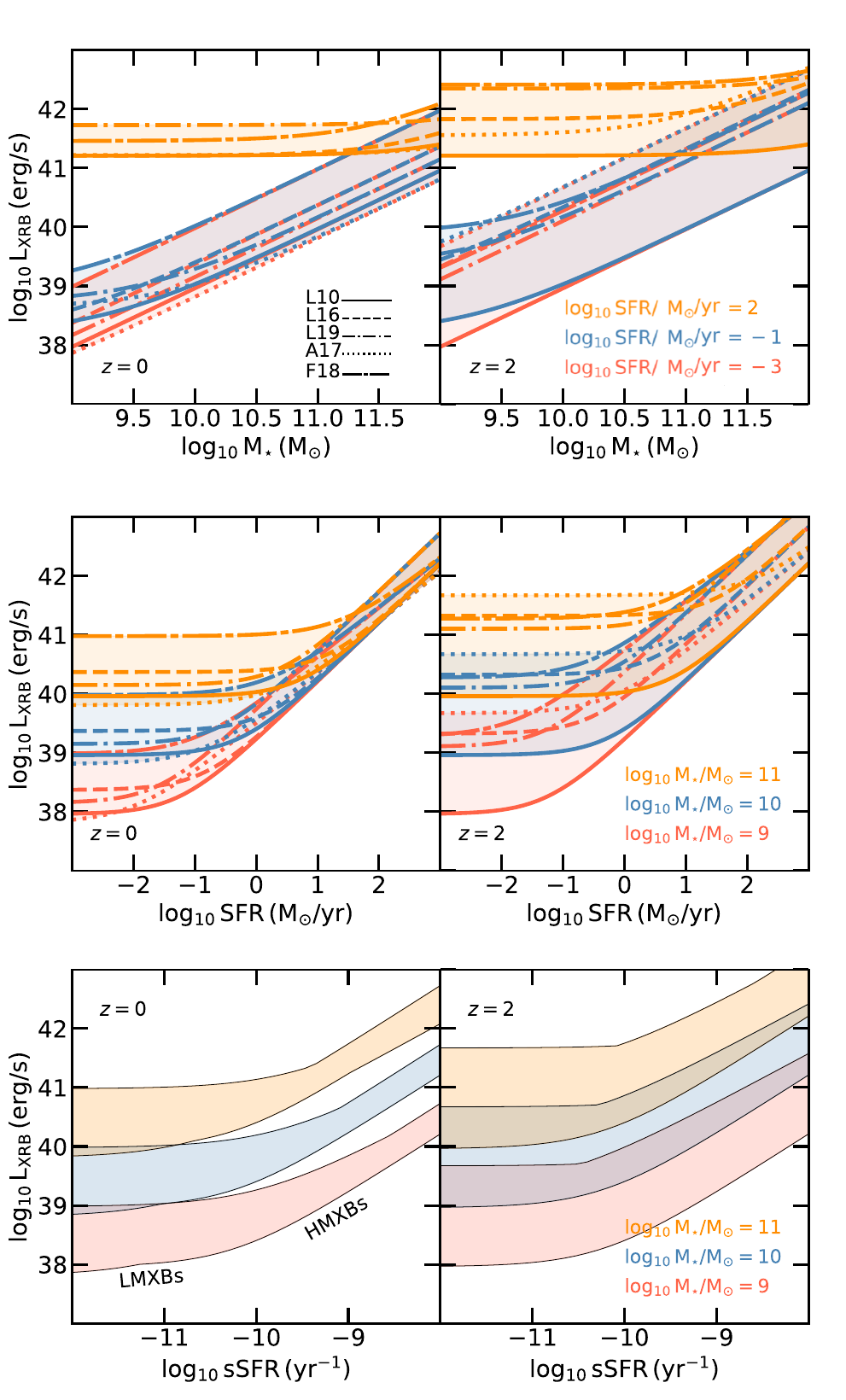}
\caption{Comparison of the XRB scaling relations from Table~\ref{tab:XRB_relations} \citep{2010ApJ...724..559L,2016ApJ...825....7L,2019ApJS..243....3L,2017MNRAS.465.3390A,2018ApJ...865...43F}, at $z=0$ (left panels) and $z=2$ (right panels). 
The difference between the models can be up to one order of magnitude in hard X-ray (2-10 keV) luminosity at $z=0$, and more at higher redshift. 
The XRB luminosity increases with $M_{\star}$ (at fixed SFR), and with SFR (at fixed $M_{\star}$), and with redshift for most of the models.
LMXBs dominate the hard X-ray (2-10 keV) luminosity of the galaxies with $\rm \log_{10} \, sSFR/yr\leqslant -10.5$. For higher sSFR, the X-ray emission is dominated by HMXRs.}
\label{fig:LXRB_theory}
\end{figure}

To study the impact of the modeling of the XRB luminosity, we use five different empirical relations \citep{2010ApJ...724..559L,2016ApJ...825....7L,2019ApJS..243....3L,2017MNRAS.465.3390A,2018ApJ...865...43F}. We report these relations in Table~\ref{tab:XRB_relations}. We compare the scaling relations in Fig.~\ref{fig:LXRB_theory}, as a function of galaxy stellar mass (top panels), SFR (middle panels), and sSFR (bottom panels). The left panels show redshift $z=0$, and the right panels $z=2$.
The difference between the different models can be up to one order of magnitude in luminosity at $z=0$, and more than an order of magnitude at higher redshift. 
At fixed SFR, the luminosity of the XRB population increases with the stellar mass of galaxies. We note that for galaxies with high SFR of $\rm \log_{10} \, SFR/(M_{\odot}/yr)\sim 2$ the luminosity is almost constant with stellar mass, for all the scaling relations studied here. The XRB population luminosity also increases with SFR, at fixed galaxy stellar mass.
The model of \citet{2010ApJ...724..559L} does not evolve with redshift. For the other models, the normalization of the scaling relations increases with increasing redshift. 
In general, the models of \citet{2019ApJS..243....3L} and \citet{2018ApJ...865...43F} provide the highest normalizations of the $L_{\rm XRB}$ relation at $z=0$. At higher redshift, the model of \citet{2017MNRAS.465.3390A} also provides high luminosities.
For low sSFR galaxies with $\rm \log_{10} \, sSFR/yr<-10.5$, the XRB population luminosity is dominated by LMXBs whose luminosity depends on the stellar mass. But for galaxies with higher sSFR of $\rm \log_{10} \, sSFR/yr>-10.5$, the XRB luminosity is dominated by HMXBs whose luminosity scales with the SFR \citep{2019ApJS..243....3L}.

\begin{figure*}
\centering
\includegraphics[scale=0.46]{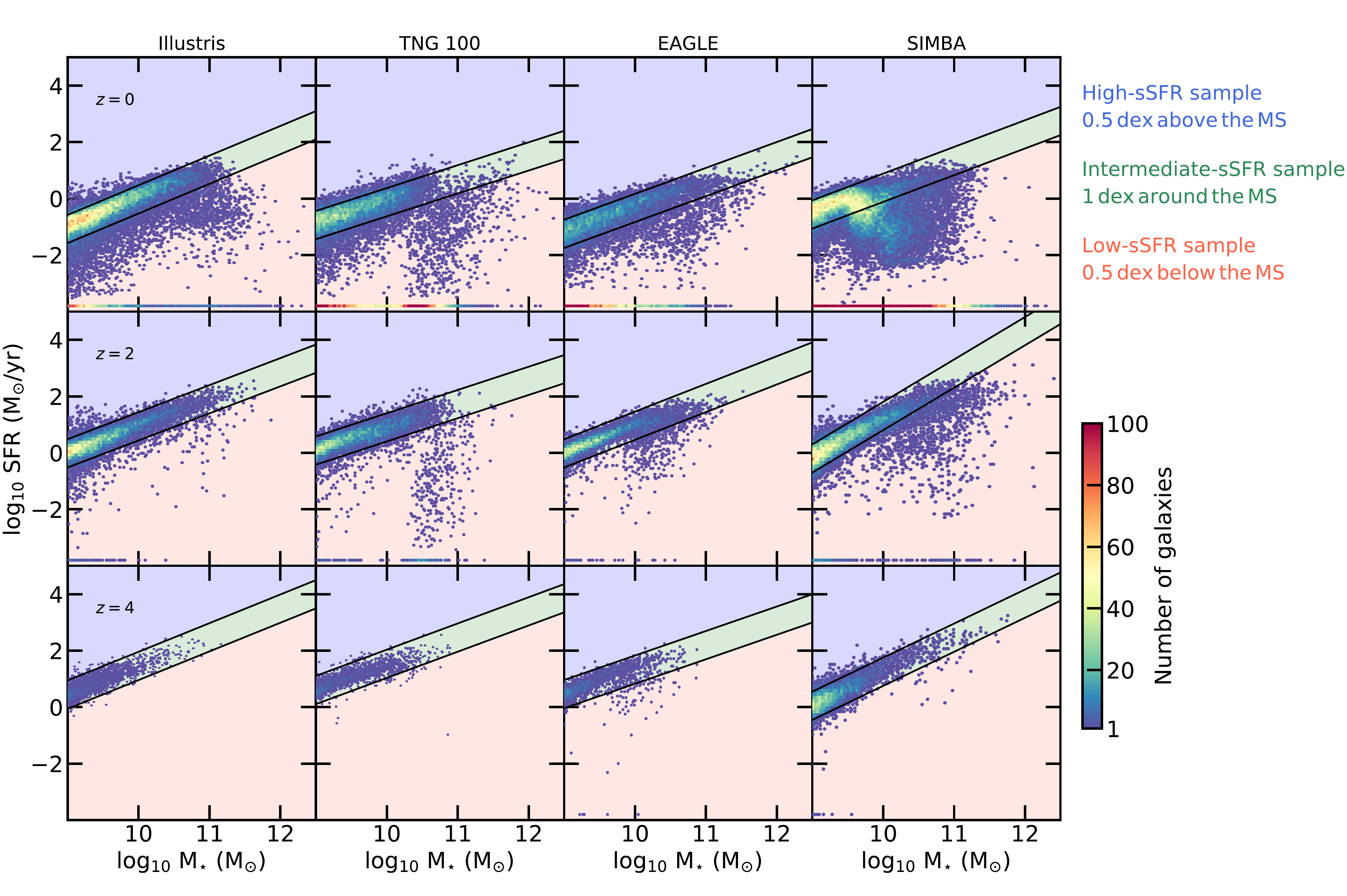}
\caption{$\rm SFR-M_{\star}$ plane with hexabins color coded by the number of galaxies in the bins.
Galaxies with SFR $\leqslant 10^{-4} \rm M_{\odot}$/yr are shown with this value. 
We define three samples of galaxies: the {\it high-sSFR} sample with galaxies 0.5 dex above the star-forming main sequence, the {\it intermediate-sSFR} sample with galaxies on the star-forming sequence, and the {\it low-sSFR} sample with galaxies below 0.5 dex of the main sequence. The {\it high-sSFR}, {\it intermediate-sSFR} and {\it low-sSFR} samples are respectively composed of starburst, main-sequence and quiescent galaxies.
}
\label{fig:M_starVsSFR_Illustris}
\end{figure*}

\subsection{Galaxy star-forming main sequence and sSFR galaxy samples}
In this paper, we will present our results as a function of galaxy properties, i.e., their stellar mass and SFR. In the following, we define the simulation star-forming main sequence and build three different galaxy samples from the simulations.
\subsubsection{Galaxy star-forming main sequence}
We show the $\rm SFR-M_{\star}$ relation for galaxies from the Illustris, TNG100, EAGLE, and SIMBA simulations in Fig.~\ref{fig:M_starVsSFR_Illustris}. 
We define the star-forming main sequence of the simulations as the mean SFR in fixed width bins (same width for all the simulations) of galaxy stellar mass for $10^{9}\leqslant M_{\star}\leqslant 10^{10}\, \rm M_{\odot}$\footnote{We use M$_{\star}\leqslant 10^{9.5}\, \rm M_{\odot}$ for SIMBA at $z=0$ because of the presence of galaxies with reduced SFR at lower stellar mass than in the other simulations.} \citep[see also][]{2014MNRAS.444.1518V,2014MNRAS.445..175G,2015MNRAS.447.3548S,2015MNRAS.450.4486F,2016MNRAS.462.2559B,2017ApJ...844..170T,2019MNRAS.485.4817D,2019MNRAS.484..915M,2019MNRAS.486.2827D}.
We purposely exclude more massive galaxies to compute the main sequence as many of these galaxies are quenched.
The main sequence can be defined as a power law:
\begin{eqnarray}
\log_{10}[ {\rm SFR_{MS}}/\rm (M_{\odot}/yr)]=\alpha + \beta \log_{10} \left(M_{\star}/\rm M_{\odot}\right),
\end{eqnarray}
with $\rm SFR_{MS}$ the SFR of the galaxies on the main sequence, in $\rm M_{\odot}/yr$. The parameters $\alpha$ and $\beta$ are computed for each simulation and redshift and are given in Table~\ref{alpha_beta_table}. 
The simulations present some differences in the galaxy population in Fig.~\ref{fig:M_starVsSFR_Illustris}, such as the exact normalization and slope of their star-forming main sequence, or the formation of the population of quenched galaxies, i.e., galaxies with low or null SFR. We note that the SIMBA simulation has the steepest main sequence, but overall we find a good agreement between the simulations for the normalization of the star-forming main sequence. 
In observations there is, at the moment, no real consensus on the exact slope and normalization of the star-forming main sequence, and the parameters depends on the observed samples. The slope of the star-forming main sequence of observed galaxies can be slightly shallower \citep{2019MNRAS.484.4413H,2019ApJ...872..160H}.

When looking at the population of quenched galaxies, we can identify some features of the AGN feedback modeling such as the sharp decrease of the SFR in TNG100 galaxies of $M_{\star}\geqslant \rm a \, few\, 10^{10}\, M_{\odot}$ \citep[but still in broad agreement with observational data, see][]{2021MNRAS.tmp.1755D}. This shows the transition at $M_{\rm BH}\sim {\rm a \, few\, } 10^{8}\, \rm M_{\odot}$ between the high accretion state and the low accretion state of the AGN feedback modeling (see Eq. \ref{Eq:chi_TNG}).
The low accretion mode feedback is efficient, and both regulates the BHs and quenches their host galaxies \citep{2018MNRAS.479.4056W,2019MNRAS.484.4413H}.
Towards $z=0$, we also note a strong decrease of the SFR of a large fraction of the galaxies with $M_{\star}\leqslant 10^{9.5}\, \rm M_{\odot}$ in all the simulations, which reflects the quenching of satellite galaxies \citep{2021MNRAS.500.4004D}. \\

\subsubsection{Defining three galaxy samples}
For the purpose of our analysis, we divide the galaxies into three subsets, which are simulation- and redshift-dependent.
The {\it high-sSFR} subset (blue background in Fig.~\ref{fig:M_starVsSFR_Illustris}) consists of starburst galaxies with SFR higher than half a dex above the main sequence.
The {\it intermediate-sSFR} subset (green background in Fig.~\ref{fig:M_starVsSFR_Illustris}) includes star-forming galaxies on the main sequence, i.e., within 1 dex. 
Finally, the {\it low-sSFR} subset (red background in Fig.~\ref{fig:M_starVsSFR_Illustris}) includes all galaxies with SFR below half a dex of the main sequence, i.e. quiescent galaxies or galaxies on their way to quiescence.

We find that a low fraction of the galaxies of $M_{\star}\geqslant 10^{9}\, \rm M_{\odot}$ are in the {\it high-sSFR} sample, typically less than $5\%$ in all the simulations and at all redshifts.
The {\it intermediate-sSFR} represents the largest fraction of galaxies ($\leqslant 90\%$, $z=4$) at high redshift in all the simulations. With time, the number of galaxies in the {\it intermediate-sSFR} samples decreases, for all the simulations.
The percentage of galaxies in the {\it low-sSFR} samples increases with time for all the simulations, as more and more galaxies quench. 
Most of the galaxies in the {\it low-sSFR} samples are massive with $M_{\star}\gtrsim 10^{10}\, \rm M_{\odot}$ and have reduced SFR because of AGN feedback. For example, in TNG there is a sharp decrease of the SFR in galaxies of $M_{\star}\gtrsim 10^{10.5}\, \rm M_{\odot}$ due to the kinetic low-accretion mode of the AGN feedback model.
Still, some of the galaxies present in the {\it low-sSFR} samples have lower masses ($M_{\star}<10^{10}\, \rm M_{\odot}$), and can have lower SFR due to gas starvation, SN feedback, environmental quenching.
The final percentage of galaxies in each sample at $z=0$ varies from simulation to simulation. 
We present these numbers in Tab.~\ref{tab:galaxy_numbers_for_different_subsets}. 
We use these three samples in Section 4.\\

From now on, we only consider galaxies of total stellar mass $M_{\star}\geqslant 10^{9}\, \rm M_{\odot}$ which host a BH. The BH can be an AGN, i.e., accreting mass, or a non-accreting BH, and in that case $L_{\rm AGN}=0$. The mass of the BHs is the mass of individual BHs in all the simulations, except in TNG and Illustris for which the BH mass is the sum of the mass all the BHs within a galaxy. In practice, only a couple of galaxies per output host several BHs at the same time.
The total stellar mass that we use here for all the simulations is not exactly computed in the same way in all the simulations, but we prefer to use each simulation definition. The total stellar mass of the galaxies is computed, e.g., as twice the stellar mass in the half mass radius for the Illustris and TNG100 simulation, within an aperture of 30 kpc in EAGLE. In SIMBA, galaxies are identified using a friends-of-friends galaxy finder, assuming a spatial linking length of 0.0056 times the mean interparticle spacing (equivalent to twice the minimum softening length). Discussion on the impact of these different definitions can be found in \citet{2018MNRAS.475..648P} for TNG. 
Finally, we do not distinguish between central and satellites galaxies, and all our results group these two types of galaxies.

\section{In which galaxies live the faint AGN?}

\begin{figure*}
\centering
\includegraphics[scale=0.34]{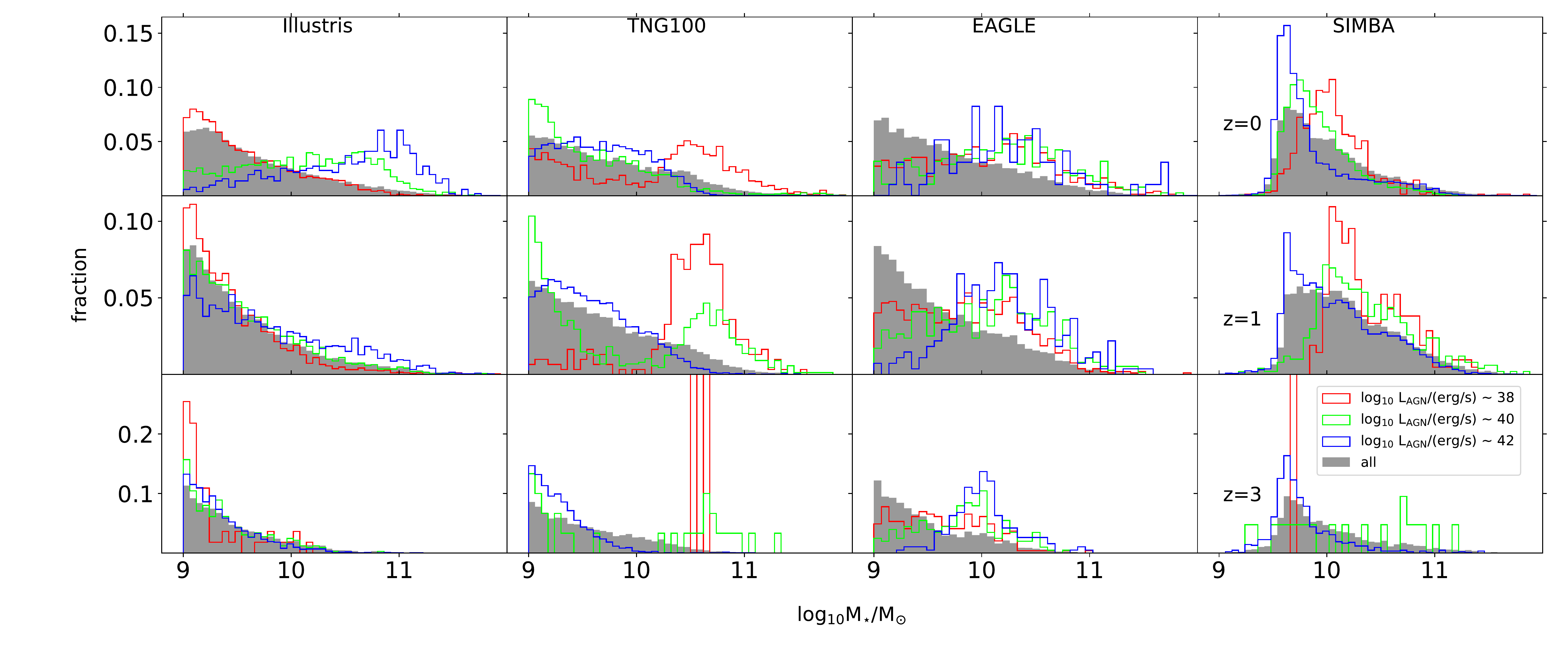}
\includegraphics[scale=0.34]{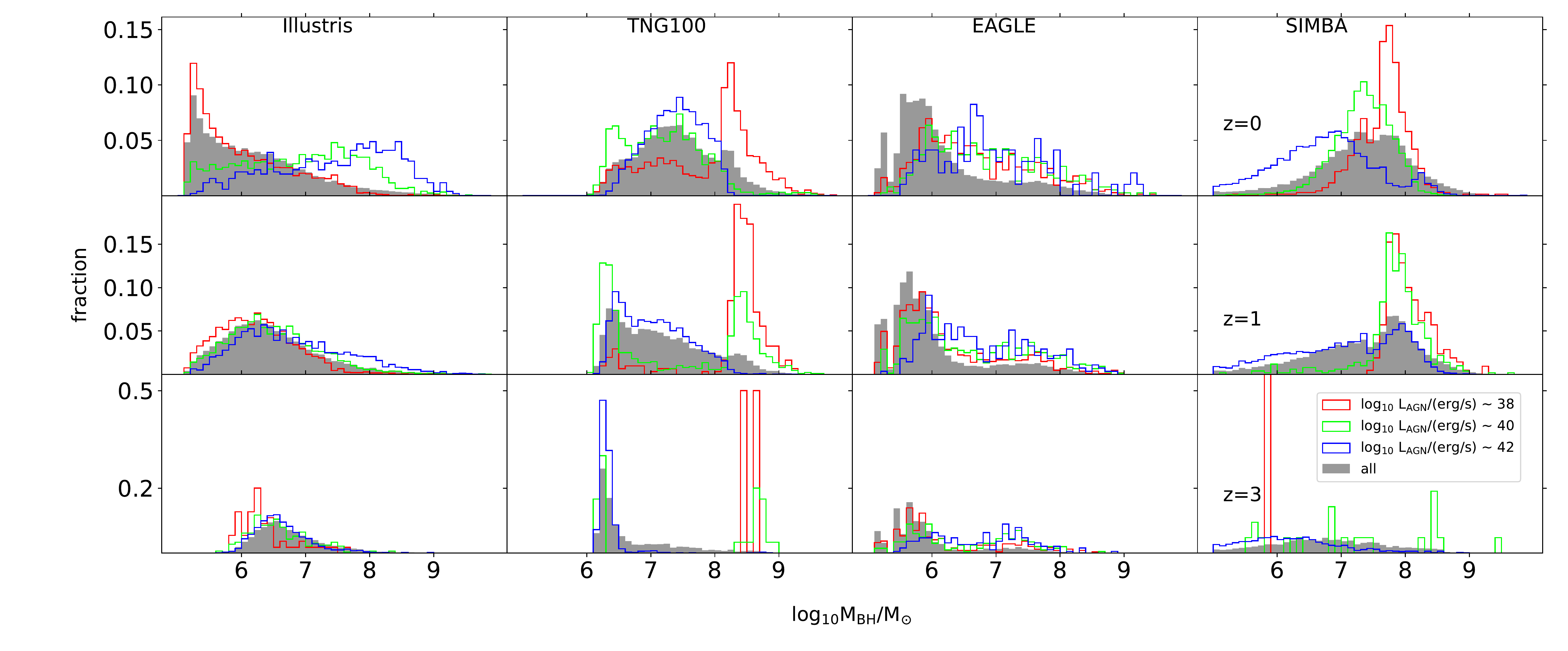}
\hbox{\hspace{0.5em} \includegraphics[scale=0.34]{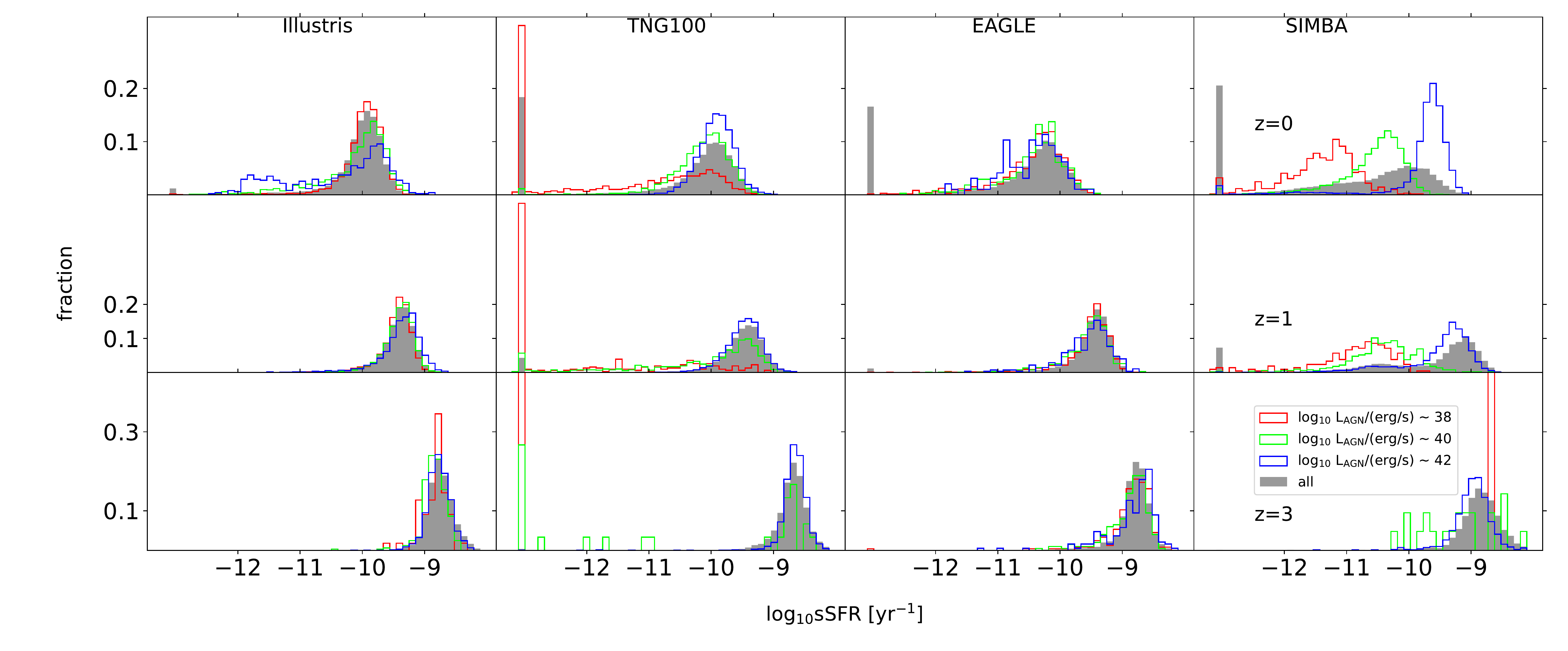}}
\caption{Normalized distributions of galaxy stellar mass, BH mass, and sSFR, for $z=0,1,3$ (top, middle, and bottom panels).
Red distributions only include galaxies with $L_{\rm AGN}\sim 10^{38} \, \rm erg/s$ (bin $10^{37.5}-10^{38.5}$ erg/s), $L_{\rm AGN}\sim 10^{40}\, \rm erg/s$ (bin $10^{39.5}-10^{40.5}$ erg/s) for the green distributions, and $L_{\rm AGN}=\sim 10^{42}\, \rm erg/s$ (bin $10^{41.5}-10^{42.5}$ erg/s) for the blue ones.
Distributions are normalized by the total number of galaxies in the specific luminosity bin. 
The grey shaded histograms show the distributions for the full galaxy population.
Faint AGN of different luminosities are powered by different BHs and reside in different stellar-mass galaxies in the different simulations. For example, the AGN with $L_{\rm AGN}\sim 10^{38}\, \rm erg/s$ reside in low-mass galaxies in Illustris and in galaxies with much higher masses in TNG100 and SIMBA.
}
\label{fig:faint_AGN_Mstar_013}
\end{figure*}

In this section, we first investigate the properties of galaxies hosting faint AGN. We divide the galaxies in subsamples depending on their AGN luminosity: L$_{\rm AGN} = 10^{37.5-38.5}$ erg/s, L$_{\rm AGN} = 10^{39.5-40.5}$ erg/s, and L$_{\rm AGN} = 10^{41.5-42.5}$ erg/s. 
For each subsample as well as for the whole galaxy sample, we show the distributions of stellar mass, BH mass and sSFR in Fig.~\ref{fig:faint_AGN_Mstar_013}. 
Galaxies with $\rm \log_{10}\, sSFR/yr^{-1}<-13$ are shown with this value.

To quantify the findings described in the following, we perform several Kolmogorov-Smirnov (KS) tests for the distributions of the faint AGN with $\log_{10}\, L_{\rm AGN}/(\rm erg/s)\sim 38$ and $\log_{10}\, L_{\rm AGN}/(\rm erg/s)\sim 42$, for all the simulations and for $M_{\star}$, $M_{\rm BH}$ and SFR (Table~\ref{tab:KS_test}). 
We can not reject the null hypothesis that two distributions were drawn from the same dsitribution for high values of $p$, i.e. $p\geqslant 0.01$. However, if $p<0.01$ the two distributions can be considered as statistically different.

\subsection{Illustris}

In the Illustris simulation, faint AGN with $\log_{10}\, L_{\rm AGN}/(\rm erg/s)\sim 38$ are mostly located in low-mass galaxies of $M_{\star}\leqslant 10^{10}\, \rm M_{\odot}$ at any redshift. These AGN also tend to be low-mass BHs of $M_{\rm BH}\leqslant 10^{8}\, \rm M_{\odot}$. 
More luminous AGN with $L_{\rm AGN}\sim 10^{42}\, \rm erg/s$ are also located in low-mass galaxies at high redshift, but with time they start deviating from the $L_{\rm AGN}\sim 10^{38}\, \rm erg/s$ AGN and are largely located in massive galaxies of $M_{\star}\sim 10^{11}\, \rm M_{\odot}$ at $z=0$. These AGN are also mostly massive BHs of $M_{\rm BH}\sim 10^{9}\, \rm M_{\odot}$ at $z=0$.

For Illustris, we find with our KS test that the distributions of $M_{\star}$, $M_{\rm BH}$, and sSFR for $\log_{10}\, L_{\rm AGN}/(\rm erg/s)\sim 38$ and $\log_{10}\, L_{\rm AGN}/(\rm erg/s)\sim 42$ are always statistically different, except the sSFR distribution at $z=3$.

\subsection{TNG100}
In the TNG100 simulation, the faint AGN are located in galaxies with generally different properties. 
From high redshift to low redshift, the AGN with $\log_{10}\, L_{\rm AGN}/(\rm erg/s)\sim 38$ are mostly located in massive galaxies of $M_{\star}\geqslant 10^{10.5}\, \rm M_{\odot}$. There is also a population of AGN with $\log_{10}\, L_{\rm AGN}/(\rm erg/s)\sim 38$ in lower-mass galaxies of $M_{\star}\sim 10^{9}\, \rm M_{\odot}$, similar to the population found in Illustris.
Brighter AGN with $\log_{10}\, L_{\rm AGN}/(\rm erg/s)\sim 42$ are mostly hosted in lower-mass galaxies with $M_{\star}\leqslant 10^{10}\, \rm M_{\odot}$. 
Since the $M_{\rm BH}-M_{\star}$ relationship is tight in TNG100 \citep[][and Fig.~\ref{fig:BH_masses} below]{2021MNRAS.503.1940H}, the $\log_{10}\, L_{\rm AGN}/(\rm erg/s)\sim 38$ AGN are also mostly massive BHs of $M_{\rm BH}\sim 10^{9}\, \rm M_{\odot}$. More luminous AGN are BHs of $M_{\rm BH}\leqslant 10^{7}\, \rm M_{\odot}$ at high redshift ($z\geqslant 2$) and BHs of $M_{\rm BH}\leqslant 10^{8}\, \rm M_{\odot}$ at $z\leqslant 2$.
The distributions of the $\log_{10}\, L_{\rm AGN}/(\rm erg/s)\sim 38$ and the $\log_{10}\, L_{\rm AGN}/(\rm erg/s)\sim 42$ AGN are statistically different (KS test). At $z=3$, there are only few AGN with $\log_{10}\, L_{\rm AGN}/(\rm erg/s)\sim 42$ and the KS test can not be performed.

The fact that faint AGN ($\log_{10}\, L_{\rm AGN}/(\rm erg/s)<40$) are found in massive galaxies of a few $10^{10}\, \rm M_{\odot}$ and that they are powered by BHs of a few $10^{8}\, \rm M_{\odot}$ is due to the efficient low-accretion mode of the TNG feedback model. These AGN are found in quenched galaxies with reduced SFR. 
The population of quenched galaxies can also be observed in the sSFR distribution.

\subsection{EAGLE}
In the EAGLE simulation, the AGN with $\log_{10}\, L_{\rm AGN}/(\rm erg/s)\sim 42$ are mostly located in galaxies of $M_{\star}\sim 10^{10}\, \rm M_{\odot}$ at $z=3$. The distribution of the stellar mass of their host galaxies becomes broader with time, with AGN located in the range $M_{\star}=10^{9.5}-10^{11}\, \rm M_{\odot}$.
At high redshift $z\geqslant 2$, the distribution of the stellar mass of the galaxies hosting AGN with $\log_{10}\, L_{\rm AGN}/(\rm erg/s)\sim 38$ is different from the $\log_{10}\, L_{\rm AGN}/(\rm erg/s)\sim 42$ AGN. The $\log_{10}\, L_{\rm AGN}/(\rm erg/s)\sim 38$ AGN are located in less massive galaxies.
However, at lower redshift ($z<1$) the $M_{\star}$ distribution of $\log_{10}\, L_{\rm AGN}/(\rm erg/s)\sim 38$ extends to more massive galaxies and become very similar to the distribution of more luminous AGN with $\log_{10}\, L_{\rm AGN}/(\rm erg/s)\sim 42$. 
For the mass of the BHs powering these AGN, we note some differences at high redshift ($z\sim3$) between AGN of $\log_{10}\, L_{\rm AGN}/(\rm erg/s)\sim 38$ and $\log_{10}\, L_{\rm AGN}/(\rm erg/s)\sim 42$, with fainter AGN being lower-mass BHs of $M_{\rm BH}\sim 10^{6}\, \rm M_{\odot}$.
The $M_{\rm BH}$ distributions have broad humps
for the two types of AGN at high redshift. At $z=0$ when the distributions are broader, the peak of the distributions are still different but the distributions are less distinguishable.

In EAGLE, the faintest AGN are found in two BH populations: predominantly in low-mass BHs of $M_{\rm BH}\leqslant 10^{6}\, \rm M_{\odot}$, but also in more massive BHs of $M_{\rm BH}\sim 10^{7}\, \rm M_{\odot}$. The first BH population does not power luminous AGN because they reside in low-mass galaxies that are regulated by SN feedback. 
The second BH population consists of more massive BHs located in more massive galaxies, and are regulated by AGN feedback.

For the EAGLE simulation, by applying a KS test we find that all the $\log_{10}\, L_{\rm AGN}/(\rm erg/s)\sim 38$ and $\log_{10}\, L_{\rm AGN}/(\rm erg/s)\sim 42$ distributions are significantly different, except the sSFR distributions at $z=0$ and $z=3$ and the M$_{\star}$ distribution at $z=0$.

\subsection{SIMBA}
The distributions of the faint AGN with $\log_{10}\, L_{\rm AGN}/(\rm erg/s)\sim 38$ and $\log_{10}\, L_{\rm AGN}/(\rm erg/s)\sim 42$ of the SIMBA simulation are qualitatively similar to the distribution of TNG100. However, the $M_{\star}$ distributions of TNG100 extend to lower values than the $M_{\star}$ distributions of SIMBA. 
Our KS test indicates that the distributions of $\log_{10}\, L_{\rm AGN}/(\rm erg/s)\sim 38$ and $\log_{10}\, L_{\rm AGN}/(\rm erg/s)\sim 42$ in SIMBA are statistically different. At $z=3$, there are only few AGN with $\log_{10}\, L_{\rm AGN}/(\rm erg/s)\sim 42$ and we can not perform the KS test.

AGN with $\log_{10}\, L_{\rm AGN}/(\rm erg/s)\sim 38$ are located in more massive galaxies than the AGN with $\log_{10}\, L_{\rm AGN}/(\rm erg/s)\sim 42$. The peak of the two $M_{\star}$ corresponding distributions are separated by less than an order of magnitude.
The $M_{\star}$ distributions are less separated at higher redshift.
For the BHs, we note that at $z=1$, the peak of the BH mass is very similar, but separate at later time. At $z=0$, simulated AGN with $\log_{10}\, L_{\rm AGN}/(\rm erg/s)\sim 42$ are found mostly as BHs of $M_{\rm BH}\sim 10^{7}\, \rm M_{\odot}$. In contrast, the distribution of $\log_{10}\, L_{\rm AGN}/(\rm erg/s)\sim 38$ AGN peaks at $M_{\rm BH}\sim 10^{8}\, \rm M_{\odot}$.

The sSFR distributions are very different for the different AGN luminosities in SIMBA. The AGN with $\log_{10}\, L_{\rm AGN}/(\rm erg/s)\sim 38$ have a peak in their sSFR distribution which is nearly two orders of magnitude smaller than the AGN with $\log_{10}\, L_{\rm AGN}/(\rm erg/s)\sim 42$. In SIMBA, the faintest AGN are in galaxies with a low sSFR, they are quenched, and the AGN with $\log_{10}\, L_{\rm AGN}/(\rm erg/s)\sim 42$ are in star forming galaxies. There is a large population of galaxies that quench with time, and with them a large population of fainter AGN at $z=0$. \\

In this first section, we demonstrated that the BHs powering the faint AGN in the Illustris, TNG100, EAGLE, and SIMBA simulations, as well as the properties of their host galaxies, are different among the different simulations. Indeed, faint AGN of $L_{\rm AGN}\sim 10^{38}\, \rm erg/s$ can be powered by relatively massive BHs and be located in massive galaxies ($M_{\star}\gtrsim 10^{10}\, \rm M_{\odot}$) with reduced SFR (TNG100, SIMBA), or be powered by lower-mass BHs in less massive galaxies ($M_{\star}\lesssim 10^{10}\, \rm M_{\odot}$) still forming stars (Illustris, EAGLE).
Both galaxy and BH mass appear to be fundamental quantities to understand the faint AGN populations in simulations.
Moreover, in some simulations we already see that the sSFR of the host galaxies can play an important role.
In the next section, we go further and divide the full simulated galaxy samples into sub-samples as a function of the galaxies distance from the star-forming main sequence.

\section{AGN hard X-ray luminosity}

\begin{figure*}
\centering
\includegraphics[scale=0.7]{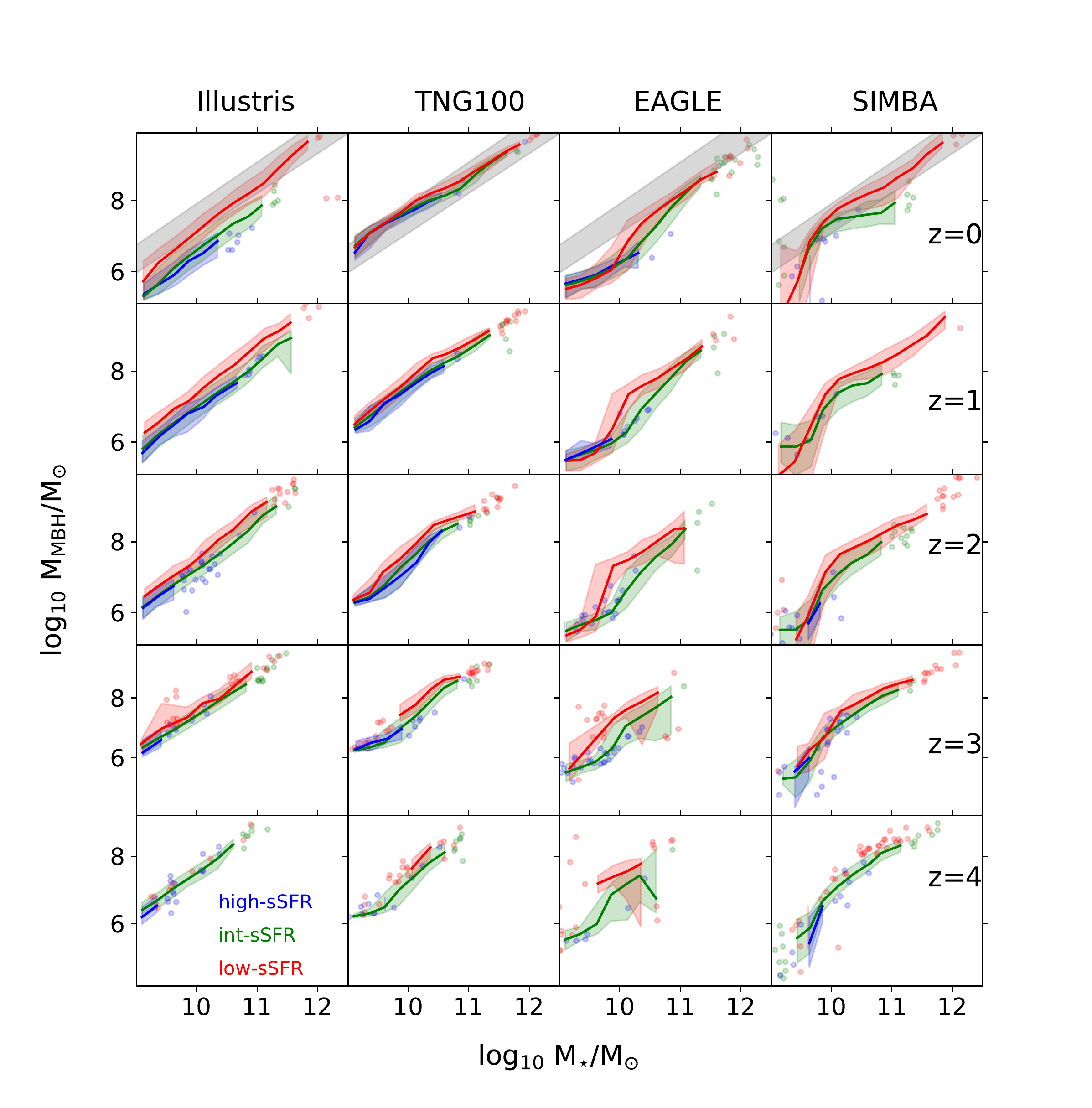}
\caption{Median of the $M_{\rm BH}-M_{\star}$ relation of the
simulations for $z=0, 1, 2, 3, 4$. 
We divide the galaxy population into the three galaxy samples ({\it high-sSFR} in blue, {\it intermediate-sSFR} in green, {\it low-sSFR} in red).
Shaded regions show the 15th-85th percentiles of the distribution at fixed total stellar mass. The width of the stellar mass bins is 0.25 dex. 
Individual galaxies are shown with dots in bins with less than 10 galaxies.
The grey region at $z=0$ encloses the empirical scaling relations of \citet{2013ARA&A..51..511K} (higher normalization of the scaling relation), \citet{2013ApJ...764..184M}, and \citet{2004ApJ...604L..89H} (lower normalization of the relation). In these simulations, the median BH mass of low-sSFR galaxies is higher than for higher sSFR galaxies, at fixed stellar mass.}
\label{fig:BH_masses}
\end{figure*}

\subsection{$M_{\rm BH}-M_{\star}$ properties in the three galaxy samples}
We show the median of the mass of the BHs located in the three galaxy samples in Fig.~\ref{fig:BH_masses}, as a function of the total stellar mass of the galaxies and for several redshifts from $z=0$ (top row) to $z=4$ (bottom row). 

On average we find that galaxies with lower SFR ({\it low-sSFR} samples, red lines) host more massive BHs at fixed stellar mass; the effect is mild and in most cases within the 15th-85th percentiles of the sample distributions.
We find this for the Illustris and TNG100 simulations at all stellar masses, and for galaxies of $M_{\star}\geqslant 10^{10}\, \rm M_{\odot}$ in the EAGLE and SIMBA simulations. 
At low redshift, the EAGLE {\it high-sSFR} galaxies with $M_{\star}< 10^{10}\, \rm M_{\odot}$ (blue lines) host BHs slightly more massive than galaxies with lower SFR. 
In SIMBA, we can not really assess the behavior of BH mass of galaxies with $M_{\star}\leqslant 10^{10}\, \rm M_{\odot}$ as the seeding takes place in galaxies of $M_{\star}\sim 10^{9.5}\, \rm M_{\odot}$.
Most simulations have a tight $M_{\rm BH}-M_{\star}$ relation \citep[not shown here, but see][]{2021MNRAS.503.1940H}, TNG100 having the tightest relation, which explains the very mild difference in the median of BH mass for the three samples at all redshifts. 
We do not show it here, but the $M_{\rm BH}-M_{\star}$ diagram of the simulations at $z=0$ is in broad agreement with the observational samples of e.g., \citet{2015arXiv150806274R}. 
To guide the eye, we show in grey shaded area in the $z=0$ panels a region enclosing several $M_{\rm BH}-M_{\rm bulge}$ empirical scaling relations \citep[e.g.,][]{2013ARA&A..51..511K,2013ApJ...764..184M,2004ApJ...604L..89H} that have been used to calibrate the simulations. It should be noted that the calibrations are done on the entire galaxy population of the simulations and not on one given sSFR subsample, whereas in this paper we are testing the simulations outcomes to deeper and more constraining details. It may hence be not surprising that some discrepancies as the ones revealed above are in place between e.g., the high-sSFR $M_{\rm BH}-M_{\star}$ relation and the empirical scaling relations.

In this section we showed that, on average, lower sSFR galaxies (quenched or on their way to quiescence) tend to host more massive BHs in cosmological simulations. Since the $M_{\rm BH}-M_{\star}$ relation is tight in simulations, the effect is small. 
This is in agreement with the results of \citet{2019MNRAS.487.5764T} for SIMBA. Moreover, this is consistent with what is found in observations \citep{2017ApJ...844..170T}, but such observational samples with estimates of both dynamical BH mass and SFR are still restrictively small. 
Our previous paper \citet{2021MNRAS.503.1940H} provides more information and comparisons between the simulated population of BHs in these simulations and observations.

\begin{figure*}
\centering
\includegraphics[scale=0.7]{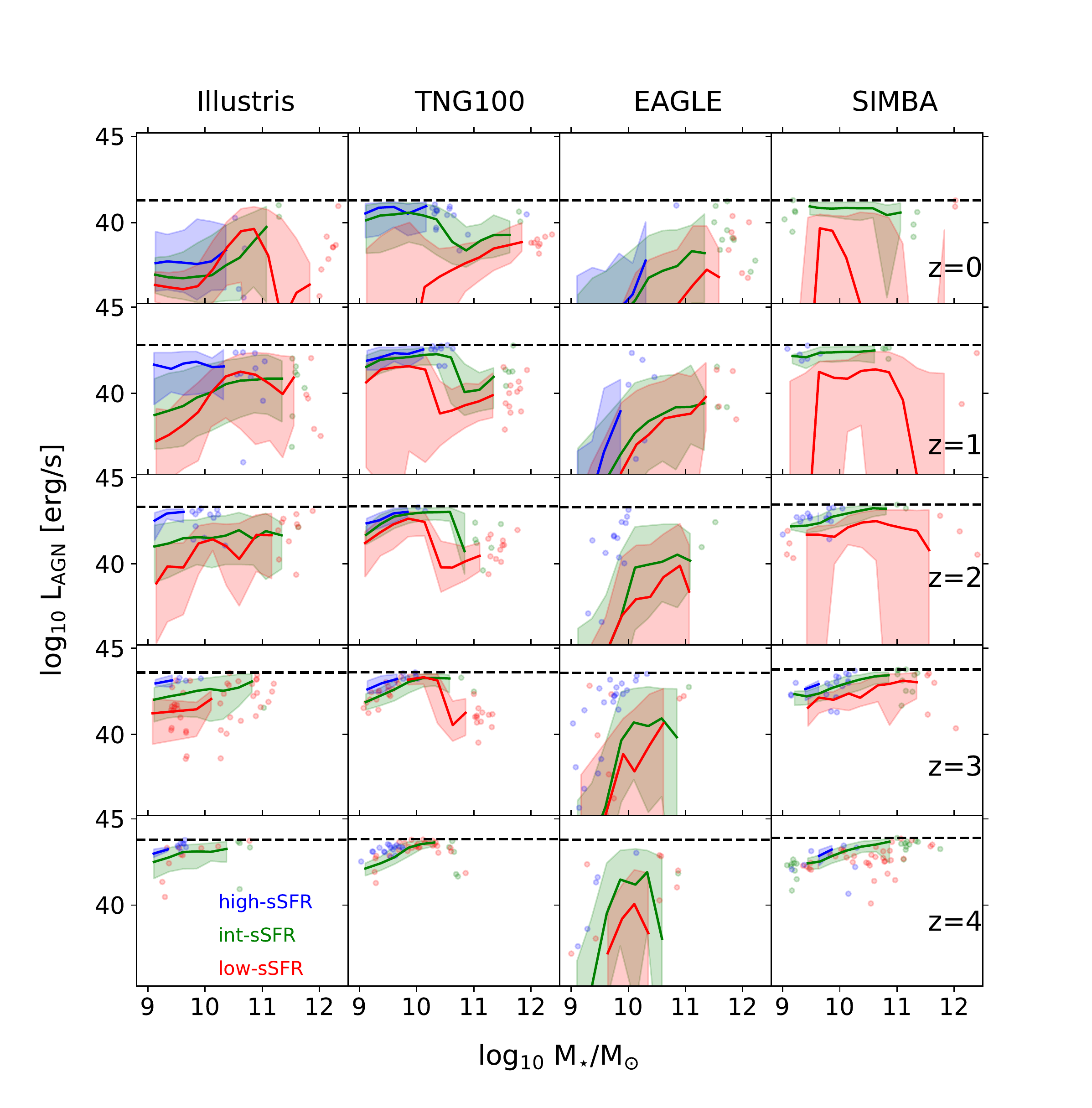}
\caption{Median AGN hard X-ray (2-10 keV) luminosity for different galaxy stellar mass bins and redshifts from $z=0$ (top row) to $z=4$ (bottom row).
Individual galaxies are shown as points in bins with less than 10 galaxies. The shaded regions indicate the 15-85th percentiles. We do not consider AGN obscuration here. All AGN that could be detected individually by the COSMOS survey are excluded (black dashed line). AGN are on average brighter in galaxies with high sSFR. AGN feedback is responsible for the decrease of luminosity in massive galaxies, in some simulations.
}
\label{fig:L_AGN_upper_cut}
\end{figure*}

\subsection{AGN X-ray (2-10 keV) luminosity in the three galaxy samples}
In this paper, we aim at deriving the galaxy total X-ray luminosity of galaxies hosting faint AGN and comparing it to observational constraints, for different types of galaxies. As a first step we derive the X-ray emission of the AGN, a crucial component of the total luminosity of galaxies.

We present in Fig.~\ref{fig:L_AGN_upper_cut} the median of the AGN luminosity as a function of the stellar mass of galaxies. In order to focus on analysis on the faint AGN, we only consider AGN with hard X-ray luminosity below $L_{\rm AGN}\sim 10^{42}\, \rm erg/s$. In practise, we use the luminosity limit of the COSMOS survey to detect AGN as individual sources. 
This allows us to evolve the luminosity limit with redshift in a way that we will be able to compare our results with observations in Section 6. In Section 6, we will also apply this limit and describe in detail how we compute it.
For our analysis we remove all AGN that are brighter than this limit. In this section, we also do not consider AGN obscuration.
The shaded regions indicate the 15th-85th percentiles of the distribution. A median value is only calculated for stellar mass bins with more than 10 galaxies, otherwise individual galaxies are shown as points.

In general, the median AGN luminosity increases with redshift. Furthermore, the galaxies with a higher sSFR (e.g., the blue lines in Fig.~\ref{fig:L_AGN_upper_cut}) have higher luminosities in the simulations than lower-sSFR galaxies (red lines), at fixed stellar masses. 
There is an exception in Illustris at redshifts $z=0,1$. Here, the AGN luminosity of the {\it low-sSFR} sample (in red) is higher for stellar masses between $M_{\star}=10^{10}$ M$_{\odot}$ and a few $10^{10}$ M$_{\odot}$.
The median AGN luminosity drops for more massive galaxies in Illustris.

In the panels of the TNG100 simulation, the low-accretion mode of the AGN feedback (kinetic mode) leads to a sharp decrease of the AGN luminosity \citep{2018MNRAS.479.4056W} for galaxies with stellar mass $\log_{10} M_{\star}/\rm M_{\odot}=10.25$ at redshifts $z=0, 1, 2, 3$ in the {\it low-sSFR} sample. 
We note here that the sharp decrease is present at $z=0$, but masked by the low $L_{\rm AGN}$ in lower mass galaxies due to the rarefaction of cold gas (lower accretion rates onto the BHs) at low redshift.

A large shaded region indicates a broad distribution of luminosities in a given stellar mass bin. The size of the shaded regions decreases with increasing redshift. Moreover, the shaded regions are large for {\it low-sSFR} galaxies, smaller for {\it intermediate-sSFR} galaxies and even smaller for {\it high-sSFR} galaxies. Many different luminosities are possible for {\it low-sSFR} galaxies depending on how much gas there is still available for the central BH. The amount of gas and its properties (e.g., temperature) can be affected by both the SN feedback in low-mass galaxies and by AGN feedback primarily in more massive galaxies \citep[e.g.,][]{2020MNRAS.499..768Z,2019MNRAS.490.3234N,2017MNRAS.468.3935H}. In general, this leads to broad distributions.

\begin{figure*}
\centering
\includegraphics[scale=0.32]{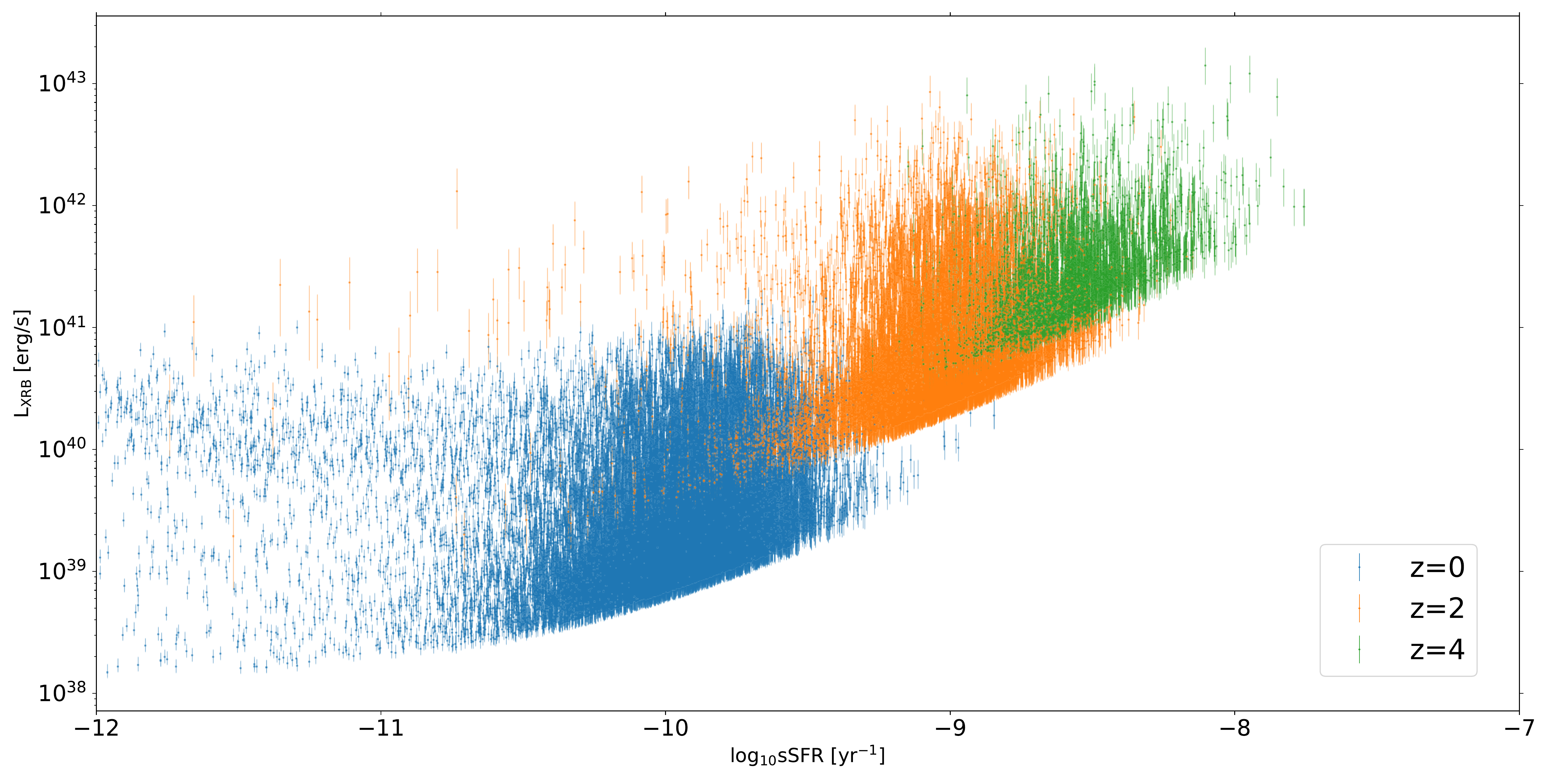}
\caption{Hard X-ray (2-10 keV) luminosity of the XRB population for the simulated galaxies of Illustris. The XRB population luminosity is estimated from the galaxy SFR, stellar mass, and redshift. Here, we use the model from \citet{2019ApJS..243....3L}. The error bars are calculated from the uncertainties of the parameters in the formula for the XRB luminosity. In the modeling, massive X-ray binaries dominate the XRB population luminosity in galaxies with high sSFR ($\rm \log_{10}\, sSFR/yr\,\geqslant -10.5$). Low-mass binaries dominate at lower sSFR ($\rm \log_{10}\, sSFR/yr\,< -10.5$).
}
\label{fig:LXRB_vs_sSFR_Illustris}
\end{figure*}

\section{Hard X-ray luminosity from X-ray binaries: Can they outshine the AGN?}
In the previous section, we derived the AGN luminosity of the host galaxies of faint AGN. In this section, we derive the luminosity of the XRB population of the simulated galaxies, and evaluate whether the XRB luminosity can be higher than the AGN luminosity, and for which galaxies.

\subsection{XRB hard X-ray (2-10 keV) luminosity as a function of galaxy properties and redshift}
In Fig.~\ref{fig:LXRB_vs_sSFR_Illustris}, we show as an example the X-ray luminosity of the XRB population in the simulated galaxies of Illustris, for different redshifts. We use the relation of \citet{2019ApJS..243....3L}, which is the most recent among the models discussed in this paper. The error bars indicate the uncertainties on the parameters in Eq.~\ref{eq:xrb_relation}. 
At high redshift ($z\sim 4$), most of the simulated galaxies form stars efficiently and have high sSFR. For the Illustris simulation and this given empirical XRB relation, this corresponds to hard X-ray luminosity of the XRB populations in the range $L_{\rm XRB}=10^{41}-10^{43}\, \rm erg/s$.
With time, there are more and more galaxies with lower SFR. As a consequence the diagram extends towards low sSFR values, and also towards lower XRB population luminosity. At $z=0$, the X-ray luminosity of the XRB population in the Illustris galaxies is $L_{\rm XRB}\leqslant 10^{41}\, \rm erg/s$.

\subsection{Relative hard X-ray luminosity contributions of the AGN and the X-ray binary populations}

\begin{figure*}
\centering
\includegraphics[scale=0.5]{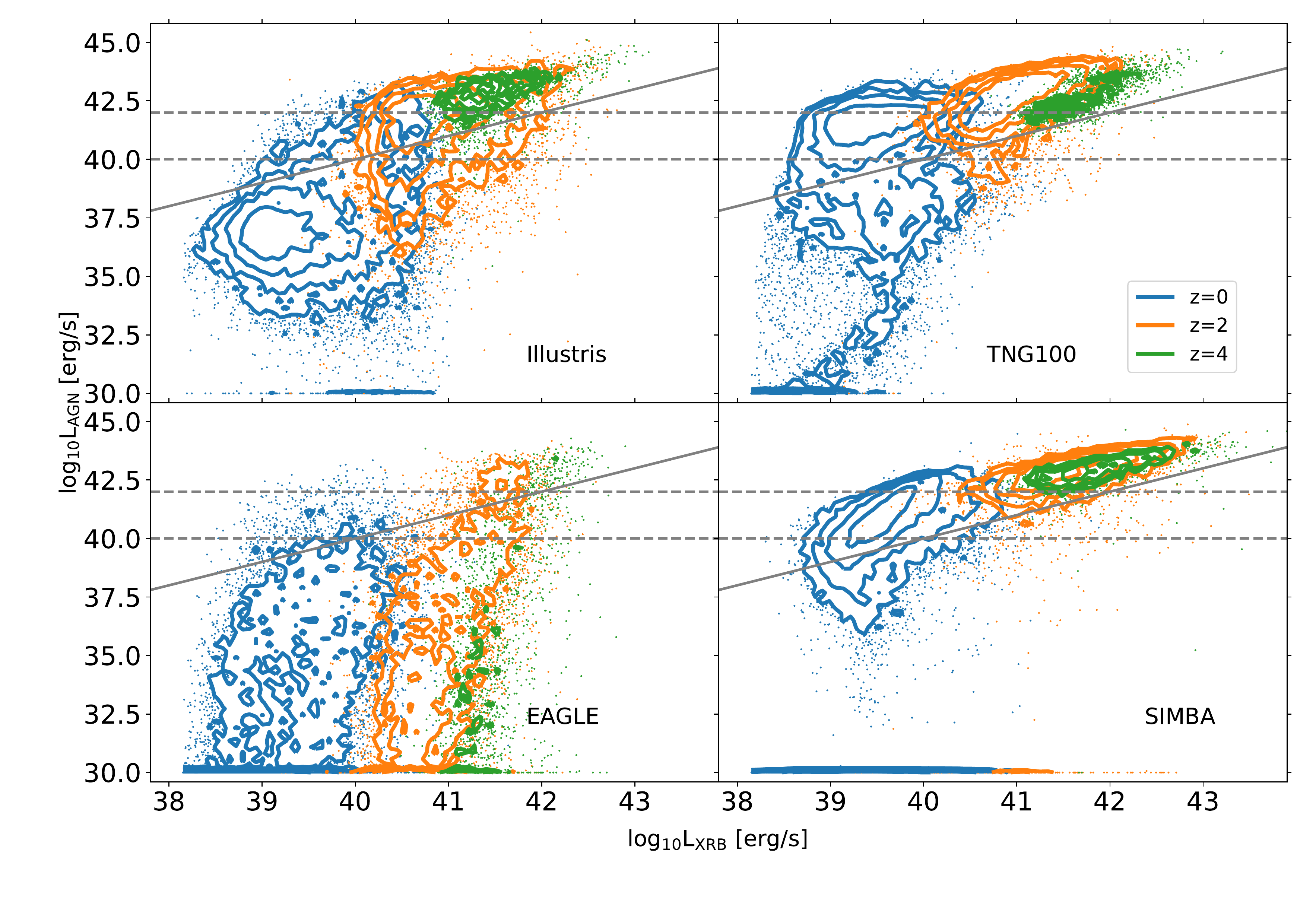}
\caption{Hard X-ray (2-10 keV) luminosity of the AGN and XRB population of the simulated galaxies.
Galaxies with AGN luminosity smaller than $10^{30}$ erg/s are plotted at that luminosity. Different colors indicate different redshifts. We only consider galaxies with M$_{\star}\geqslant 10^9$ M$_{\odot}$.
We draw a grey solid line to show equal luminosities from the XRB population and the AGN. 
The fraction of galaxies 
with $L_{\rm XRB}>L_{\rm AGN}$ increases at lower redshifts. 
In the faint AGN regime (illustrated by the regions below the horizontal dashed lines) more and more galaxies have higher XRB luminosity than AGN luminosity with time.}
\label{fig:XRB_vs_AGN}
\end{figure*}

Fig.~\ref{fig:XRB_vs_AGN} compares the luminosity contributions from the AGN to the luminosity contribution from XRBs. Here, we use the model of \citet{2019ApJS..243....3L}.
We show the contributions for different redshift with different colors.
There are 50 bins (per dimension) in the 2D histogram and a contour line is drawn if at least 5 galaxies are present. 
We draw a grey solid line to show equal contributions from the XRB population and the AGN. Since we only consider galaxies with $M_{\star}\geqslant 10^{9}$ M$_{\odot}$, the minimum XRB luminosity is $L_{\rm XRB}=10^{38.15}$ erg/s (minimum possible value from the empirical XRB scaling relations).

At high redshift ($z=4$), we find that in the Illustris, TNG100, and SIMBA simulations the X-ray luminosity of the AGN are higher than the luminosity of the XRB population for most of the galaxy population. 
At $z=4$, almost all the galaxies are on the main sequence, independently of their stellar mass.
The emission of the XRB population is more important for galaxies with higher sSFR, at fixed stellar mass (see Fig.~\ref{fig:LXRB_theory}). 
The XRB populations have high luminosities of $L_{\rm XRB}=10^{41}-10^{43}\, \rm erg/s$.
In these high-redshift star-forming galaxies, there is also gas available to feed the central BHs, and therefore, we also find bright AGN in these galaxies with $L_{\rm AGN}=10^{41}-10^{45}\, \rm erg/s$, on average. 
At $z=4$, $>90 \%$ of the galaxies in Illustris, TNG100, and SIMBA have a higher relative AGN luminosity $L_{\rm AGN}>L_{\rm XRB}$.
We note that the picture is different in the EAGLE simulation. The XRB population in the EAGLE simulation covers the same luminosity range. However, there is a larger diversity of luminosity for the AGN, and they have (much) lower luminosity than the other simulations, on average. As a result, at 
$z=4$ a large fraction of the galaxies have $L_{\rm XRB}>L_{\rm AGN}$ (below the grey line). Only $13\%$ of the galaxies have high AGN luminosities compared to their XRB populations.

With time, the luminosity of the XRB population generally decreases, as well as the luminosity of the AGN. In all the simulations, more and more galaxies reach a regime with $L_{\rm XRB}>L_{\rm AGN}$. At $z=0$, most of the galaxies have $L_{\rm XRB}>L_{\rm AGN}$ in EAGLE (only $5\%$ of galaxies have $L_{\rm AGN}>L_{\rm XRB}$). This is also the case for a large fraction of the Illustris simulation with $10\%$ of galaxies with $L_{\rm AGN}>L_{\rm XRB}$.
Finally, we note that in TNG100 and SIMBA a large number of galaxies still host bright AGN ($\sim 63\%$ for both simulations), and therefore are still in the $L_{\rm AGN}>L_{\rm XRB}$ regime (above the grey line).

Now if we only consider faint AGN in Fig.~\ref{fig:XRB_vs_AGN}, i.e. by only considering the galaxies below an horizontal line at $L_{\rm AGN}\leqslant 10^{40}\, \rm erg/s$ or $L_{\rm AGN}\leqslant 10^{42}\, \rm erg/s$ (shown as grey dashed lines in Fig.~\ref{fig:XRB_vs_AGN}), we find that a significant fraction of the galaxies are dominated by the XRB populations. We quantify this in Table~\ref{tab:galaxy_Lagn_sup_Lxrb}.
The percentage of XRB dominated galaxies with AGN luminosities $L_{\rm AGN}\leq 10^{42}$ erg/s is 92\% in Illustris at redshift $z=0$ (compared to 90\% in the full sample). This increases further if we only include galaxies with AGN luminosities $L_{\rm AGN}\leq 10^{40}\, \rm erg/s$ (98\%). We find similar trends for TNG100, EAGLE, and SIMBA.\\

\begin{table}
  \begin{center}
    \caption{Percentage ($\%$) of all the simulated galaxies with $M_{\star}\geqslant 10^{9}\, \rm M_{\odot}$ with $L_{\rm XRB}> L_{\rm AGN}$, at redshift $z=4,2,0$. In the second and third columns, we only consider galaxies with a hard X-ray (2-10 keV) AGN luminosity $\leq 10^{42}$ erg/s or $\leq 10^{40}$ erg/s and show the percentages of galaxies with $L_{\rm XRB}> L_{\rm AGN}$ again. We neglect AGN obscuration here.
    }
    \scalebox{0.85}{
    \begin{tabular}{ll|ccc}
    \hline
    && \multicolumn{3}{c}{Percentage of galaxies with $L_{\rm XRB}> L_{\rm AGN}$ ($\%$)}\\
    &&&&\\
    & & all galaxies & $L_{\rm AGN}\leqslant 10^{42}\, \rm erg/s$ & $L_{\rm AGN}\leqslant 10^{40} \, \rm erg/s$\\
    \hline
    \hline
    Illustris & $z=0$ & 90  &92&98\\
    Illustris & $z=2$ & 39  &56&$>99$\\
    Illustris & $z=4$ & 8 &44&100\\
    \hline
    TNG100 & $z=0$ & 37  &42&85\\
    TNG100 & $z=2$ & 6 &17&99\\
    TNG100 & $z=4$ & 2 &11&-\\
    \hline
    EAGLE & $z=0$ & 95 &95&99\\
    EAGLE & $z=2$ & 93  &98&100\\
    EAGLE & $z=4$ & 87  &98&100\\
    \hline
    SIMBA & $z=0$ & 37  &39&78\\
    SIMBA & $z=2$ & 9 &42&100\\
    SIMBA & $z=4$ &3 &37&100\\

    \hline
    \end{tabular}
    }

  \label{tab:galaxy_Lagn_sup_Lxrb}    
  \end{center}
\end{table}

\begin{figure*}
\centering
\includegraphics[scale=0.65]{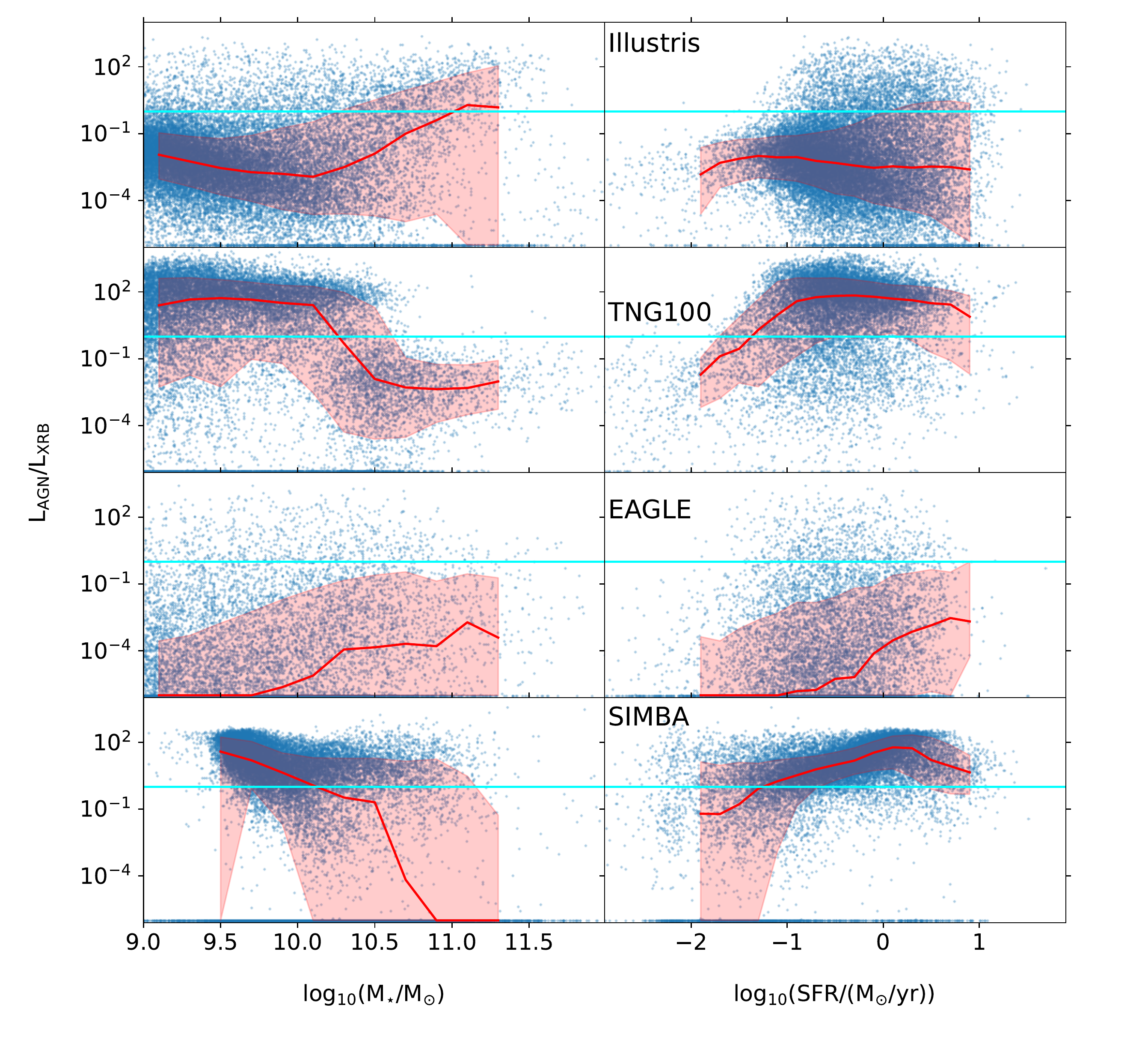}
\caption{Ratio between AGN and XRB luminosities  
at $z=0$. The blue lines indicate equal contributions and red lines show the running median in bins of width 0.2 dex. The shaded regions are the 15-85th percentiles of the distributions. Ratios below $10^{-6}$ are displayed at this value. The median $L_{\rm AGN}/L_{\rm XRB}$ varies from one simulation to another. 
While there is a significant population of galaxies with $L_{\rm AGN}<L_{\rm XRB}$ in Illustris and EAGLE for all $M_{\star}$, we find that AGN dominate the luminosity for $M_{\star}\leqslant 10^{10.5}\, \rm M_{\odot}$, and also $\rm log_{10}\, SFR/(M_{\odot}/yr)\geqslant -1$ in TNG and SIMBA.
}
\label{fig:XRB_vs_AGN_galaxyprop}
\end{figure*}

\begin{table*}
 \caption{The percentages of all galaxies dominated by XRB emission ($L_{\rm XRB}>L_{\rm AGN}$) is given for different sSFR and stellar mass samples. 
 Percentages of galaxies hosting faint AGN with L$_{\rm AGN}\leqslant 10^{42}$ erg/s with $L_{\rm XRB}>L_{\rm AGN}$ are also given.
 Percentages for subsamples with less than 10 galaxies are written in parentheses. We do not consider AGN obscuration here.
 }
  \begin{center}
    \scalebox{1.1}{
    \begin{tabular}{lll|ll|ll|ll}
    \hline
    \multicolumn{9}{c}{Percentage of all galaxies and faint AGN hosts with $L_{\rm XRB}> L_{\rm AGN}$ ($\%$)}\\
    
    &&&&&&&&\\
    & redshift &$M_{\star}\, \rm (M_{\odot})$&
    \multicolumn{2}{c|}{{\it low-sSFR}} & 
    \multicolumn{2}{c|}{{\it intermediate-sSFR}} & 
    \multicolumn{2}{c}{{\it high-sSFR}}\\
    
    && & all galaxies & faint AGN hosts & all gal & faint hosts & all gal  & faint hosts\\
    \hline
    \hline
    Illustris & $z=0$ &$10^{9}-10^{9.5}$ & $>99$ & $>99$ &96 &96&81 &83\\
    &&$10^{9.5}-10^{10.5}$ & 93 &93  &91 &92&75 &83\\
    &&$10^{10.5}-10^{11.5}$ & 50 &57 &74 &78&(67) &(100)\\
     & $z=3$ &$10^{9}-10^{9.5}$& 28&37 &17 &38&3 &(100)\\
    &&$10^{9.5}-10^{10.5}$& 32 &52  &22& 70&0 &-\\
    &&$10^{10.5}-10^{11.5}$& 16 &(100)  &12& 100&- &-\\
    \hline
    
    TNG100 & $z=0$ &$10^{9}-10^{9.5}$& 82 &82 &15& 17&7 &10\\
    &&$10^{9.5}-10^{10.5}$& 85 &86 &12 &16 &6 &11\\
    &&$10^{10.5}-10^{11.5}$ & 99& 99  &64 &78&17 &36\\
     & $z=3$ &$10^{9}-10^{9.5}$& 0 &(0)&3 &6&4 &(50)\\
    &&$10^{9.5}-10^{10.5}$ & 6& 70  &3& 33&0 &-\\
    &&$10^{10.5}-10^{11.5}$& 56& 83&1 &(100)&(0) &-\\
    \hline
    
    EAGLE & $z=0$ &$10^{9}-10^{9.5}$& $>99$&$>99$ &96& 96&95 &95\\
    &&$10^{9.5}-10^{10.5}$& 96 &96 &93 &93&94 &97\\
    &&$10^{10.5}-10^{11.5}$& 91 &92 &89 &90&(50) &(50)\\
     & $z=3$ &$10^{9}-10^{9.5}$ & 92 &100 &99 &99&91 &100\\
    &&$10^{9.5}-10^{10.5}$ & 82 &96 &81 &97&24 &(80)\\
    &&$10^{10.5}-10^{11.5}$ & 57 &86 &72 &100&- &-\\
    \hline

    SIMBA&$z=0$&$10^{9.5}-10^{10.5}$ & 50 &50  &2 &2&5 &(67)\\
    &&$10^{10.5}-10^{11.5}$ &73 &76 &25 &28&- &-\\
    &$z=3$&$10^{9.5}-10^{10.5}$& 8 &21 &4 &27&10 &(100)\\
    &&$10^{10.5}-10^{11.5}$ & 22 &97 &6 &(100)&- &-\\
    \hline
    \hline
    \end{tabular}
    }
   
    \label{tab:L_as_a_function_of galaxy_properties}
  \end{center}
\end{table*}

\subsection{Comparison between the XRB hard X-ray luminosity and the AGN luminosity as a function of galaxy properties}
In the previous section, we have considered all galaxies independently of their properties ($M_{\star}$, SFR). In the following, we investigate the relative luminosities of the AGN compared to the XRB populations as a function of galaxy properties.

For illustration we show in Fig.~\ref{fig:XRB_vs_AGN_galaxyprop} the $L_{\rm AGN}/L_{\rm XRB}$ ratio between AGN and XRB luminosity as function of stellar mass and SFR for all simulations at $z=0$. The median $L_{\rm AGN}/L_{\rm XRB}$ vary from one simulation to another.
In Illustris and EAGLE, there is a significant population of galaxies for which the XRB population outshine the AGN, and even more so for lower mass galaxies and for galaxies with lower SFR. In TNG and SIMBA, instead, AGN dominate the galaxy X-ray luminosity for $M_{\star}\leqslant 10^{10.5}\, \rm M_{\odot}$ and also in galaxies with $\rm log_{10}\, SFR/(M_{\odot}/yr)\geqslant -1$.

We quantify our findings and the redshift evolution below, and in Table \ref{tab:L_as_a_function_of galaxy_properties}.
In Illustris, almost all low-mass galaxies have a higher XRB luminosity than AGN luminosity, for all the sSFR galaxy groups.
Indeed, we find that $81\%$ of the low-mass {\it high-sSFR} galaxies with $M_{\star}=10^{9}-10^{9.5}$ M$_{\odot}$ have a higher XRB luminosity than AGN luminosity. We find 96\% in the {\it intermediate-sSFR} sample and $>99\%$ in the {\it low-sSFR} sample.
For more massive galaxies in the range $M_{\star}=10^{9.5}-10^{10.5}$ M$_{\odot}$, the XRB contribution dominates in $75\%$ in the {\it high-sSFR} sample, $91\%$ in the {\it intermediate-sSFR} sample and $93\%$ in the {\it low-sSFR} sample). 
The corresponding values for the other simulations are given in Table \ref{tab:L_as_a_function_of galaxy_properties}. 
At high redshift, the galaxy total luminosity is dominated by the AGN emission in general for all the simulations, except EAGLE which produces fainter AGN.
More galaxies are dominated by the XRB emission with time, for all the Illustris, TNG100, EAGLE, and SIMBA simulations.
In more detail, the number of Illustris and EAGLE galaxies dominated by XRB at low redshift decreases at higher galaxy stellar mass, for all the sSFR galaxy subsets.
However, in TNG100 the number of galaxies with $L_{\rm XRB}>L_{\rm AGN}$ increases for more massive galaxies. This is because at low redshift a significant fraction of massive galaxies experiences quenching.
In general for all the simulations, we find that the XRB emission is more likely to dominate the total galaxy emission in low-sSFR
galaxies than in higher-sSFR galaxies.

In this section we demonstrated that while the simulations can have similar trends (more {\it low-sSFR} galaxies with  $L_{\rm XRB}>L_{\rm AGN}$), the XRB emission can dominate already at high redshift (EAGLE), or not (Illustris, TNG100, SIMBA), can dominate more in low-redshift low-mass galaxies (Illustris, EAGLE)
or more in massive galaxies (TNG100).

Our findings here have a large implication for the detection of AGN in dwarf galaxies. As indicated in Table~\ref{tab:L_as_a_function_of galaxy_properties}, we find that all the simulations presented here have a very large fraction of low-mass galaxies ($M_{\star}\leqslant 10^{9.5}\, \rm M_{\odot}$) with $L_{\rm XRB}>L_{\rm AGN}$ at $z=0$. 
This makes the confirmation of the presence of an AGN in these galaxies challenging, and is in agreement with results from the current search for these AGN \citep{2013ApJ...775..116R,2015ApJ...809L..14B,2016ApJ...817...20M,2018MNRAS.478.2576M,2020ARA&A..58..257G,2019MNRAS.488..685M,2020ApJ...888...36R}. 
Interestingly, some simulations predict a large fraction of $L_{\rm XRB}>L_{\rm AGN}$ galaxies independently of the SFR of these galaxies (Illustris, EAGLE), some other simulations predict that AGN emission dominates in star-forming main-sequence and starburst galaxies (TNG100, SIMBA). 
Faint AGN in low-mass galaxies should be more detectable in hard X-rays in galaxies forming stars more efficiently. 
If the trend identified in TNG100 is correct, an enhancement of X-ray emission due to the AGN would be observed more often in low-redshift star-forming galaxies than in more quiescent galaxies.

\section{Galaxy total hard X-ray luminosity of faint AGN hosts in simulations and comparison to observations}

In the previous sections, we have analyzed the luminosity of the faint AGN, but also the luminosity of the XRB population in their host galaxies, as a function of galaxy properties. In this section, we predict the average total hard X-ray (2-10 keV) luminosity of the faint AGN host galaxies, from the Illustris, TNG100, EAGLE, and SIMBA simulations. We compare our predictions to recent observations of stacked galaxies \citep{2018ApJ...865...43F}.

\subsection{Total hard X-ray luminosity of faint AGN host galaxies in simulations}
We compare the total X-ray luminosity of the galaxies, defined as $L_{\rm total}=L_{\rm AGN}+L_{\rm XRB}$, in the different simulations. We do not include the emission from the hot gas.
Here, we purposely work in the hard X-ray band (2-10 keV), in which the emission from the hot gas is thought to be the smallest contribution and much lower than the XRB contribution \citep{2016ApJ...825....7L}. The hot gas contributes to the X-ray emission with a diffuse, soft thermal component \citep{2018ApJ...865...43F}. We address the contribution of the hot gas further in the discussion section.\\

\begin{figure*}
\centering
\includegraphics[scale=0.85,]{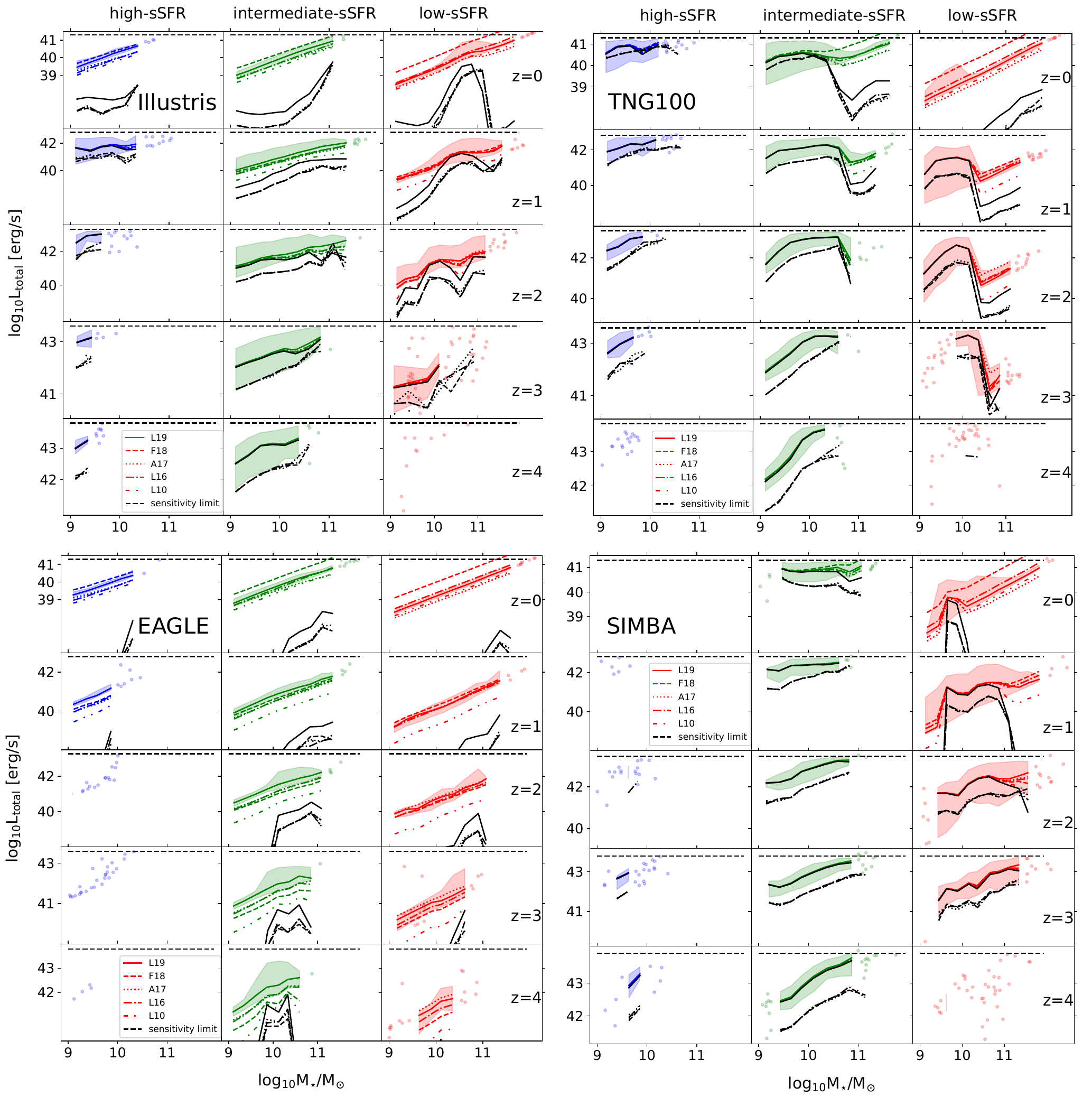}
\caption{Total galaxy hard X-ray (2-10 keV) luminosity ($L_{\rm total}=L_{\rm AGN}+L_{\rm XRB}$) as a function of galaxy stellar mass 
for different sSFR samples. We only include galaxies hosting too faint AGN to be detected by current X-ray facilities as individual sources, i.e. below the COSMOS Legacy detection limit.
Shaded regions indicate the 15-85th percentiles. Percentiles and individual galaxies are only shown for the XRB correction from \citet{2019ApJS..243....3L}. 
Different line styles of the colored lines indicate different XRB empirical scaling relations, as explained in Table~\ref{tab:XRB_relations}.  
Black solid lines indicate the median AGN luminosity (without obscuration) as shown in Fig.~\ref{fig:L_AGN_upper_cut}, and entering in the computation of $L_{\rm total}$. We also show the effect of AGN obscuration with the other black lines: (dashed for model M1, dotted for M2, dashed-dotted for M3, and loosely dotted for M4). A larger fraction of obscured faint AGN leads to a more linear shape of the $L_{\rm total}-M_{\star}$ relation, because the total luminosity is dominated by the XRB contribution.
}
\label{fig:L_total_Illustris_new}
\end{figure*}

In Fig.~\ref{fig:L_total_Illustris_new}, we show the median of the total X-ray luminosity of the simulated galaxies for different stellar mass bins for the four simulations in colored lines. We show the median AGN luminosity (without obscuration) as black solid lines.
Similarly as in Fig.~\ref{fig:L_AGN_upper_cut}, we only consider the galaxies hosting faint AGN, below the individual detection limit of the X-ray COSMOS survey. 
In the regime of faint AGN host galaxies, we have demonstrated in the previous section that the luminosity of the XRB populations could overshine those of the AGN. This, of course, also depends on the X-ray scaling relations that we use to compute the XRB population luminosity. 
To understand the impact of these different relations, we show the median values of the galaxy total luminosity after applying the different scaling relations (indicated by different linestyles for the colored lines). 
The shaded regions and individual galaxies (plotted as points) are only given for the XRB scaling relation of \citet{2019ApJS..243....3L}.\\

In general, the median galaxy total hard X-ray luminosity increases with stellar mass. 
The shape of the total luminosity as a function of the galaxy stellar mass follows the shape of the median AGN luminosity, when the latter is brighter on average than the XRB population. 
In Illustris and EAGLE, the median total luminosity increases with stellar mass, up to the most massive galaxies. 
In TNG100, we find a strong evidence for the efficient low-accretion mode AGN feedback effect in massive galaxies, especially in the low-SFR sample. We find the same feature in the SIMBA populations. Depending on the scaling relations for the XRB that we employ, the decrease due to AGN feedback of the galaxy total hard X-ray luminosity in massive galaxies can be washed out by high luminosities of the XRB or can also still be there in case of lower XRB luminosities. \\

Different XRB scaling relations have a strong influence on the total luminosity, changing it by more than one order of magnitude depending on the relations employed.
In Illustris, TNG100 and SIMBA, the choice of the XRB scaling relation particularly affects the low redshifts $z\leqslant 1$. We note that it affects TNG100 and SIMBA even more in the low-sSFR galaxies (red lines): particularly in massive galaxies of $M_{\star}\geqslant 10^{10.2}\, \rm M_{\odot}$ in TNG100 at $z\geqslant 1$ and all galaxies at $z=0$, and both in low-mass galaxies of $M_{\star}\leqslant 10^{9.5}\, \rm M_{\odot}$ and massive galaxies of $M_{\star}\geqslant 10^{10.5}\, \rm M_{\odot}$ in SIMBA at all redshifts. 
In EAGLE, the choice of the XRB scaling relation impacts all galaxy types ({\it high-sSFR}, {\it intermediate-sSFR}, and {\it low-sSFR} samples), at all redshifts.
The differences that we find here in the impact of the scaling relation 
are inversely related to the median of the AGN luminosity (black lines in Fig.~\ref{fig:L_total_Illustris_new}). If the AGN are luminous and brighter than the XRB population the choice of the scaling relation does not make a huge difference, because the XRBs do not contribute significantly to the total luminosity of the galaxies. This is the case in most of the simulations at high redshift. Instead, if the median AGN luminosity is small, as it is the case in EAGLE at all redshifts, the contribution of the XRBs becomes more important, and the choice of the XRB scaling relation as well.

The obscuration of a given fraction of the faint AGN could reinforce the contribution of the XRB in the total X-ray luminosity of the galaxies. In Fig.~\ref{fig:L_total_Illustris_new}, we show the median hard X-ray luminosity of the AGN populations with different models for their obscuration in dashed (model M1), dotted (M2), dashed-dotted (M3), and loosely dotted (M4) lines. We only plot the models for which we decrease the luminosity of the obscured faint AGN by one order of magnitude. 
For the Illustris, TNG100, and SIMBA simulations, the largest impact of our obscuration models is found at the highest redshift ($z=4$). At this redshift the obscuration can lead to almost one order of magnitude decrease in the AGN median luminosity. 
In general, we find very little difference between our models M1, M2, M3 and M4 in the median of the AGN luminosity. 
At lower redshifts, the effect of obscuration slightly diminishes but is still significant. In some simulations, the effect of obscuration decreases with the host galaxy stellar mass.
In practice, if there is a large fraction of obscured AGN among the faint AGN, the relative contribution of the XRB population to the observable galaxy X-ray luminosity would be larger. Consequently the shape of the $L_{\rm total}-M_{\star}$ relation would be almost completely driven by the XRB in these simulations, for all galaxy sSFR groups, and at any stellar mass $M_{\star}$ (except e.g., the low-mass end of TNG100 and SIMBA). 

\subsection{Comparison of the galaxy total hard X-ray luminosity with recent observations}
Here, we compare the population of simulated faint AGN to observations of stacked galaxies presented in \citet{2018ApJ...865...43F}. The observational study presents the total galaxy X-ray luminosity extracted from $\sim 75000$ stacked star-forming galaxies in the redshift range $0.1<z<5$, using the {\it Chandra} COSMOS Legacy survey.

\subsubsection{Mimicking the observation analysis: defining new sSFR galaxy samples}
We follow \citet{2018ApJ...865...43F}, and we divide the simulated galaxy populations into three new subsets. The {\it high-sSFR} subset now consists of galaxies with a sSFR higher than $10^{-8.5}$ yr$^{-1}$. The {\it intermediate-sSFR} subset includes galaxies whose sSFR range from $10^{-9.5}$ yr$^{-1}$ to $10^{-8.5}$ yr$^{-1}$. Finally, the {\it low-sSFR} subset includes all galaxies with a sSFR lower than $10^{-9.5}$ yr$^{-1}$. To mimic the observation selection, we also now exclude galaxies with $\rm sSFR\leqslant 10^{-11}\, \rm M_{\odot}/yr$ from our low-sSFR sample. 
For the simulations, this means that we remove galaxies with very low sSFR, which in some simulations corresponds to a non-negligible number of quenched galaxies.
The new samples are shown in Fig.~\ref{fig:paper_main_sequence_faintAGN} for all the simulations. Compared to our previous samples, the most important difference is that the SFR range covered by galaxies in a given sample will not change with redshift. For example, galaxies in the {\it intermediate-sSFR} sample are mostly on the star-forming main sequence of the simulations at $z=4$ or $z=2$, while they are clearly among the most star-forming galaxies at $z=0$. We provide a summary of the sample definitions employed in this section and in the previous sections in Table~\ref{tab:table_cuts}.

\begin{table*}
    \caption{Different galaxy samples and different cuts applied to the simulations.}
    \begin{tabular}{l|l}
    \hline
    \hline
    \multicolumn{2}{l}{Analysis of the large-scale cosmological simulations Illustris, TNG100, EAGLE, and SIMBA.}\\
    \multicolumn{2}{l}{Used in sections 2, 4, 5, 6.1 and Fig. 3, 5, 6, 10}\\
    \hline
    \hline
        \multicolumn{2}{l}{}\\
    \multicolumn{2}{l}{Galaxy samples defined as redshift- and simulation-dependent.}\\
        \multicolumn{2}{l}{}\\
    {\it High-sSFR} sample & 0.5 dex above the star-forming main sequence.\\
    {\it Intermediate-sSFR} sample & 1 dex around the star-forming main sequence.\\
    {\it Low-sSFR} sample & 0.5 dex below the star-forming main sequence.\\
    \multicolumn{2}{l}{}\\
    \hline
    \multicolumn{2}{l}{}\\
    \multicolumn{2}{l}{}\\
    \multicolumn{2}{l}{}\\
    \multicolumn{2}{l}{}\\
    \hline
    \hline
    \multicolumn{2}{l}{Comparison to \citet{2018ApJ...865...43F}}\\
    \multicolumn{2}{l}{Used in Section 6.2 and Fig. 11, 12, 13, B1, C1, C2}\\ 
    \hline
    \hline    
    \multicolumn{2}{l}{} \\
    \multicolumn{2}{l}{Galaxy samples identical to \citet{2018ApJ...865...43F}.}\\
    \multicolumn{2}{l}{}\\
    {\it High-sSFR} sample & $\rm sSFR/yr >10^{-8.5}$\\
    {\it Intermediate-sSFR} sample & $\rm 10^{-9.5}<sSFR/yr<10^{-8.5}$\\
    {\it Low-sSFR} sample & $\rm 10^{-11}<sSFR/yr<10^{-9.5}$\\
    \multicolumn{2}{l}{}\\
    \hline
    \multicolumn{2}{l}{}\\
    \multicolumn{2}{l}{Cuts to mimic the study of \citet{2018ApJ...865...43F}}\\
    \multicolumn{2}{l}{}\\
    sSFR cut & $\rm sSFR\geqslant 10^{-11} \, \rm yr^{-1}$ \\
    & \\
    Upper luminosity cut &  Detection of individual AGN with the COSMOS survey in the 2-10 keV band. \\
    & Depends on redshift and volume of the simulation.\\
    & AGN brighter than this upper limit are removed from the samples.\\

    \hline

    \end{tabular}

    \label{tab:table_cuts}
\end{table*}

\subsubsection{Mimicking the observation analysis: the COSMOS sensitivity curve and upper luminosity limits}
As before, we only include galaxies whose AGN could not be detected as a point source by the X-ray instrument \textit{Chandra} (COSMOS survey), i.e. galaxies with AGN luminosities smaller than the sensitivity limit of the COSMOS survey. 

The COSMOS survey luminosity sensitivity to detect objects depends a set of parameters, including their distances to the instrument (i.e., their redshift), and the sensitivity of the instrument. 
Here, we 
use the flux-area sensitivity curve from \citet{2016ApJ...819...62C} (their Fig. 16, for the $0.5-2\, \rm keV$ band). 
The sensitivity limit of the COSMOS survey depends on the area of sky covered by the instrument, with a lower sensitivity at the edges of the pointings that are combined to build the survey. 
We compute the $0.5-2\, \rm keV$ flux sensitivity of the COSMOS survey for the Illustris, TNG100, EAGLE, and SIMBA simulations at redshifts $z=0,1,2,3,4$ (using the simulations cosmology). 
We convert these sensitivity limits from the $0.5-2\, \rm keV$ band to the $2-10\, \rm keV$ band with the following k-correction:
\begin{equation}
    F_{\rm 2-10 \, keV}=F_{\rm 0.5-2\, keV} \frac{10^{2-\gamma}-2^{2-\gamma}}{2^{2-\gamma}-0.5^{2-\gamma}} (1+z)^{\gamma -2}
\end{equation}
with $\gamma = 1.4$ \citep{2004A&A...419..837D}. For redshift $z=0$, we directly use the same flux sensitivity limit as \cite{2018ApJ...865...43F}.
The flux received from a given object with an intrinsic luminosity $L$ depends on the redshift of this object: $F=L\,/\, 4 \pi d_{\rm L}^2$, with $d_{\rm L}$ the luminosity distance.
The luminosity distance $d_{\rm L}$ is computed from the angular diameter distance with $d_{\rm L} = (1+z)^2 d_{\rm A}$.
We invert the flux/luminosity equation and convolve it with the instrument flux sensitivity limit to compute the luminosity limit for individual AGN detection in the COSMOS survey. We use this luminosity upper limit in the following: we only keep galaxies with AGN fainter than this limit.
At $z=0$, the hard X-ray COSMOS limit is $L_{\rm AGN}\sim 2\times 10^{41}\, \rm erg/s$ for all the simulations. The limit is $L_{\rm AGN}\sim 6.2-6.5\times 10^{42}, \, 1.9-2.8\times 10^{43}, \, 3.8-5.9\times 10^{43}, \, 6.2-8.0\times 10^{43}\, \rm erg/s$ for $z=1,2,3,4$, respectively. The exact value for each simulation depends on the cosmology and volume of the given simulation, but the variations do not affect our findings.

\begin{figure*}
\centering
\includegraphics[scale=0.35]{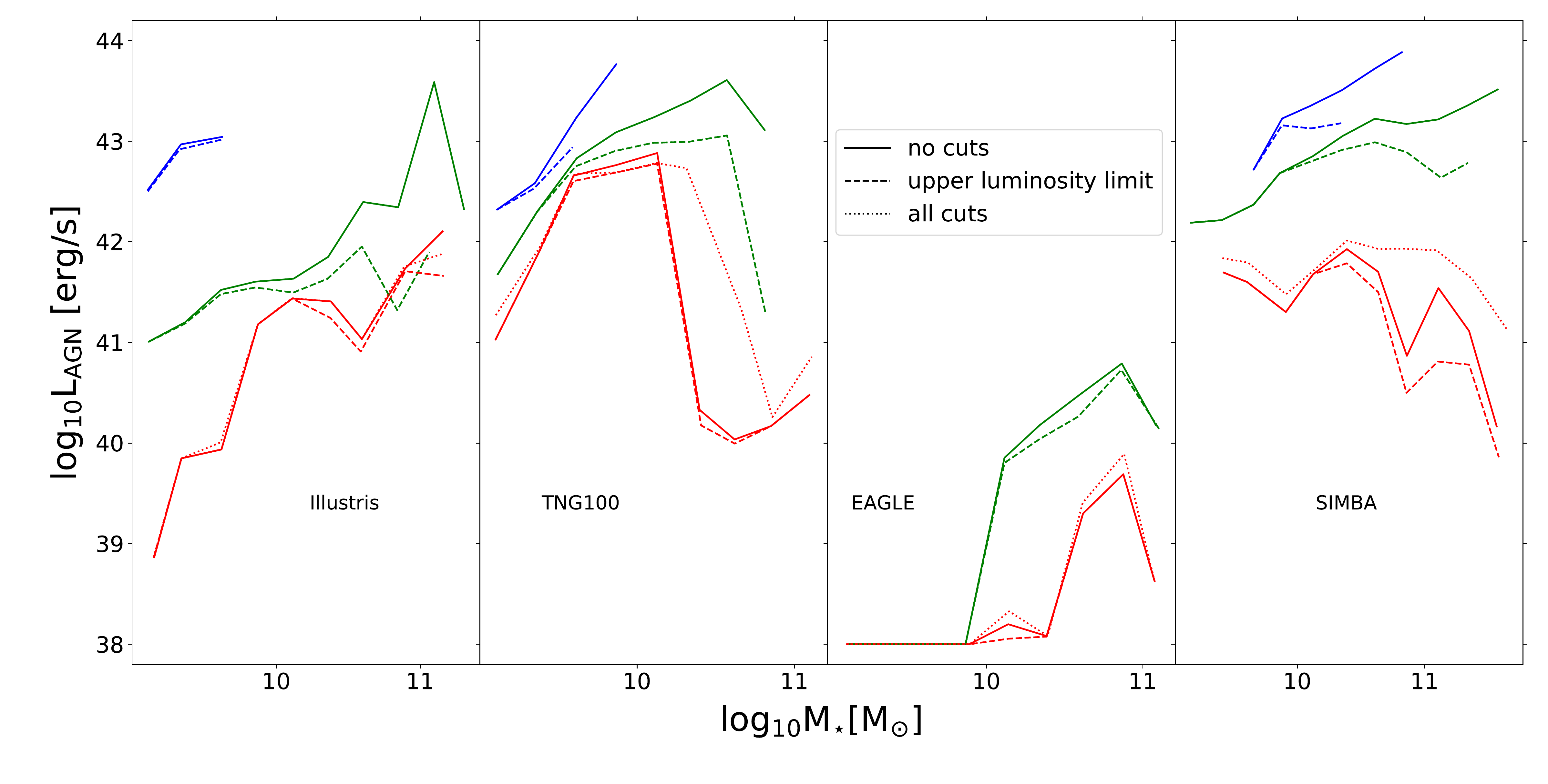}
\caption{Impact of the COSMOS sensitivity cut for individual AGN detection (dashed lines) and the sSFR cut (dotted lines) on the median AGN hard X-ray luminosities (2-10 keV, solid lines) for the three galaxy samples of all the simulations (blue for {\it high-sSFR} galaxies, green for {\it intermediate-sSFR} ones, and red for {\it low-sSFR} galaxies). Only galaxies with $\rm sSFR\geqslant 10^{-11}\, \rm yr^{-1}$, and hosted AGN fainter than individual detections in the COSMOS surveys are considered here.  
Median luminosity values lower than $10^{38}$ erg/s are shown at this value. 
We only show the results for $z=2$, for which we have statistics in the three galaxy samples, but we find similar results for other redshifts. The COSMOS sensitivity cut leads to a lower median AGN luminosity and the sSFR cut leads to a higher median AGN luminosity in the {\it low-sSFR} sample.
}
\label{fig:cuts_comparison}
\end{figure*}

The effect of the different cuts that we use is shown in Fig.~\ref{fig:cuts_comparison}, at $z=2$ for illustration. Solid lines indicate the median AGN luminosity in the three galaxy samples {\it high-sSFR}, {\it intermediate-sSFR}, and {\it low-sSFR}, without any cuts.
We first apply the upper limit for individual detection in the COSMOS survey (dashed lines).
In Illustris, this cut leads to a decrease of the median luminosity for the {\it intermediate-sSFR} galaxies (green lines) with $\log_{10} M_{\star}/\rm M_{\odot} \gtrsim 10$ up to one order of magnitude at $\log_{10} M_{\star}/\rm M_{\odot} \approx 10.8$. 
In TNG100, the median luminosity decreases for $\log_{10} M_{\star}/\rm M_{\odot} \gtrsim 9.5$. The effect of this cut increases with the galaxy stellar mass. This cut does not affect significantly the {\it low-sSFR} galaxies. 
For the {\it high-sSFR} and {\it intermediate-sSFR} galaxies, the cut has a simular effect in the SIMBA simulation. We note a stronger effect for the {\it low-sSFR} galaxies than in the TNG100.
The effect of the upper luminosity cut is the smallest in the EAGLE simulation.

The lower $\rm sSFR\geqslant 10^{-11}\,yr^{-1}$cut only affects the median values for low-sSFR galaxies (red lines) and its effect can be seen as the difference between the dashed line (when we apply the COSMOS cut) and the dotted line (COSMOS cut + sSFR cut). 
In Illustris, the sSFR cut does not have any effect on the median luminosity. 
In TNG100, the sSFR cut leads to an increase of two orders of magnitude of the median AGN luminosity in the stellar mass bin $10.25<\log_{10}M_{\star}/\rm M_{\odot}<10.5$ and to an increase of one order of magnitude in the stellar mass bin $10.5<\log_{10} M_{\star}/\rm M_{\odot}<10.75$. 
This corresponds to the stellar mass ranges in which a significant fraction of these massive galaxies have reduced SFR due to the efficient quenching of the kinetic AGN feedback mode in the TNG model (see Fig.~\ref{fig:M_starVsSFR_Illustris}). Many {\it low-sSFR} galaxies with small luminosities are excluded which leads to an increase in the median luminosity. We find a similar behavior in the SIMBA simulation.
There is almost no impact of the sSFR cut on the median AGN luminosity in the EAGLE, only a small increase of the luminosity for the {\it low-sSFR} galaxies.

\subsubsection{Comparison of the galaxy total X-ray luminosity to observations}
\begin{figure*}
\centering
\includegraphics[scale=0.7]{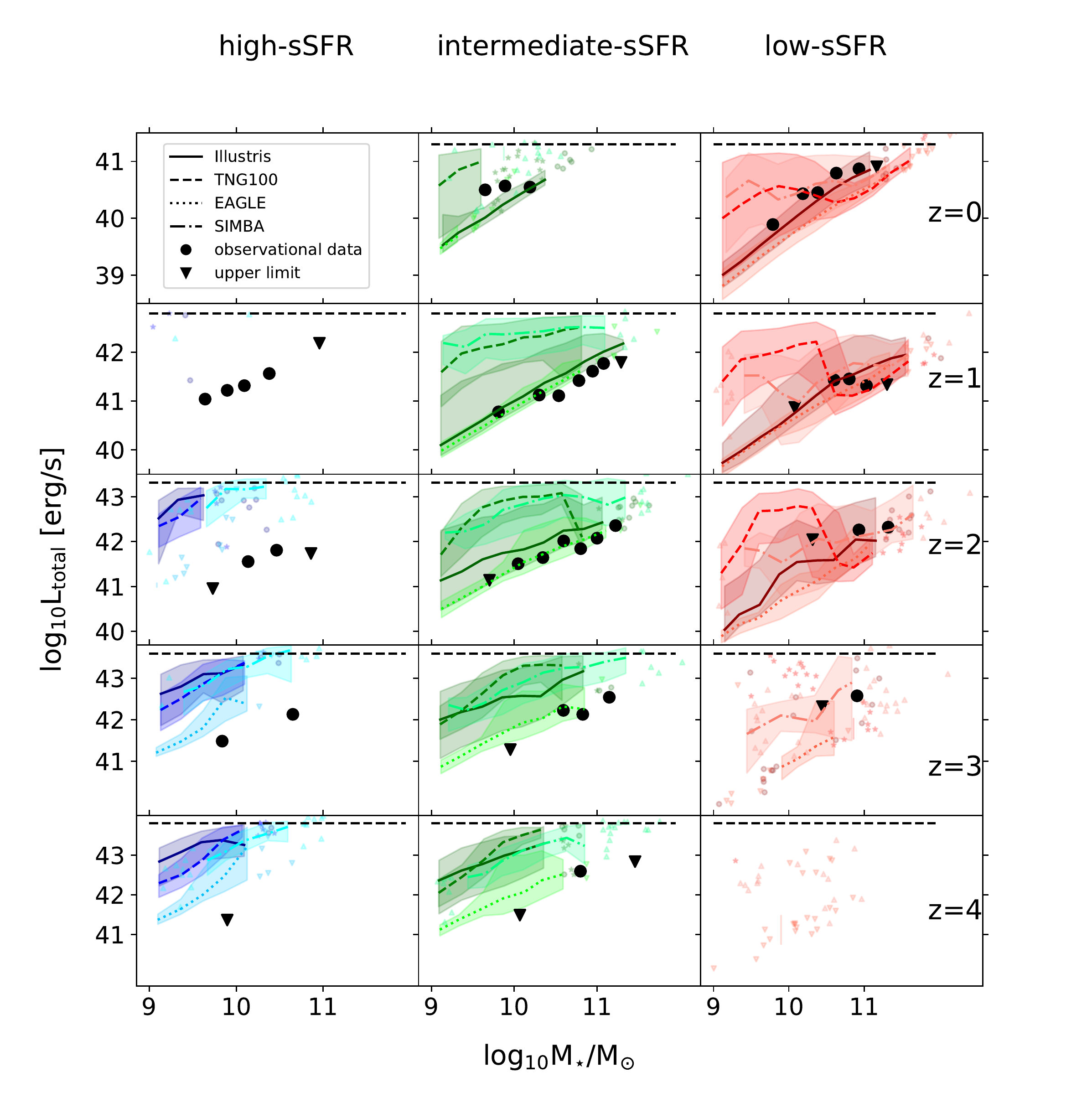}
\caption{Median galaxy total hard X-ray (2-10 keV) luminosities ($L_{\rm total}=L_{\rm AGN}+L_{\rm XRB}$). Different line styles are used for the Illustris, TNG100, EAGLE, and SIMBA simulations. 
The galaxies are divided into three different sSFR subsets: red for sSFR$<10^{-9.5}$yr$^{-1}$, green for $10^{-9.5}$yr$^{-1}<$sSFR$<10^{-8.5}$yr$^{-1}$ and blue for sSFR$>10^{-8.5}$yr$^{-1}$. Furthermore, galaxies with sSFR$<10^{-11}$yr$^{-1}$ are excluded as well as galaxies with AGN fainter than the individual COSMOS detection limit which is indicated by the black dashed line (we only show the limit computed for the Illustris simulation for simplicity). The lines indicate the median values for stellar mass bins with at least 10 galaxies, otherwise we show the individual galaxies (colored circles for Illustris, stars for TNG100, downward triangles for EAGLE and upward triangles for SIMBA). The shaded regions indicate the 15-85th percentile range of each subset. Here, the XRB correction from \citet{2019ApJS..243....3L} is used. 
The results are compared to the observed total luminosities of stacked galaxies from \citet{2018ApJ...865...43F} (black dots). The triangle symbols indicate upper limits of the observations. The uncertainty of the observational data is smaller than the width of the symbols used here.}
\label{fig:L_total_comparison}
\end{figure*}

We show the median of the total X-ray luminosity of the galaxies $L_{\rm total}=L_{\rm AGN}+L_{\rm XRB}$ (excluding the hot gas component) in Fig.~\ref{fig:L_total_comparison} for the {\it high-sSFR} sample (blue, left panels), the {\it intermediate-sSFR} galaxies (green, middle panels), and the {\it low-sSFR} galaxies (red, right panels). Here, we decided to work with the median of the galaxy luminosity but the mean luminosity provides almost identical values.
All the limits/cuts explained above are applied\footnote{Galaxies with sSFR$<10^{-11}$ yr$^{-1}$ are excluded as well as galaxies with AGN fainter than the individual COSMOS detection limit which is indicated by the black dashed line in all the panels. The COSMOS upper limits (dashed black lines) are shown for Illustris only in Fig.~\ref{fig:L_total_comparison}, which is why points for other simulations may be higher.}.

The results are compared to the observed total luminosities of stacked galaxies from \citet{2018ApJ...865...43F} (black dots). The contribution from the hot gas is included in the total X-ray luminosity in \citet{2018ApJ...865...43F}, and not in the simulations. The triangle symbols indicate upper limits of the observations. 
The observed X-ray luminosities are corrected for obscuration based on the measured X-ray hardness ratios which provide a rough measure, and mostly impact the low-redshift data \citet{2018ApJ...865...43F}.
We discuss this further in the following section.

In the observations, the total X-ray luminosity of the galaxies increases with their stellar mass for both the galaxies in the {\it high-sSFR} and {\it intermediate-sSFR} samples. For the {\it low-sSFR} sample, the correlation is less obvious (except at $z=0$, for which an increase with $M_{\star}$ is favored), and the total X-ray luminosity is more or less constant for massive galaxies with $M_{\star}\geqslant 10^{10.5}\, \rm M_{\odot}$.
In the following, we describe our general findings, first for these models ignoring the impact of obscuration, on the agreement between the simulations and the observations for each of the galaxy sSFR subsets.\\

\noindent {\it High-sSFR} samples ($\rm sSFR>10^{-8.5}\, yr^{-1}$):\\
\noindent We find that all the simulations have a higher median of the X-ray total luminosity of their galaxy population for the {\it high-sSFR} samples, compared to the observational constraints of \citet{2018ApJ...865...43F}. The luminosity for the simulations does not include the hot gas component, and therefore, the difference with the observations could be potentially larger. 
All the simulations have an increasing galaxy total luminosity with the stellar mass of the galaxies, in agreement with the trend in the observations.\\

\noindent {\it Intermediate-sSFR} samples ($\rm 10^{-9.5}\, yr^{-1}<sSFR<10^{-8.5}\, yr^{-1}$):\\
\noindent The trend obtained for the {\it intermediate-sSFR} sample of simulated galaxies is also similar to the observed one. We find a good agreement between the EAGLE simulation and the observations, for all redshift except $z=0$.
The other simulations overpredict the total X-ray luminosity of their {\it intermediate-sSFR} galaxies, on average. The overprediction is stronger towards less massive galaxies, and can be more than one order of magnitude.\\

\noindent {\it Low-sSFR} samples ($\rm 10^{-11}\, yr^{-1}<sSFR<10^{-9.5}\, yr^{-1}$):\\
\noindent
Finally, for the {\it low-sSFR} sample galaxies, we find that on average, the TNG100 and SIMBA simulations overpredict the total luminosity for the faint AGN low-mass host galaxies.  A better general agreement is found for the EAGLE and SIMBA simulations, which show an increasing total luminosity with increasing stellar mass.
In more detail, we find that the best agreement at $z=3, 2$ is obtained for the SIMBA simulation. At $z=1$, the Illustris, EAGLE, and SIMBA simulations provide a good agreement with the observations. While the Illustris and EAGLE have increasing galaxy total luminosity with stellar mass, the X-ray median luminosity is decreasing in SIMBA up to $M_{\star}\sim 10^{10}\, \rm M_{\odot}$, and then increasing slightly to $M_{\star}\sim 10^{11}\, \rm M_{\odot}$. For more massive galaxies, the median luminosity of SIMBA stays constant, as in the observations.
At $z=0$, the best agreement is found for the Illustris simulation. This is particularly true for the low-mass galaxies $M_{\star}\leqslant 10^{10}\, \rm M_{\odot}$ where the median luminosity of the Illustris simulation is lower than the SIMBA simulation.
\\

For different redshifts and galaxy types, we find that the highest/lowest galaxy total X-ray luminosity is not always produced by the same simulations. We develop this in the following.
For the three galaxy samples and all the redshifts, we find that the EAGLE simulation produces the lowest median total X-ray luminosity (dotted lines). This is because EAGLE is the simulation with the largest population of faint AGN \citep{2016MNRAS.462..190R}. The AGN X-ray luminosity function of EAGLE is below all the other simulations \citep[e.g.,][]{2015MNRAS.452..575S,2019MNRAS.487.5764T,2019MNRAS.484.4413H}. The luminosity function is in agreement with \citet{2015MNRAS.451.1892A} and \citet{2015ApJ...802...89B} in the range $\log_{10}L_{\rm x,\, 2-10\, keV}\leqslant 10^{44}\,\rm  erg/s$ at high redshift for $z\geqslant 2$, in the range $\log_{10} L_{\rm x,\, 2-10\, keV}\leqslant 10^{43}\, \rm erg/s$ at $z=2$, and is underestimated otherwise. At $z=0$, the EAGLE simulation is below the constraints of \citet{2015MNRAS.451.1892A} and \citet{2015ApJ...802...89B}. These constraints do not cover AGN with fainter luminosity than $\log_{10} L_{\rm x,\, 2-10\, keV}=10^{42}\, \rm erg/s$.
We find that for the {\it high-sSFR} samples the highest median is in Illustris for $M_{\star}\leqslant 10^{10}\, \rm M_{\odot}$, and TNG100 for $M_{\star}\sim 10^{10}\, \rm M_{\odot}$, and in SIMBA for more massive galaxies.
For the {\it intermediate-sSFR} samples, TNG100 has the highest median for galaxies with $10^{9.5}\, \rm M_{\odot}\leqslant M_{\star}\leqslant 10^{10.5}\, \rm M_{\odot}$ at $z=2, 3, 4$, while SIMBA has the highest median at $z=1$ and for galaxies with $M_{\star}\geqslant 10^{10.5}\, \rm M_{\odot}$ at $z=1, 2, 3, 4$.
TNG100 produces the highest total luminosity median for {\it low-sSFR} sample galaxies of $M_{\star}\leqslant \,\rm a \, few\, 10^{10}\, M_{\odot}$, and SIMBA and Illustris for more massive galaxies.

Interestingly, we know from Fig.~\ref{fig:cuts_comparison} that the fact that observations here exclude galaxies with $\rm sSFR<10^{-11}\, yr^{-1}$ can artificially boost the galaxy total X-ray luminosity for the {\it low-sSFR} galaxies. This would particularly affect the median in the bins containing massive galaxies. 
In our analysis we use the same sSFR cut as in the observations of \citet{2018ApJ...865...43F}, however this does not ensure that the sSFR distributions of the simulated and observed samples are the same. 
Depending on the number of galaxies with sSFR close to the limit $\rm sSFR<10^{-11}\,  yr^{-1}$ in the observations, we may still be counting too many simulated galaxies with very low sSFR. This could explain the lower total luminosity in massive galaxies, particularly for the TNG100, and SIMBA simulations (as they were already affected by the $\rm sSFR<10^{-11}\, yr^{-1}$ cut in Fig.~\ref{fig:cuts_comparison}).

\begin{figure*}
\centering
\includegraphics[scale=0.8]{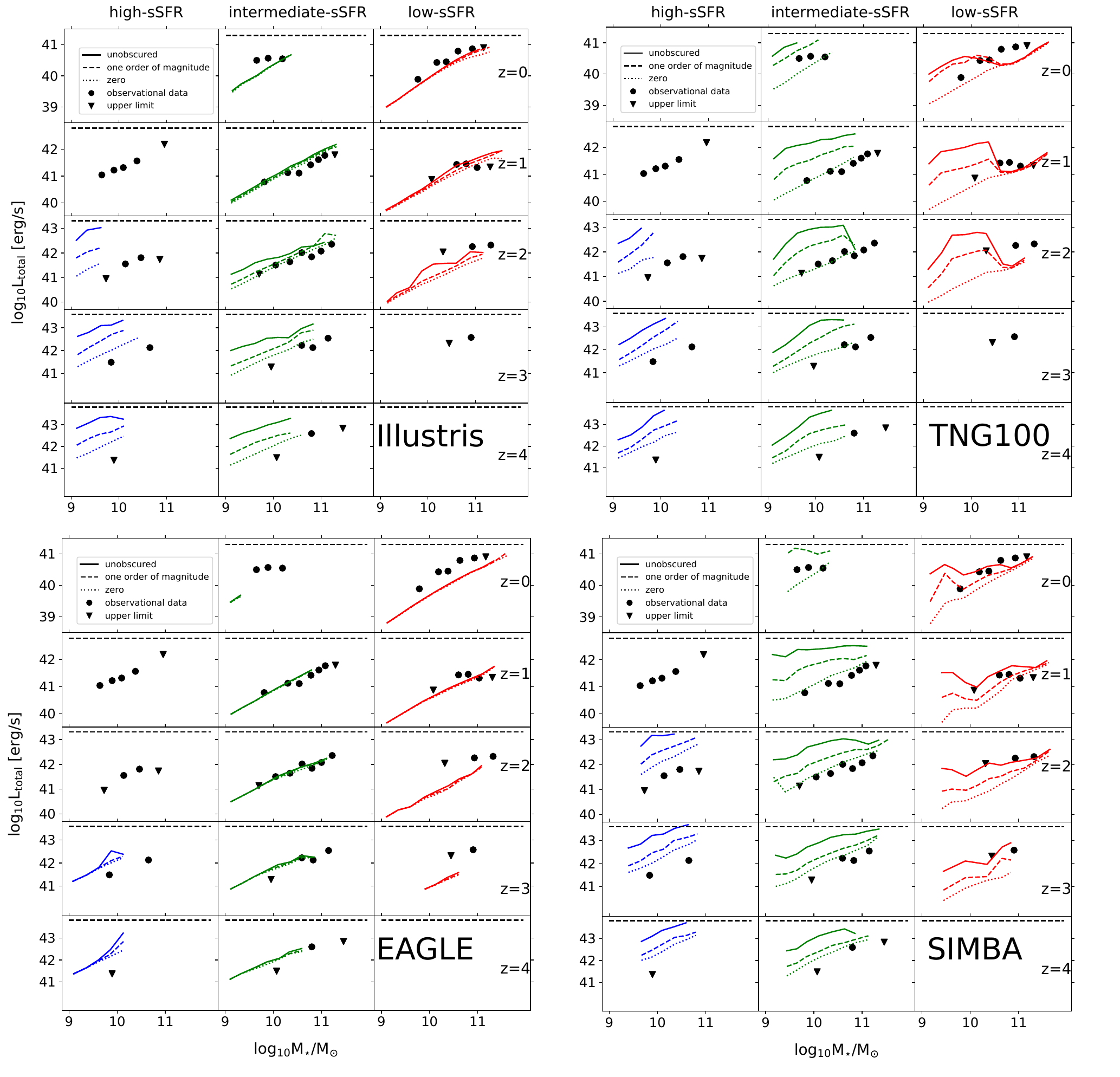}
\caption{Median galaxy total hard X-ray (2-10 keV) luminosities ($L_{\rm total}=L_{\rm AGN}+L_{\rm XRB}$). Same as Fig.~\ref{fig:L_total_comparison} but here we account for given fractions of obscured AGN (see Fig.~\ref{fig:obscuration}). Solid lines show $L_{\rm total}$ without considering that any AGN is obscured.
The dashed lines indicate the model where the luminosity of obscured AGN is reduced by one order of magnitude ({\it fainter AGN } M1 model) and the dotted lines indicate the model where the luminosity of obscured AGN is set to zero ({\it missed AGN} M1 model).
The results are compared to the observed total luminosities of stacked galaxies from \citet{2018ApJ...865...43F} (black symbols).
Typically, the median (or mean) $L_{\rm total}$ can be affected by about half an order of magnitude when we account for AGN obscuration ({\it fainter AGN } model), and even more if we assume that the obscured AGN would be completely missed by the observations ({\it missed AGN} model). 
With strong AGN obscuration models, the AGN features that were visible in the galaxy total luminosity (i.e., nonlinearity of the $L_{\rm tot}-M_{\star}$ relation, peak due to large fraction of bright AGN, decrease of $L_{\rm tot}$ because of AGN feedback) can be completely erased. Here, we used the XRB scaling relation of \citet{2019ApJS..243....3L} which predicts the highest XRB luminosity among all XRB models. Using other XRB models or having more obscured AGN, leads to smaller total galaxy luminosity.
}
\label{fig:L_total_comparison_obscuration}
\end{figure*}

Observing faint systems is challenging. We can not rule out that the faintest systems that we see in simulations are actually missed by the observations. Observations are always limited when trying to observe fainter and fainter systems. 
The completeness of the observational samples  is  particularly  difficult  to  address  at  the  faint  end  of the  galaxy  distribution.  This  can  potentially  create  significant discrepancies with the samples of simulated galaxies, since in the latter we have access to the faintest galaxies and AGN.
Therefore, in a separate test, we applied different lower luminosity cuts to our samples of simulated galaxies. In practice, to mimic what could take place in the observations we assumed that (a) galaxies with total X-ray luminosity below $10^{40}$ or $10^{41}\, \rm erg/s$ \citep[for $z=0,1$ and $z=2, 3, 4$, respectively, following][]{2018ApJ...865...43F} are not detected, or (b) that galaxies with total luminosity two or three orders of magnitude below the COSMOS Legacy limit are not detected. We found that missing the faintest galaxies when {\it observing} the simulations would mostly impact the total luminosity of low-mass galaxies with $M_{\star}\leqslant 10^{10.5}\, \rm M_{\odot}$ for the strongest luminosity cut. 
The median/mean luminosity of the detected galaxies would be increased compared to the median/mean luminosity of the full galaxy population with the same physical characteristics,
but it would not strongly impact our main conclusions regarding the agreement between simulations and the observations of \citet{2018ApJ...865...43F}.

\subsubsection{Comparison of the galaxy total X-ray luminosity to observations when accounting for AGN obscuration}

The observations of \citet{2018ApJ...865...43F} already include some rough estimate for obscuration, assuming that the XRB and the AGN are subject to the same obscuration \citep[which may or may not be a valid assumption, see the discussion in][]{2018ApJ...865...43F}.
Their correction mostly impacts the total luminosity at low redshift. An upper limit on the impact of the correction is $\sim 0.4$ dex difference for $z<0.6$, 0.2 dex difference in the range $0.6<z<2.3$, and negligible differences at higher redshift. Given the uncertainties on this {\it average} correction applied to the XRB and AGN, and in order to quantify the role of AGN obscuration on the total luminosity, we apply our correction to the simulations. 

In Fig.~\ref{fig:L_total_comparison_obscuration} we show the same figure as Fig.~\ref{fig:L_total_comparison}, i.e., the median of the galaxy hard X-ray luminosity, but this time we account for obscured AGN.
We note that there are no strong differences between the model variation M1, M2, M3, M4 in our results, which means that the fractions of obscured AGN with $L_{\rm AGN}\leqslant 10^{41}\, \rm erg/s$ does not impact the total galaxy luminosity. In practice, this is because the median of the galaxy X-ray luminosity is often higher than $10^{41}\, \rm erg/s$.
We apply the M1 obscuration model to the simulated AGN samples. More precisely, we show two sets of obscuration models, for which we either decrease by one order of magnitude the AGN luminosity ({\it fainter AGN} models), or either assume $L_{\rm AGN}=0$ ({\it missed AGN}). 
We describe below the impact of our two sets of obscuration models on the total galaxy luminosity, and re-examine our previous conclusions.

In general, we find that obscuration can decrease the total galaxy luminosity by up to one order of magnitude for the {\it fainter AGN} models.
When we assume that we would completely miss the obscured AGN ({\it missed AGN} models), the impact can be higher and reach two orders of magnitudes at stellar masses for which simulations produce a large fraction of AGN.
This is the case in the TNG100 and SIMBA simulations, which have a large fraction of AGN for galaxies with stellar masses of $M_{\star}\leqslant 10^{10.5}\, \rm M_{\odot}$. When we set $L_{\rm AGN}=0$ for the obscured AGN, the total galaxy luminosity is fully driven by the XRB population.

For the {\it high-sSFR} galaxies, adding the obscuration models does not impact our previous conclusions: all the simulations overpredict the total galaxy luminosity. For all the simulations except EAGLE, the galaxy luminosity of the {\it high-sSFR} subsets is dominated by the AGN luminosity, which explains the decrease of the luminosity when we obscure some of the AGN. We note that even when we fully remove the obscured AGN, i.e., when the galaxy luminosity is dominated by the XRB population, the luminosity is higher than for the observations. This could mean that the empirical scaling relations for the luminosity of the XRB population overpredict the median/mean hard X-ray galaxy luminosity.

For galaxies in the {\it intermediate-sSFR} group and for most of the simulations, our fiducial model with the XRB scaling relation of \citet{2019ApJS..243....3L} and the {\it fainter-AGN} obscuration models still produce higher galaxy luminosity (see Fig.~\ref{fig:L_total_comparison_obscuration}). A better agreement is found if we assume lower XRB scaling relations \citep{2010ApJ...724..559L,2016ApJ...825....7L,2017MNRAS.465.3390A,2018ApJ...865...43F} 
or that the obscured faint AGN are completely missed by the observations.

For galaxies in the {\it low-sSFR} subsets in Illustris and EAGLE, we find that the total galaxy luminosity is in general smaller than in the observations. For these two simulations, the discrepancy would be larger for other XRB scaling relations (as shown in Fig.~\ref{fig:L_total_Illustris_new}) or obscuration models accounting for more obscured AGN.
In SIMBA, the agreement between the total galaxy luminosity from simulations (with the L19 XRB relation and without AGN obscuration) and observations is good for $z>1$. We note some differences at $z=0$, with a higher galaxy luminosity median for simulated galaxies with $M_{\star}\leqslant 10^{10}\,\rm M_{\star}$, and lower median for $M_{\star}\geqslant 10^{10.5}\,\rm M_{\star}$ than in observations. Obscuration models assuming more obscured AGN would in general lower the good agreement of SIMBA with observations. 
TNG100 has a particular shape of the total galaxy luminosity median, which is not clearly visible in observations, with a higher luminosity median in low-mass galaxies, and lower luminosity median in massive galaxies, compared to observations. Assuming more obscured AGN in the simulations reduces the differences with observations.

From our analysis we find that for some simulations such as Illustris or EAGLE, whose total luminosity at low redshift is dominated by XRB, are not impacted by our modeling of AGN obscuration. For the other simulations with a higher contribution of the AGN luminosity to the total galaxy luminosity, without considering our obscuration correction we often find an excess of luminosity in the simulations compared to the observations. We have demonstrated that this excess can be reduced by obscuration at low redshift for all {\it intermediate-sSFR} galaxies and {\it high-sSFR} galaxies with $M_{\star}\leqslant 10^{10.5} \, \rm M_{\odot}-10^{11} \, \rm M_{\odot}$ (depending on the simulation). Similarly, the excess could be due to an overestimate of the obscuration in the observation data \citep{2018ApJ...865...43F} at low redshift.

\section{Discussion}
In this section, we discuss different aspects that could impact our results, and our comparison with current observations of galaxy total hard X-ray luminosity.

\subsection{Calibration of the cosmological simulations and their AGN X-ray luminosity functions}
The BH subgrid models of these simulations have been broadly calibrated against one of the empirical $M_{\rm BH}-M_{\rm bulge}$ relations available in the literature. While the Illustris, TNG100, and EAGLE simulations adjust the efficiency parameter of the AGN feedback model, the simulation SIMBA instead calibrates the accretion efficiency. None of the simulations studied here were calibrated against the AGN luminosity function. 

At $z=0$, these simulations 
have X-ray luminosity functions in good overall agreement with the constraints of, e.g., \citet{2015ApJ...802...89B}.  
TNG100 has a higher normalization of the luminosity function \citep{2019MNRAS.484.4413H}, and EAGLE a lower normalization, for $\log_{10}\, L_{\rm AGN}/{\rm (erg/s)}\leqslant 43$ \citep{2016MNRAS.462..190R}. Illustris and SIMBA lie within the constraints of \citet{2015ApJ...802...89B} for this luminosity regime \citep[see][for the analysis of AGN properties with observational constraints]{2015MNRAS.452..575S,2019MNRAS.487.5764T}.
At higher redshift, all the simulations except EAGLE overpredict the observational constraints in the range $\log_{10}\, L_{\rm AGN}/{\rm (erg/s)}\leqslant 42.5-43$ \citep{2021MNRAS.503.1940H}.
The regime that we investigate here ($\log_{10}\, L_{\rm AGN}/{\rm (erg/s)}\leqslant 42$) is below the range which is usually constrained by observations of the X-ray luminosity function. The differences between the simulations and the observational constraints\footnote{See also \citet{2015MNRAS.452..575S,2018MNRAS.479.4056W,2016MNRAS.460.2979V,2016MNRAS.462..190R,2019MNRAS.487.5764T} for studies of the BH populations in the different simulations.} on the X-ray luminosity function could affect the galaxy total X-ray luminosity in a non-trivial way, for the different galaxy sSFR samples.

\subsection{Comparison to observations}

To make the analysis of the simulations consistent with the observations from \citet{2018ApJ...865...43F} (stacked galaxies from the \textit{Chandra} COSMOS Legacy survey), we have adopted the same sSFR cut (excluding simulated galaxies with $\rm sSFR\leqslant 10^{-11}\, \rm yr^{-1}$), and the same upper luminosity limit to exclude galaxies with AGN detectable by the \textit{Chandra} COSMOS Legacy survey.
These same cuts do not ensure that the sSFR distributions of the observation and simulation samples are similar. For example, if a {\it high-sSFR} simulation sample has a distribution peaking at larger sSFR values than the corresponding observational sample, this will lead to higher total galaxy luminosities in the simulations than in observations.
Therefore, we can not exclude that some discrepancies between the observations and the simulations are due to different distributions of their respectives sSFR samples.

Two additional caveats for our comparison in this paper are the X-ray emission from hot gas and the obscuration. 

\subsubsection{Hot gas contribution}
In our analysis, we have neglected the contribution to the X-ray emission from the hot
interstellar medium (ISM). This contribution was, however, studied in the observations considered here. 
The hot gas is expected to contribute to the X-ray emission with a diffuse, soft thermal component \citep{2018ApJ...865...43F}. The hot gas component is thought to be significant and to dominate the X-ray emission over the XRB emission for rest-frame energy of $<$ 1.5 keV (\cite{2016ApJ...825....7L}, their Fig. 4). For high energies of $>$ 1.5 keV, as it is the case in our analysis, the XRB emission is expected to dominate \citep{2016ApJ...825....7L}. There is currently no definitive quantification (from observations or theory) across galaxy masses, galaxy apertures, types and redshifts. 
The contribution of the hot gas could not only depend on energy band, but could also decrease with increasing redshift, and could be important mostly for $z<1$ (\cite{2016ApJ...825....7L}, results obtained assuming that the spectral energy distribution (SED) of star-forming galaxies do not change strongly with redshift). 
According to \cite{2012MNRAS.426.1870M}, the hot gas X-ray emission could increase with the SFR of a galaxy. In that case, the influence of the hot gas emission would be higher in the high-sSFR galaxy sample at a fixed stellar mass and at higher stellar masses at fixed sSFR. 
From a numerical and theoretical perspective, definitive assessments of the X-ray emission from the hot gas
is still needed, but some preliminary estimates of the gas X-ray emission within galaxies in the soft bands have been derived from the Illustris and TNG simulations \citep{2020MNRAS.494..549T}. 
We postpone the task of including the contribution of the X-ray emitting hot gas within galaxies to future work, as this requires a careful assessment of the dependence on X-ray wavelength and on aperture within which the mock or real observations are taken.

\subsubsection{Obscuration}
Gas and dust obscuration can play a crucial role when comparing observations to simulations.
It is not clear yet if the obscuration originates from material close to the galaxies nuclear region. In that case, the luminosity of the XRB population distributed within the galaxies would not be strongly obscured, but the AGN would be. The obscuration could also originate from material in the whole galaxy, and in that case it would lead to the obscuration of both the XRB population and the AGN.
\cite{Buchner:2016hzt} find that the gas on the galaxy scale is only responsible for a part of the Compton-thin AGN, and does not provide Compton-thick lines of sight. The heavily obscured (Compton-thick) AGN would therefore mostly result from obscuration in the nuclear region. If this is the case, the hard X-ray emission from the galaxy-wide XRB population would be less impacted than the emission from the AGN.

Simulations do not consistently capture the obscuration of the AGN, as obscuration can arise from regions close to the AGN on spatial scales below the simulation resolution. Therefore, we have tested the role of AGN obscuration by applying four different models to the simulated AGN samples. 
Our models depend on redshift and hard X-ray (2-10 keV) AGN luminosity for AGN with $L_{\rm AGN}\geqslant 10^{41} \,\rm erg/s$ \citep[][for more details]{2019MNRAS.484.4413H}, and only on hard X-ray AGN luminosity for fainter AGN. Obscuration could also depend on galaxy SFR, a parameter that we do not consider here.
We have not applied any further obscuration model to the XRB populations.

In practice, the galaxy total hard X-ray luminosity can be impacted significantly depending on the fraction of obscured AGN that we assume.
We find that the more the faint AGN are obscured the more the shape of the total galaxy luminosity as a function of galaxy stellar mass is driven by the XRB luminosity (since we do not apply any obscuration model to the XRB emission). 
If the linearity of the XRB empirical $L_{\rm XRB}-M_{\star}$ scaling relations is a good estimate \citep[as found in e.g.,][]{2019ApJS..243....3L} and in the presence of a large population of obscured faint AGN (or just a population of very faint AGN, such as in EAGLE), we should observe a linear $L_{\rm total}-M_{\star}$ relation in the observations, independently of the galaxy sSFR. In that case, we find that any deviation from a linear $L_{\rm total}-M_{\star}$ relation would be due to features of the AGN populations.

\subsection{Detection of AGN in dwarf galaxies}
It is now demonstrated that AGN can exist in dwarf galaxies \citep{2013ApJ...775..116R,2015ApJ...809L..14B,2016ApJ...817...20M,2018MNRAS.478.2576M,2020ARA&A..58..257G,2019MNRAS.488..685M,2020ApJ...888...36R}. 
Quantifying the fraction of galaxies hosting BHs and the BH mass distribution in dwarf galaxies can constrain the theoretical models of BH formation \citep{2012NatCo...3E1304G}. Since these galaxies have not evolved much over cosmic times compared to their massive counterparts, they could have retained the initial properties of BH formation: initial BH mass, and initial BH formation efficiency.
While the BH occupation fraction in low-mass galaxies in such large-scale simulations may not be relevant/accurate because of the simple seeding of BHs in massive galaxies or halos, the AGN occupation fraction is fundamental to address the connection between the AGN and their host galaxies (e.g., the correlations between AGN activity and the SFR of their galaxies).

The AGN found in dwarf galaxies in observations can generally be qualified as faint AGN \citep[][and references therein]{2016ApJ...817...20M,2018ApJ...863....1C,2020ApJ...898L..30M}. In this work, we have shown that in the Illustris, TNG100, and EAGLE
simulations the XRB population in galaxies of $M_{\star}\leqslant 10^{9.5}\, \rm M_{\odot}$ can outshine the AGN emission in hard X-rays. This is a significant issue when trying to detect an AGN in X-rays; detection in X-ray is one of the most common method to detect low-mass AGN in low-mass galaxies today. 
What is interesting is that the simulations do not predict the same trend with SFR. In Illustris and EAGLE, the XRB population outshine the AGN in $>90\%$ of the galaxies, whether these galaxies form stars efficiently (starburst) or not (below the star-forming main sequence). However, in TNG100, 
the XRB population outshine the AGN only in galaxies below the main sequence, but not in main-sequence galaxies or starburst galaxies. There, AGN activity is enhanced when SFR activity is enhanced.
Confronting current and future observations of AGN in dwarf galaxies to our results on the AGN population predicted by cosmological simulations will help us to understand the observations and at the same time to constrain our modeling of BH and galaxy physics in simulations.

\section{Conclusion}
We have analyzed the properties of the faint AGN (in the hard X-ray band 2-10 keV, $L_{\rm AGN}\leqslant 10^{42}\, \rm erg/s$) and their host galaxies of the four large-scale cosmological simulations Illustris, TNG100, EAGLE, and SIMBA. We have modeled the contribution from the XRB population and from the AGN (including their possible obscuration) to the total galaxy hard X-ray luminosity. 
We summarize below our main findings.

\begin{itemize}

    \item The properties of the faint AGN host galaxies
    vary from simulation to simulation (Fig.~\ref{fig:faint_AGN_Mstar_013}). Faint AGN of $L_{\rm AGN}\sim 10^{38}\, \rm erg/s$ can be powered by relatively massive BHs and be located in massive galaxies ($M_{\star}\gtrsim 10^{10}\, \rm M_{\odot}$) with reduced SFR (TNG100, SIMBA), or be powered by lower-mass BHs in less massive galaxies ($M_{\star}\lesssim 10^{10}\, \rm M_{\odot}$) still forming stars (Illustris, EAGLE).

    \item We find that the two possible behaviors described above depend on the effectiveness of AGN feedback in massive galaxies. In TNG100 and SIMBA, the efficient feedback taking place in massive galaxies reduces both their sSFR (Fig.~\ref{fig:faint_AGN_Mstar_013}), but also the ability of their BHs to accrete efficiently. AGN feedback drives the build up of the faint AGN population in these simulated galaxies.
    
    \item 
    In all the simulations, except EAGLE, most of the galaxies have brighter AGN than the XRB population, at high redshift ($z>2$, Fig.~\ref{fig:XRB_vs_AGN}). With time, the AGN number density decreases and consequently more and more galaxies have a brighter XRB population than their AGN.
    The general fainter population of AGN in EAGLE (at all redshifts) compared to the other simulations leads to a significant number of galaxies with a brighter XRB population than AGN.

    \item The relative contribution of the AGN and the XRB population to the XRB+AGN total galaxy hard X-ray luminosity depends on the stellar mass and the SFR of the galaxies (Fig.~\ref{fig:L_total_Illustris_new}). 
    Starburst galaxies host brighter AGN in all simulations and across redshift: the AGN luminosity dominates over the XRB luminosity in most of these galaxies.
    At low redshifts ($z\leqslant 3$), the efficient AGN feedback in TNG100 and SIMBA leads to a strong decrease of the median AGN luminosity in massive galaxies ($M_{\star}\geqslant 10^{10}\, \rm M_{\odot}$) with reduced sSFR, and more galaxies are dominated by XRB emission.
    
    \item In low-mass galaxies of $M_{\star}\leqslant 10^{9.5}\, \rm M_{\odot}$ at $z=0$, we find that the XRB emission always outshines the AGN emission in low-sSFR galaxies in all the simulations (Table~\ref{tab:L_as_a_function_of galaxy_properties}). The XRB still dominates in main-sequence and startburst galaxies in Illustris and EAGLE, but does not outshine the AGN in TNG100.
    This has important implications for the search of AGN in dwarf galaxies.
    
    \item The total AGN+XRB hard X-ray luminosity of faint AGN host galaxies (i.e., neglecting the hot ISM X-ray emission) increases with increasing $M_{\star}$, for all redshifts and all the simulations (Fig.~\ref{fig:L_total_Illustris_new}).
    We note a turnover for the massive TNG100 and SIMBA galaxies for which the lower AGN median luminosity (due to AGN feedback) propagates to the total galaxy hard X-ray luminosity.
    
    \item We find that a nonlinear $L_{\rm total}-M_{\star}$ relation in faint AGN galaxies (Fig.~\ref{fig:L_total_Illustris_new} and Fig.~\ref{fig:L_total_comparison_obscuration}) is explained by a nonlinear $L_{\rm XRB}-M_{\star}$ scaling relation (in that case XRB luminosity models need to be updated), or by peaks of AGN activity at some stellar masses. We find that the obscuration of faint AGN can completely erase these AGN signatures in the $L_{\rm total}-M_{\star}$ relation (see Fig.~\ref{fig:L_total_comparison_obscuration}). In that case, the shape of the $L_{\rm total}-M_{\star}$ relation would be fully driven by the XRB emission. 
    
    \item The simulations, with our modeling of AGN and XRB luminosity, tend to overestimate the total AGN+XRB galaxy X-ray luminosity in the high-sSFR sample and for most of the simulations in the intermediate-sSFR sample (neglecting the hot gas ISM X-ray emission) compared to the observations of the COSMOS Legacy stacked galaxies \citep{2018ApJ...865...43F}.
    Simulated galaxies with $\rm sSFR > 10^{-9.5}\, yr^{-1}$ are too bright.
    Galaxies with with $\rm 10^{-9.5}\, yr^{-1}< sSFR < 10^{-8.5}\, yr^{-1}$ are also too bright, except a good agreement for EAGLE.
    For low-sSFR galaxies of $\rm 10^{-11}\, yr^{-1}< sSFR < 10^{-9.5}\, yr^{-1}$, we find that some simulations underestimate or overestimate the median galaxy luminosity (Fig.~\ref{fig:L_total_comparison}).

    \item In both simulations and observations \citep{2018ApJ...865...43F}, 
    {\it high-sSFR} galaxies have higher total galaxy X-ray luminosity than {\it low-sSFR} galaxies at fixed stellar mass, in general (Fig.~\ref{fig:L_total_comparison}). 
    
    \item The empirically-driven XRB scaling relations used in this work span 0.5 dex in luminosity (at fixed $M_{\star}$), which is about the same order of magnitude as some of our obscuration models (Fig.~\ref{fig:L_total_comparison_obscuration}). These two aspects are highly degenerate and further observational constraints will be needed to disentangle them.

\end{itemize}

Our work and predictions pave the way for upcoming and concept space missions such as Athena, AXIS, and Lynx. These missions will increase by several orders of magnitude the sensitivity of the current X-ray instruments, and will allow us to make promising progress on our understanding of faint AGN, a luminosity regime that as we have demonstrated can be dominated by X-ray binaries for specific sSFR and $M_{\star}$ regimes.

\section*{Acknowledgment}
We thank Mar Mezcua for a fruitful discussion. We thank Nhut Truong for sharing his catalogs of TNG galaxy hot gas X-ray emission with us.
RSK acknowledges financial support from the German Research Foundation (DFG) via the Collaborative Research Center (SFB 881, Project-ID 138713538) 'The Milky Way System' (subprojects A1, B1, B2, and B8). He also thanks for funding from the Heidelberg Cluster of Excellence STRUCTURES in the framework of Germany's Excellence Strategy (grant EXC-2181/1 - 390900948) and for funding from the European Research Council via the ERC Synergy Grant ECOGAL (grant 855130). DAA acknowledges support by NSF grant AST-2009687 and by the Flatiron Institute, which is supported by the Simons Foundation.

\section*{Data availability}
The data from the Illustris and the TNG100 simulations can be found on their respective websites: https://www.illustris-project.org, https://www.tng-project.org. The data from the EAGLE simulation can be obtained upon request to the EAGLE team at their website: http://icc.dur.ac.uk/Eagle/. 
The data from the SIMBA simulation can
be found on the website: http://simba.roe.ac.uk.

\appendix

\section{Definitions of the main sequence and galaxy number densities of the galaxy sSFR samples}

\subsection{Definition of the main sequence of all the simulations}

In Table~\ref{alpha_beta_table}, we present the parameters $\alpha$ and $\beta$ of the main sequence of each simulation for different redshifts as discussed in Section 2.4.1.

\begin{table}
  \begin{center}
    \caption{Parameters $\alpha$ and $\beta$ of the main sequence of each simulation for different redshifts. The galaxy populations and respective main sequences are shown in Fig.~\ref{fig:M_starVsSFR_Illustris}.
    }
    \scalebox{1}{
    \begin{tabular}{llll}
    &$z$&$\alpha$&$\beta$\\
    \hline
    Illustris & 0&-10.52 & 1.05\\
    &2&-8.63&0.96\\
    &4&-8.69&1.01\\
    \hline
    TNG100&0&-8.23&0.81\\
    &2&-7.33&0.82\\
    &4&-7.75&0.93\\
    \hline
    EAGLE&0&-9.50&0.92\\
    &2&-8.84&0.98\\
    &4&-7.35&0.87\\
    \hline
    SIMBA&0&-9.07&0.94\\
    &2&-13.73&1.50\\
    &4&-10.80&1.21\\

    \hline
    \label{alpha_beta_table}
    \end{tabular}
    }

  \end{center}
\end{table}

\subsection{Definitions and galaxy number densities of the galaxy sSFR samples}

In Table~\ref{tab:galaxy_numbers_for_different_subsets}, we present the percentages and number densities for the three different subsets that we used in Section 2, 4, 5, and 6. We only consider galaxies that are well resolved in all the simulations, i.e. $M_{\star}\geqslant 10^{9}\, \rm M_{\odot}$.
The {\it high-sSFR} sample groups starburst galaxies 0.5 dex above the main sequence. The {\it intermediate-sSFR} galaxies represent the galaxies on the main sequence (within 1 dex). Finally, {\it low-sSFR} galaxies are galaxies below 0.5 dex of the main sequence.

\begin{figure*}
\centering
\includegraphics[scale=0.46]{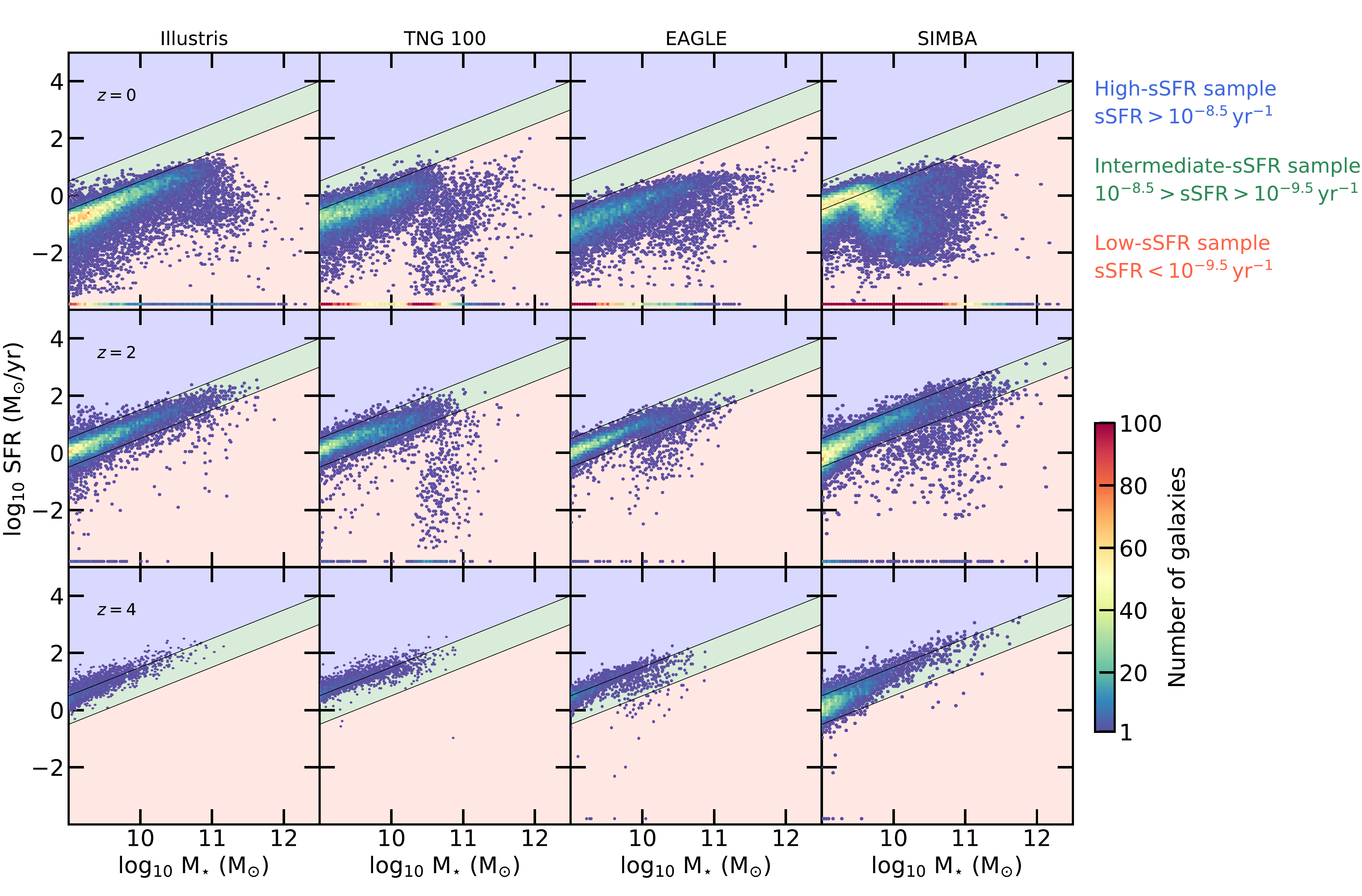}
\caption{SFR as a function of galaxy stellar mass for the Illustris, TNG100, EAGLE, and SIMBA simulations. 
Hexabins are color coded by the number of galaxies in bins. Finally, we show the three samples that we use with background colors: blue for galaxies with $\rm sSFR>10^{-8.5}$ yr$^{-1}$, green for galaxies with $\rm 10^{-9.5}<{\rm sSFR}<10^{-8.5}$ yr$^{-1}$, and red for $\rm 10^{-11}<sSFR<10^{-9.5}$ yr$^{-1}$. Galaxies with $sSFR<10^{-11}\, \rm yr^{-1}$ are shown at $\log_{10} $SFR$/(\rm M_{\odot}/yr)=-4$, but are not considered in the comparison with observations.
These definitions follows \citet{2018ApJ...865...43F}, and are different from the ones used in the first sections of the paper and showed in Fig.~\ref{fig:M_starVsSFR_Illustris}, especially for $z<2$.}
\label{fig:paper_main_sequence_faintAGN}
\end{figure*}

To compare the simulations to the results from observations of \citet{2018ApJ...865...43F}, we have changed the definition of our galaxy sSFR samples. 
{\it High-sSFR} galaxies are defined by $\rm sSFR/ yr>10^{-8.5}$, {\it intermediate-sSFR} galaxies by $\rm 10^{-9.5} <sSFR/ yr<10^{-8.5}$, and {\it low-sSFR} galaxies by $\rm 10^{-11} <sSFR/ yr<10^{-9.5}$.
Fig.~\ref{fig:paper_main_sequence_faintAGN} shows the SFR as a function of the stellar mass with these new definitions. 
The background colors show the three sSFR samples.
Contrary to Fig.~\ref{fig:M_starVsSFR_Illustris}, the division does not depend on redshift or simulation and galaxies on the star-forming main sequence of the simulations are not always included in the {\it intermediate-sSFR} sample.
In Table~\ref{tab:galaxy_numbers_for_different_subsets}, we add the percentages and number densities of the different subsets for our second set of definitions.

\begin{table*}
 \caption{Percentage of galaxies ($\%$) 
 and number density of galaxies $n$ ($10^{-5} \, \rm cMpc^{-3}$) in the three {\it high-sSFR}, {\it intermediate-sSFR}, and {\it low-sSFR} samples, for redshift $z=0,1,2,3,4$. Only galaxies with M$_{\star}\geqslant 10^{9}\, \rm M_{\odot}$ are included. This table refers to the division shown in Fig.~\ref{fig:M_starVsSFR_Illustris} and Fig.~\ref{fig:paper_main_sequence_faintAGN}.}
  \begin{center}
    \scalebox{1.1}{
    \begin{tabular}{ll|rr|rr|rr|rr|rr|rr}
    \hline
    &&\multicolumn{6}{c|}{Figure 3 (simulations)} & \multicolumn{6}{c}{Figure A1 (comparison to observations)}\\
    \hline
    & &\multicolumn{2}{c|}{{\it High-sSFR}} & \multicolumn{2}{c|}{{\it Interm-sSFR}} & \multicolumn{2}{c|}{{\it Low-sSFR}} &\multicolumn{2}{c|}{{\it High-sSFR}} & \multicolumn{2}{c|}{{\it Interm-sSFR}} & \multicolumn{2}{c}{{\it Low-sSFR}}\\
    &$z$ &\%&$n$ &\%&$n$&\%&$n$ &\%&$n$ &\%&$n$&\%&$n$\\
    \hline
    \hline
    Illustris & 0  & 3.19 & 72.27 & 79.58 & 1801.4 & 17.23 & 390.08 &0.0 & 0.0 & 2.07 & 46.77 & 97.93 & 2216.98\\
    &1 & 2.04 & 34.27 & 88.65 & 1492.36 & 9.31 & 156.71 &0.21 & 3.56 & 65.29 & 1099.14 & 34.49 & 580.65\\
    &2 & 1.79 & 17.22 & 91.99 & 885.47 & 6.22 & 59.85 &1.78 & 17.14 & 91.69 & 882.57 & 6.53 & 62.83\\
    &3 & 1.32 & 6.04 & 96.36 & 440.08 & 2.32 & 10.6 &11.18 & 51.08 & 88.06 & 402.17 & 0.76 & 3.48\\
    &4 & 2.57 & 4.97 & 96.79 & 187.18 & 0.64 & 1.24&36.47 & 70.53 & 63.53 & 122.85 & 0.0 & 0.0\\
    \hline
    TNG100&0 & 4.67 & 63.1 & 60.69 & 820.52 & 34.64 & 468.31&0.0 & 0.0 & 3.5 & 47.25 & 96.5 & 1304.69\\
    &1 & 3.21 & 36.78 & 80.63 & 924.83 & 16.16 & 185.39  &0.08 & 0.96 & 51.73 & 593.33 & 48.19 & 552.72 \\
    &2 & 2.28 & 17.32 & 89.36 & 679.58 & 8.36 & 63.62  &1.77 & 13.49 & 87.8 & 667.71 & 10.43 & 79.32 \\
    &3 & 1.14 & 4.42 & 94.88 & 369.09 & 3.98 & 15.48   &16.6 & 64.57 & 81.75 & 318.01 & 1.65 & 6.41\\
    &4 & 0.96 & 1.62 & 96.91 & 163.94 & 2.14 & 3.61  &63.14 & 106.81 & 36.73 & 62.14 & 0.13 & 0.22\\
    \hline
    EAGLE&0 & 3.4 & 45.2 & 61.95 & 823.9 & 34.65 & 460.9 &0.0 & 0.0 & 0.41 & 5.5 & 99.59 & 1324.5 \\
    &1 & 1.11 & 14.0 & 87.49 & 1101.5 & 11.4 & 143.5 &0.01 & 0.1 & 59.34 & 747.1 & 40.65 & 511.8 \\
    &2 & 0.47 & 3.8 & 93.21 & 749.3 & 6.32 & 50.8  &0.34 & 2.7 & 92.71 & 745.3 & 6.95 & 55.9\\
    &3 &0.73 & 2.9 & 94.34 & 377.0 & 4.93 & 19.7&5.11 & 20.4 & 91.27 & 364.7 & 3.63 & 14.5\\
    &4 & 0.39 & 0.7 & 94.03 & 170.0 & 5.59 & 10.1&35.56 & 64.3 & 61.95 & 112.0 & 2.49 & 4.5\\
    \hline
    SIMBA&0 & 1.68 & 18.52 & 48.05 & 529.75 & 50.28 & 554.34 &0.0 & 0.0 & 18.02 & 198.66 & 81.98 & 903.96 \\
    &1 & 1.14 & 7.51 & 66.22 & 434.83 & 32.64 & 214.32&0.23 & 1.54 & 67.65 & 444.26 & 32.11 & 210.86\\
    &2 & 2.81 & 10.82 & 81.78 & 315.0 & 15.41 & 59.36&3.26 & 12.55 & 85.31 & 328.61 & 11.43 & 44.02\\
    &3 & 3.9 & 9.97 & 89.45 & 228.37 & 6.65 & 16.98&5.3 & 13.52 & 87.92 & 224.47 & 6.79 & 17.33 \\
    &4 & 3.06 & 5.41 & 92.09 & 162.59 & 4.84 & 8.55 &10.51 & 18.55 & 87.78 & 154.98 & 1.71 & 3.02\\

    \hline
    \end{tabular}
    }
   
    \label{tab:galaxy_numbers_for_different_subsets}
  \end{center}
\end{table*}

\section{In which galaxies live the faint AGN: Number densities of galaxies present in the three AGN luminosity subsamples.}
In Table \ref{tab:numbers_luminosity_subsamples}, we report the number densities of the galaxies in the three subsets with different AGN luminosity ranges (L$_{\rm AGN} = 10^{37.5}$ erg/s - $10^{38.5}$ erg/s, L$_{\rm AGN} = 10^{39.5}$ erg/s - $10^{40.5}$ erg/s, L$_{\rm AGN} = 10^{41.5}$ erg/s - $10^{42.5}$ erg/s). From redshift $z=3$  to $z=1$, all number densities increase for all simulations. However, from $z=1$ to $z=0$, only the number densities for the galaxies with L$_{\rm AGN} \sim 10^{38}$ erg/s increases in Illustris. In TNG100 and SIMBA, the number densities of galaxies with L$_{\rm AGN} \sim 10^{38}$ erg/s and L$_{\rm AGN} \sim 10^{40}$ erg/s both increase.\\

\begin{table}
    \caption{The number densities (in $10^{-5}\rm cMpc^{-3}$) of galaxies present in the three AGN luminosity subsamples of Fig.~\ref{fig:faint_AGN_Mstar_013}, i.e. with hard X-ray luminosity of $L_{\rm AGN}\sim 10^{38}, 10^{40}, 10^{42}\, \rm erg/s$, for the redshifts $z=0, 1, 3$. We also report the number densities of galaxies in the full sample of galaxies, showed as grey histograms in Fig.~\ref{fig:faint_AGN_Mstar_013}.}
  \begin{center}
    \scalebox{1}{
    \begin{tabular}{llrrrr}
    \hline
      $z$ & $L_{\rm AGN} \, (\rm erg/s)$ &Illustris & TNG100 & EAGLE & SIMBA\\
      \hline
      \hline
      0&$10^{37.5}$-$10^{38.5}$ &333.21&102.02&73.3&21.70\\
      &$10^{39.5}$-$10^{40.5}$ &100.17&136.96&37.8&149.54\\
      &$10^{41.5}$-$10^{42.5}$ &42.47&314.76&9.7&107.47\\
      &all  &1975.74&1241.73&1245.5&725.20\\
      \hline
      1&$10^{37.5}$-$10^{38.5}$ &239.83&22.56&107.1&6.60\\
      &$10^{39.5}$-$10^{40.5}$ &256.30&64.87&75.8&15.82\\
      &$10^{41.5}$-$10^{42.5}$ &181.13&487.63&27.4&122.63\\
      &all  &1551.72&1091.28&1229.0&379.27\\
      \hline
      3&$10^{37.5}$-$10^{38.5}$ &4.55&0.15&24.4&0.03\\
      &$10^{39.5}$-$10^{40.5}$ &26.91&2.21&20.1&0.66\\
      &$10^{41.5}$-$10^{42.5}$ &150.91&167.63&19.0&26.32\\
      &all  &438.76&380.81&395.1&78.61\\

    \hline
    \end{tabular}
    }

    \label{tab:numbers_luminosity_subsamples}
  \end{center}
\end{table}

In Table~\ref{tab:KS_test}, we present the p values from the Kolmogorov-Smirnov test with
    distributions of the galaxy samples with L$_{\rm AGN} \sim 10^{38}$ erg/s and
    L$_{\rm AGN} \sim 10^{42}$ erg/s. These values are used in Section 3.

\begin{table}
  \begin{center}
    \caption{p values from the Kolmogorov-Smirnov test with the
    distributions of the galaxy samples with L$_{\rm AGN} \sim 10^{38}$ erg/s and
    L$_{\rm AGN} \sim 10^{42}$ erg/s. We only use the test for samples with at
    least 50 galaxies. The null hypothesis can be rejected if p$<0.01$,
    meaning that the distributions of different AGN luminosities are
    statistically different. We find that this is the case for almost all
    simulations and redshifts here.}
    \scalebox{1}{
    \begin{tabular}{ll|ccc}
    &&M$_{\star}$&M$_{\rm BH}$&sSFR\\
    \hline
    Simulation&$z$&p&p&p\\
    \hline
    Illustris & 0 &$<10^{-5}$&$<10^{-5}$&$<10^{-5}$\\
    &1&$<10^{-5}$&$<10^{-5}$&$<10^{-5}$\\
    &3&0.0005&$<10^{-5}$&0.0391\\
    \hline
    TNG100&0&$<10^{-5}$&$<10^{-5}$&$<10^{-5}$\\
    &1&$<10^{-5}$&$<10^{-5}$&$<10^{-5}$\\
    &3&-&-&-\\
    \hline
    EAGLE&0&0.0196&0.0017&0.1504\\
    &1&$<10^{-5}$&$<10^{-5}$&0.0035\\
    &3&$<10^{-5}$&$<10^{-5}$&0.0137\\
    \hline
    SIMBA&0&$<10^{-5}$&$<10^{-5}$&0\\
    &1&$<10^{-5}$&$<10^{-5}$&$<10^{-5}$\\
    &3&-&-&-\\
    \hline
    \end{tabular}}

  \label{tab:KS_test}    
  \end{center}
\end{table}

\bibliographystyle{mn2e}
\bibliography{biblio_complete,literature,biblio_complete-2}

\begin{thebibliography}{119}
\expandafter\ifx\csname natexlab\endcsname\relax\def\natexlab#1{#1}\fi

\bibitem[{{Aird}, {Coil} \& {Georgakakis}(2017){Aird}, {Coil}, \&
  {Georgakakis}}]{2017MNRAS.465.3390A}
{Aird} J., {Coil} A.~L., {Georgakakis} A., 2017, \mnras, 465, 3390

\bibitem[{{Aird}, {Coil} \& {Georgakakis}(2019){Aird}, {Coil}, \&
  {Georgakakis}}]{2019MNRAS.484.4360A}
{Aird} J., {Coil} A.~L., {Georgakakis} A., 2019, \mnras, 484, 4360

\bibitem[{{Aird} {et~al}\mbox{.}(2015){Aird}, {Coil}, {Georgakakis}, {Nandra},
  {Barro}, \& {P{\'e}rez-Gonz{\'a}lez}}]{2015MNRAS.451.1892A}
{Aird} J., {Coil} A.~L., {Georgakakis} A., {Nandra} K., {Barro} G.,
  {P{\'e}rez-Gonz{\'a}lez} P.~G., 2015, \mnras, 451, 1892

\bibitem[{{Angl{\'e}s-Alc{\'a}zar}
  {et~al}\mbox{.}(2017{\natexlab{a}}){Angl{\'e}s-Alc{\'a}zar}, {Dav{\'e}},
  {Faucher-Gigu{\`e}re}, {{\"O}zel}, \& {Hopkins}}]{2017MNRAS.464.2840A}
{Angl{\'e}s-Alc{\'a}zar} D., {Dav{\'e}} R., {Faucher-Gigu{\`e}re} C.-A.,
  {{\"O}zel} F., {Hopkins} P.~F., 2017{\natexlab{a}}, \mnras, 464, 2840

\bibitem[{{Angl{\'e}s-Alc{\'a}zar}
  {et~al}\mbox{.}(2017{\natexlab{b}}){Angl{\'e}s-Alc{\'a}zar},
  {Faucher-Gigu{\`e}re}, {Quataert}, {Hopkins}, {Feldmann}, {Torrey}, {Wetzel},
  \& {Kere{\v{s}}}}]{2017MNRAS.472L.109A}
{Angl{\'e}s-Alc{\'a}zar} D., {Faucher-Gigu{\`e}re} C.-A., {Quataert} E.,
  {Hopkins} P.~F., {Feldmann} R., {Torrey} P., {Wetzel} A., {Kere{\v{s}}} D.,
  2017{\natexlab{b}}, \mnras, 472, L109

\bibitem[{{Baldassare} {et~al}\mbox{.}(2015){Baldassare}, {Reines}, {Gallo}, \&
  {Greene}}]{2015ApJ...809L..14B}
{Baldassare} V.~F., {Reines} A.~E., {Gallo} E., {Greene} J.~E., 2015, \apjl,
  809, L14

\bibitem[{{Basu-Zych} {et~al}\mbox{.}(2016){Basu-Zych}, {Lehmer},
  {Hornschemeier}, {Fragos}, {Zezas}, {Yukita}, \&
  {Ptak}}]{2016HEAD...1540203B}
{Basu-Zych} A., {Lehmer} B., {Hornschemeier} A.~E., {Fragos} T., {Zezas} A.,
  {Yukita} M., {Ptak} A., 2016, in AAS/High Energy Astrophysics Division \#15,
  AAS/High Energy Astrophysics Division, p. 402.03

\bibitem[{{Basu-Zych} {et~al}\mbox{.}(2013{\natexlab{a}}){Basu-Zych}, {Lehmer},
  {Hornschemeier}, \& {Ptak}}]{2013HEAD...1330104B}
{Basu-Zych} A., {Lehmer} B., {Hornschemeier} A.~E., {Ptak} A.,
  2013{\natexlab{a}}, in AAS/High Energy Astrophysics Division, Vol.~13,
  AAS/High Energy Astrophysics Division \#13, p. 301.04

\bibitem[{{Basu-Zych} {et~al}\mbox{.}(2013{\natexlab{b}}){Basu-Zych}, {Lehmer},
  {Hornschemeier}, {Bouwens}, {Fragos}, {Oesch}, {Belczynski}, {Brandt},
  {Kalogera}, {Luo}, {Miller}, {Mullaney}, {Tzanavaris}, {Xue}, \&
  {Zezas}}]{2013ApJ...762...45B}
{Basu-Zych} A.~R. {et~al.}, 2013{\natexlab{b}}, \apj, 762, 45

\bibitem[{{Bluck} {et~al}\mbox{.}(2016){Bluck}, {Mendel}, {Ellison}, {Patton},
  {Simard}, {Henriques}, {Torrey}, {Teimoorinia}, {Moreno}, \&
  {Starkenburg}}]{2016MNRAS.462.2559B}
{Bluck} A. F.~L. {et~al.}, 2016, \mnras, 462, 2559

\bibitem[{{Bonfield} {et~al}\mbox{.}(2011){Bonfield}, {Jarvis}, {Hardcastle},
  {Cooray}, {Hatziminaoglou}, {Ivison}, {Page}, {Stevens}, {de Zotti}, {Auld},
  {Baes}, {Buttiglione}, {Cava}, {Dariush}, {Dunlop}, {Dunne}, {Dye}, {Eales},
  {Fritz}, {Hopwood}, {Ibar}, {Maddox}, {Micha{\l}owski}, {Pascale}, {Pohlen},
  {Rigby}, {Rodighiero}, {Serjeant}, {Smith}, {Temi}, \& {van der
  Werf}}]{2011MNRAS.416...13B}
{Bonfield} D.~G. {et~al.}, 2011, \mnras, 416, 13

\bibitem[{{Boroson}, {Kim} \& {Fabbiano}(2011){Boroson}, {Kim}, \&
  {Fabbiano}}]{2011ApJ...729...12B}
{Boroson} B., {Kim} D.-W., {Fabbiano} G., 2011, \apj, 729, 12

\bibitem[{{Brorby}(2016)}]{2016cxo..prop.4835B}
{Brorby} M., 2016, {Metallicity effects on high mass X-ray binary formation}.
  Chandra Proposal

\bibitem[{Buchner \& Bauer(2017)}]{Buchner:2016hzt}
Buchner J., Bauer F.~E., 2017, Mon. Not. Roy. Astron. Soc., 465, 4348

\bibitem[{{Buchner} {et~al}\mbox{.}(2015){Buchner}, {Georgakakis}, {Nandra},
  {Brightman}, {Menzel}, {Liu}, {Hsu}, {Salvato}, {Rangel}, {Aird}, {Merloni},
  \& {Ross}}]{2015ApJ...802...89B}
{Buchner} J. {et~al.}, 2015, \apj, 802, 89

\bibitem[{{{\c{C}}atmabacak} {et~al}\mbox{.}(2020){{\c{C}}atmabacak},
  {Feldmann}, {Angl{\'e}s-Alc{\'a}zar}, {Faucher-Gigu{\`e}re}, {Hopkins}, \&
  {Kere{\v{s}}}}]{2020arXiv200712185C}
{{\c{C}}atmabacak} O., {Feldmann} R., {Angl{\'e}s-Alc{\'a}zar} D.,
  {Faucher-Gigu{\`e}re} C.-A., {Hopkins} P.~F., {Kere{\v{s}}} D., 2020, arXiv
  e-prints, arXiv:2007.12185

\bibitem[{{Cheng} {et~al}\mbox{.}(2018){Cheng}, {Li}, {Xu}, {Li}, {Zhu}, \&
  {Fang}}]{2018ApJ...869...52C}
{Cheng} Z., {Li} Z., {Xu} X., {Li} X., {Zhu} Z., {Fang} T., 2018, \apj, 869, 52

\bibitem[{{Chilingarian} {et~al}\mbox{.}(2018){Chilingarian}, {Katkov},
  {Zolotukhin}, {Grishin}, {Beletsky}, {Boutsia}, \&
  {Osip}}]{2018ApJ...863....1C}
{Chilingarian} I.~V., {Katkov} I.~Y., {Zolotukhin} I.~Y., {Grishin} K.~A.,
  {Beletsky} Y., {Boutsia} K., {Osip} D.~J., 2018, \apj, 863, 1

\bibitem[{{Choi} {et~al}\mbox{.}(2012){Choi}, {Ostriker}, {Naab}, \&
  {Johansson}}]{2012ApJ...754..125C}
{Choi} E., {Ostriker} J.~P., {Naab} T., {Johansson} P.~H., 2012, \apj, 754, 125

\bibitem[{{Churazov} {et~al}\mbox{.}(2005){Churazov}, {Sazonov}, {Sunyaev},
  {Forman}, {Jones}, \& {B{\"o}hringer}}]{Churazov2005}
{Churazov} E., {Sazonov} S., {Sunyaev} R., {Forman} W., {Jones} C.,
  {B{\"o}hringer} H., 2005, MNRAS, 363, L91

\bibitem[{{Civano} {et~al}\mbox{.}(2016){Civano}, {Marchesi}, {Comastri},
  {Urry}, {Elvis}, {Cappelluti}, {Puccetti}, {Brusa}, {Zamorani}, {Hasinger},
  {Aldcroft}, {Alexand er}, {Allevato}, {Brunner}, {Capak}, {Finoguenov},
  {Fiore}, {Fruscione}, {Gilli}, {Glotfelty}, {Griffiths}, {Hao}, {Harrison},
  {Jahnke}, {Kartaltepe}, {Karim}, {LaMassa}, {Lanzuisi}, {Miyaji}, {Ranalli},
  {Salvato}, {Sargent}, {Scoville}, {Schawinski}, {Schinnerer}, {Silverman},
  {Smolcic}, {Stern}, {Toft}, {Trakhtenbrot}, {Treister}, \&
  {Vignali}}]{2016ApJ...819...62C}
{Civano} F. {et~al.}, 2016, \apj, 819, 62

\bibitem[{{Colbert} {et~al}\mbox{.}(2004){Colbert}, {Strickland}, {Veilleux},
  \& {Weaver}}]{2004AAS...20515501C}
{Colbert} E.~J.~M., {Strickland} D.~K., {Veilleux} S., {Weaver} K.~A., 2004, in
  American Astronomical Society Meeting Abstracts, Vol. 205, American
  Astronomical Society Meeting Abstracts, p. 155.01

\bibitem[{{Crain} {et~al}\mbox{.}(2015){Crain}, {Schaye}, {Bower}, {Furlong},
  {Schaller}, {Theuns}, {Dalla Vecchia}, {Frenk}, {McCarthy}, {Helly},
  {Jenkins}, {Rosas-Guevara}, {White}, \& {Trayford}}]{2015MNRAS.450.1937C}
{Crain} R.~A. {et~al.}, 2015, \mnras, 450, 1937

\bibitem[{{Dav{\'e}} {et~al}\mbox{.}(2019){Dav{\'e}}, {Angl{\'e}s-Alc{\'a}zar},
  {Narayanan}, {Li}, {Rafieferantsoa}, \& {Appleby}}]{2019MNRAS.486.2827D}
{Dav{\'e}} R., {Angl{\'e}s-Alc{\'a}zar} D., {Narayanan} D., {Li} Q.,
  {Rafieferantsoa} M.~H., {Appleby} S., 2019, \mnras, 486, 2827

\bibitem[{{Dav{\'e}}, {Thompson} \& {Hopkins}(2016){Dav{\'e}}, {Thompson}, \&
  {Hopkins}}]{2016MNRAS.462.3265D}
{Dav{\'e}} R., {Thompson} R., {Hopkins} P.~F., 2016, \mnras, 462, 3265

\bibitem[{{De Luca} \& {Molendi}(2004)}]{2004A&A...419..837D}
{De Luca} A., {Molendi} S., 2004, \aap, 419, 837

\bibitem[{{Donnari} {et~al}\mbox{.}(2021{\natexlab{a}}){Donnari}, {Pillepich},
  {Joshi}, {Nelson}, {Genel}, {Marinacci}, {Rodriguez-Gomez}, {Pakmor},
  {Torrey}, {Vogelsberger}, \& {Hernquist}}]{2021MNRAS.500.4004D}
{Donnari} M. {et~al.}, 2021{\natexlab{a}}, \mnras, 500, 4004

\bibitem[{{Donnari} {et~al}\mbox{.}(2021{\natexlab{b}}){Donnari}, {Pillepich},
  {Nelson}, {Marinacci}, {Vogelsberger}, \& {Hernquist}}]{2021MNRAS.tmp.1755D}
{Donnari} M., {Pillepich} A., {Nelson} D., {Marinacci} F., {Vogelsberger} M.,
  {Hernquist} L., 2021{\natexlab{b}}, \mnras

\bibitem[{{Donnari} {et~al}\mbox{.}(2019){Donnari}, {Pillepich}, {Nelson},
  {Vogelsberger}, {Genel}, {Weinberger}, {Marinacci}, {Springel}, \&
  {Hernquist}}]{2019MNRAS.485.4817D}
{Donnari} M. {et~al.}, 2019, \mnras, 485, 4817

\bibitem[{{Douna} {et~al}\mbox{.}(2015){Douna}, {Pellizza}, {Mirabel}, \&
  {Pedrosa}}]{2015A&A...579A..44D}
{Douna} V.~M., {Pellizza} L.~J., {Mirabel} I.~F., {Pedrosa} S.~E., 2015, \aap,
  579, A44

\bibitem[{{Dray}(2006)}]{2006MNRAS.370.2079D}
{Dray} L.~M., 2006, \mnras, 370, 2079

\bibitem[{Fabbiano(2006)}]{Fabbiano:2005pj}
Fabbiano G., 2006, Ann. Rev. Astron. Astrophys., 44, 323

\bibitem[{{Fornasini} {et~al}\mbox{.}(2018){Fornasini}, {Civano}, {Fabbiano},
  {Elvis}, {Marchesi}, {Miyaji}, \& {Zezas}}]{2018ApJ...865...43F}
{Fornasini} F.~M., {Civano} F., {Fabbiano} G., {Elvis} M., {Marchesi} S.,
  {Miyaji} T., {Zezas} A., 2018, \apj, 865, 43

\bibitem[{{Fornasini}, {Civano} \& {Suh}(2020){Fornasini}, {Civano}, \&
  {Suh}}]{2020MNRAS.495..771F}
{Fornasini} F.~M., {Civano} F., {Suh} H., 2020, \mnras, 495, 771

\bibitem[{{Fornasini} {et~al}\mbox{.}(2019){Fornasini}, {Kriek}, {Sanders},
  {Shivaei}, {Civano}, {Reddy}, {Shapley}, {Coil}, {Mobasher}, {Siana}, {Aird},
  {Azadi}, {Freeman}, {Leung}, {Price}, {Fetherolf}, {Zick}, \&
  {Barro}}]{2019ApJ...885...65F}
{Fornasini} F.~M. {et~al.}, 2019, \apj, 885, 65

\bibitem[{{Fragos} {et~al}\mbox{.}(2013){Fragos}, {Lehmer}, {Naoz}, {Zezas}, \&
  {Basu-Zych}}]{2013ApJ...776L..31F}
{Fragos} T., {Lehmer} B.~D., {Naoz} S., {Zezas} A., {Basu-Zych} A., 2013,
  \apjl, 776, L31

\bibitem[{{Furlong} {et~al}\mbox{.}(2015){Furlong}, {Bower}, {Theuns},
  {Schaye}, {Crain}, {Schaller}, {Dalla Vecchia}, {Frenk}, {McCarthy}, {Helly},
  {Jenkins}, \& {Rosas-Guevara}}]{2015MNRAS.450.4486F}
{Furlong} M. {et~al.}, 2015, \mnras, 450, 4486

\bibitem[{{Gaskin} {et~al}\mbox{.}(2018){Gaskin}, {Dominguez}, {Gelmis},
  {Mulqueen}, {Swartz}, {McCarley}, {{\"O}zel}, {Vikhlinin}, {Schwartz},
  {Tananbaum}, {Blackwood}, {Arenberg}, {Purcell}, \&
  {Allen}}]{2018SPIE10699E..0NG}
{Gaskin} J.~A. {et~al.}, 2018, in Society of Photo-Optical Instrumentation
  Engineers (SPIE) Conference Series, Vol. 10699, \procspie, p. 106990N

\bibitem[{{Genel} {et~al}\mbox{.}(2014){Genel}, {Vogelsberger}, {Springel},
  {Sijacki}, {Nelson}, {Snyder}, {Rodriguez-Gomez}, {Torrey}, \&
  {Hernquist}}]{2014MNRAS.445..175G}
{Genel} S. {et~al.}, 2014, \mnras, 445, 175

\bibitem[{{Georgakakis} {et~al}\mbox{.}(2015){Georgakakis}, {Aird}, {Buchner},
  {Salvato}, {Menzel}, {Brandt}, {McGreer}, {Dwelly}, {Mountrichas}, {Koki},
  {Georgantopoulos}, {Hsu}, {Merloni}, {Liu}, {Nandra}, \&
  {Ross}}]{2015MNRAS.453.1946G}
{Georgakakis} A. {et~al.}, 2015, \mnras, 453, 1946

\bibitem[{{Georgantopoulos} {et~al}\mbox{.}(2017){Georgantopoulos},
  {Pouliasis}, {Bonanos}, {Sokolovsky}, {Yang}, {Hatzidimitriou}, {Bellas},
  {Gavras}, \& {Spetsieri}}]{2017xru..conf...88G}
{Georgantopoulos} I. {et~al.}, 2017, in The X-ray Universe 2017, {Ness} J.-U.,
  {Migliari} S., eds., p.~88

\bibitem[{{Gilfanov}(2004)}]{2004MNRAS.349..146G}
{Gilfanov} M., 2004, \mnras, 349, 146

\bibitem[{{Gilli} {et~al}\mbox{.}(2014){Gilli}, {Norman}, {Vignali},
  {Vanzella}, {Calura}, {Pozzi}, {Massardi}, {Mignano}, {Casasola}, {Daddi},
  {Elbaz}, {Dickinson}, {Iwasawa}, {Maiolino}, {Brusa}, {Vito}, {Fritz},
  {Feltre}, {Cresci}, {Mignoli}, {Comastri}, \&
  {Zamorani}}]{2014A&A...562A..67G}
{Gilli} R. {et~al.}, 2014, \aap, 562, A67

\bibitem[{{Greene}(2012)}]{2012NatCo...3E1304G}
{Greene} J.~E., 2012, Nature Communications, 3, 1304

\bibitem[{{Greene}, {Strader} \& {Ho}(2020){Greene}, {Strader}, \&
  {Ho}}]{2020ARA&A..58..257G}
{Greene} J.~E., {Strader} J., {Ho} L.~C., 2020, \araa, 58, 257

\bibitem[{{Grimm}, {Gilfanov} \& {Sunyaev}(2003){Grimm}, {Gilfanov}, \&
  {Sunyaev}}]{2003ChJAS...3..257G}
{Grimm} H.-J., {Gilfanov} M., {Sunyaev} R., 2003, Chinese Journal of Astronomy
  and Astrophysics Supplement, 3, 257

\bibitem[{{G{\"u}ltekin} {et~al}\mbox{.}(2009){G{\"u}ltekin}, {Richstone},
  {Gebhardt}, {Lauer}, {Tremaine}, {Aller}, {Bender}, {Dressler}, {Faber},
  {Filippenko}, {Green}, {Ho}, {Kormendy}, {Magorrian}, {Pinkney}, \&
  {Siopis}}]{2009ApJ...698..198G}
{G{\"u}ltekin} K. {et~al.}, 2009, \apj, 698, 198

\bibitem[{{Habouzit} {et~al}\mbox{.}(2019){Habouzit}, {Genel}, {Somerville},
  {Kocevski}, {Hirschmann}, {Dekel}, {Choi}, {Nelson}, {Pillepich}, {Torrey},
  {Hernquist}, {Vogelsberger}, {Weinberger}, \&
  {Springel}}]{2019MNRAS.484.4413H}
{Habouzit} M. {et~al.}, 2019, \mnras, 484, 4413

\bibitem[{{Habouzit} {et~al}\mbox{.}(2021){Habouzit}, {Li}, {Somerville},
  {Genel}, {Pillepich}, {Volonteri}, {Dav{\'e}}, {Rosas-Guevara}, {McAlpine},
  {Peirani}, {Hernquist}, {Angl{\'e}s-Alc{\'a}zar}, {Reines}, {Bower},
  {Dubois}, {Nelson}, {Pichon}, \& {Vogelsberger}}]{2021MNRAS.503.1940H}
{Habouzit} M. {et~al.}, 2021, \mnras, 503, 1940

\bibitem[{{Habouzit}, {Volonteri} \& {Dubois}(2017){Habouzit}, {Volonteri}, \&
  {Dubois}}]{2017MNRAS.468.3935H}
{Habouzit} M., {Volonteri} M., {Dubois} Y., 2017, \mnras, 468, 3935

\bibitem[{{Hahn} {et~al}\mbox{.}(2019){Hahn}, {Starkenburg}, {Choi},
  {Dav{\'e}}, {Dickey}, {Geha}, {Genel}, {Hayward}, {Maller}, {Mandyam},
  {Pandya}, {Popping}, {Rafieferantsoa}, {Somerville}, \&
  {Tinker}}]{2019ApJ...872..160H}
{Hahn} C. {et~al.}, 2019, \apj, 872, 160

\bibitem[{{H{\"a}ring} \& {Rix}(2004)}]{2004ApJ...604L..89H}
{H{\"a}ring} N., {Rix} H.-W., 2004, \apjl, 604, L89

\bibitem[{{Hopkins}(2015)}]{2015MNRAS.450...53H}
{Hopkins} P.~F., 2015, \mnras, 450, 53

\bibitem[{{Hopkins}(2017)}]{2017arXiv171201294H}
{Hopkins} P.~F., 2017, arXiv e-prints, arXiv:1712.01294

\bibitem[{{Hopkins} \& {Quataert}(2011)}]{2011MNRAS.415.1027H}
{Hopkins} P.~F., {Quataert} E., 2011, \mnras, 415, 1027

\bibitem[{{Hopkins}, {Richards} \& {Hernquist}(2007){Hopkins}, {Richards}, \&
  {Hernquist}}]{Hop_bol_2007}
{Hopkins} P.~F., {Richards} G.~T., {Hernquist} L., 2007, ApJ, 654, 731

\bibitem[{{Kaaret}(2014)}]{2014MNRAS.440L..26K}
{Kaaret} P., 2014, \mnras, 440, L26

\bibitem[{{Kim} \& {Fabbiano}(2010)}]{2010ApJ...721.1523K}
{Kim} D.-W., {Fabbiano} G., 2010, \apj, 721, 1523

\bibitem[{{Kormendy} \& {Ho}(2013)}]{2013ARA&A..51..511K}
{Kormendy} J., {Ho} L.~C., 2013, \araa, 51, 511

\bibitem[{{Koudmani}, {Henden} \& {Sijacki}(2021){Koudmani}, {Henden}, \&
  {Sijacki}}]{2021MNRAS.503.3568K}
{Koudmani} S., {Henden} N.~A., {Sijacki} D., 2021, \mnras, 503, 3568

\bibitem[{{Koudmani} {et~al}\mbox{.}(2019){Koudmani}, {Sijacki}, {Bourne}, \&
  {Smith}}]{2019MNRAS.484.2047K}
{Koudmani} S., {Sijacki} D., {Bourne} M.~A., {Smith} M.~C., 2019, \mnras, 484,
  2047

\bibitem[{{Lehmer} {et~al}\mbox{.}(2010){Lehmer}, {Alexander}, {Bauer}, {Brand
  t}, {Goulding}, {Jenkins}, {Ptak}, \& {Roberts}}]{2010ApJ...724..559L}
{Lehmer} B.~D., {Alexander} D.~M., {Bauer} F.~E., {Brand t} W.~N., {Goulding}
  A.~D., {Jenkins} L.~P., {Ptak} A., {Roberts} T.~P., 2010, \apj, 724, 559

\bibitem[{{Lehmer} {et~al}\mbox{.}(2016){Lehmer}, {Basu-Zych}, {Mineo}, {Brand
  t}, {Eufrasio}, {Fragos}, {Hornschemeier}, {Luo}, {Xue}, {Bauer}, {Gilfanov},
  {Ranalli}, {Schneider}, {Shemmer}, {Tozzi}, {Trump}, {Vignali}, {Wang},
  {Yukita}, \& {Zezas}}]{2016ApJ...825....7L}
{Lehmer} B.~D. {et~al.}, 2016, \apj, 825, 7

\bibitem[{{Lehmer} {et~al}\mbox{.}(2014){Lehmer}, {Berkeley}, {Zezas}, {Alexand
  er}, {Basu-Zych}, {Bauer}, {Brandt}, {Fragos}, {Hornschemeier}, {Kalogera},
  {Ptak}, {Sivakoff}, {Tzanavaris}, \& {Yukita}}]{2014ApJ...789...52L}
{Lehmer} B.~D. {et~al.}, 2014, \apj, 789, 52

\bibitem[{{Lehmer} {et~al}\mbox{.}(2007){Lehmer}, {Brandt}, {Alexander},
  {Bell}, {McIntosh}, {Bauer}, {Hasinger}, {Mainieri}, {Miyaji}, {Schneider},
  \& {Steffen}}]{2007ApJ...657..681L}
{Lehmer} B.~D. {et~al.}, 2007, \apj, 657, 681

\bibitem[{{Lehmer} {et~al}\mbox{.}(2019){Lehmer}, {Eufrasio}, {Tzanavaris},
  {Basu-Zych}, {Fragos}, {Prestwich}, {Yukita}, {Zezas}, {Hornschemeier}, \&
  {Ptak}}]{2019ApJS..243....3L}
{Lehmer} B.~D. {et~al.}, 2019, \apjs, 243, 3

\bibitem[{{Linden} {et~al}\mbox{.}(2010){Linden}, {Kalogera}, {Sepinsky},
  {Prestwich}, {Zezas}, \& {Gallagher}}]{2010ApJ...725.1984L}
{Linden} T., {Kalogera} V., {Sepinsky} J.~F., {Prestwich} A., {Zezas} A.,
  {Gallagher} J.~S., 2010, \apj, 725, 1984

\bibitem[{{Liu} {et~al}\mbox{.}(2020){Liu}, {Veilleux}, {Canalizo}, {Rupke},
  {Manzano-King}, {Bohn}, \& {U}}]{2020ApJ...905..166L}
{Liu} W., {Veilleux} S., {Canalizo} G., {Rupke} D. S.~N., {Manzano-King} C.~M.,
  {Bohn} T., {U} V., 2020, \apj, 905, 166

\bibitem[{{Lutz} {et~al}\mbox{.}(2008){Lutz}, {Sturm}, {Tacconi}, {Valiante},
  {Schweitzer}, {Netzer}, {Maiolino}, {Andreani}, {Shemmer}, \&
  {Veilleux}}]{2008ApJ...684..853L}
{Lutz} D. {et~al.}, 2008, \apj, 684, 853

\bibitem[{{Madau} \& {Fragos}(2017)}]{2017ApJ...840...39M}
{Madau} P., {Fragos} T., 2017, \apj, 840, 39

\bibitem[{{Manzano-King}, {Canalizo} \& {Sales}(2019){Manzano-King},
  {Canalizo}, \& {Sales}}]{2019ApJ...884...54M}
{Manzano-King} C.~M., {Canalizo} G., {Sales} L.~V., 2019, \apj, 884, 54

\bibitem[{{Marinacci} {et~al}\mbox{.}(2018){Marinacci}, {Vogelsberger},
  {Pakmor}, {Torrey}, {Springel}, {Hernquist}, {Nelson}, {Weinberger},
  {Pillepich}, {Naiman}, \& {Genel}}]{2018MNRAS.480.5113M}
{Marinacci} F. {et~al.}, 2018, \mnras, 480, 5113

\bibitem[{{Matthee} \& {Schaye}(2019)}]{2019MNRAS.484..915M}
{Matthee} J., {Schaye} J., 2019, \mnras, 484, 915

\bibitem[{{McConnell} \& {Ma}(2013)}]{2013ApJ...764..184M}
{McConnell} N.~J., {Ma} C.-P., 2013, \apj, 764, 184

\bibitem[{{Merloni} {et~al}\mbox{.}(2014){Merloni}, {Bongiorno}, {Brusa},
  {Iwasawa}, {Mainieri}, {Magnelli}, {Salvato}, {Berta}, {Cappelluti},
  {Comastri}, {Fiore}, {Gilli}, {Koekemoer}, {Le Floc'h}, {Lusso}, {Lutz},
  {Miyaji}, {Pozzi}, {Riguccini}, {Rosario}, {Silverman}, {Symeonidis},
  {Treister}, {Vignali}, \& {Zamorani}}]{2014MNRAS.437.3550M}
{Merloni} A. {et~al.}, 2014, \mnras, 437, 3550

\bibitem[{{Mezcua} {et~al}\mbox{.}(2016){Mezcua}, {Civano}, {Fabbiano},
  {Miyaji}, \& {Marchesi}}]{2016ApJ...817...20M}
{Mezcua} M., {Civano} F., {Fabbiano} G., {Miyaji} T., {Marchesi} S., 2016,
  \apj, 817, 20

\bibitem[{{Mezcua} {et~al}\mbox{.}(2018){Mezcua}, {Civano}, {Marchesi}, {Suh},
  {Fabbiano}, \& {Volonteri}}]{2018MNRAS.478.2576M}
{Mezcua} M., {Civano} F., {Marchesi} S., {Suh} H., {Fabbiano} G., {Volonteri}
  M., 2018, \mnras, 478, 2576

\bibitem[{{Mezcua} \& {Dom{\'\i}nguez S{\'a}nchez}(2020)}]{2020ApJ...898L..30M}
{Mezcua} M., {Dom{\'\i}nguez S{\'a}nchez} H., 2020, \apjl, 898, L30

\bibitem[{{Mezcua}, {Suh} \& {Civano}(2019){Mezcua}, {Suh}, \&
  {Civano}}]{2019MNRAS.488..685M}
{Mezcua} M., {Suh} H., {Civano} F., 2019, \mnras, 488, 685

\bibitem[{{Mineo}, {Gilfanov} \& {Sunyaev}(2012{\natexlab{a}}){Mineo},
  {Gilfanov}, \& {Sunyaev}}]{2012MNRAS.419.2095M}
{Mineo} S., {Gilfanov} M., {Sunyaev} R., 2012{\natexlab{a}}, \mnras, 419, 2095

\bibitem[{{Mineo}, {Gilfanov} \& {Sunyaev}(2012{\natexlab{b}}){Mineo},
  {Gilfanov}, \& {Sunyaev}}]{2012MNRAS.426.1870M}
{Mineo} S., {Gilfanov} M., {Sunyaev} R., 2012{\natexlab{b}}, \mnras, 426, 1870

\bibitem[{{Mor} {et~al}\mbox{.}(2012){Mor}, {Netzer}, {Trakhtenbrot},
  {Shemmer}, \& {Lira}}]{2012ApJ...749L..25M}
{Mor} R., {Netzer} H., {Trakhtenbrot} B., {Shemmer} O., {Lira} P., 2012, \apjl,
  749, L25

\bibitem[{{Mushotzky} {et~al}\mbox{.}(2019){Mushotzky}, {Aird}, {Barger},
  {Cappelluti}, {Chartas}, {Corrales}, {Eufrasio}, {Fabian}, {Falcone},
  {Gallo}, {Gilli}, {Grant}, {Hardcastle}, {Hodges-Kluck}, {Kara}, {Koss},
  {Li}, {Lisse}, {Loewenstein}, {Markevitch}, {Meyer}, {Miller}, {Mulchaey},
  {Petre}, {Ptak}, {Reynolds}, {Russell}, {Safi-Harb}, {Smith}, {Snios},
  {Tombesi}, {Valencic}, {Walker}, {Williams}, {Winter}, {Yamaguchi}, {Zhang},
  {Arenberg}, {Brand t}, {Burrows}, {Georganopoulos}, {Miller}, {Norman}, \&
  {Rosati}}]{2019BAAS...51g.107M}
{Mushotzky} R. {et~al.}, 2019, in Bulletin of the American Astronomical
  Society, Vol.~51, p. 107

\bibitem[{{Naiman} {et~al}\mbox{.}(2018){Naiman}, {Pillepich}, {Springel},
  {Ramirez-Ruiz}, {Torrey}, {Vogelsberger}, {Pakmor}, {Nelson}, {Marinacci},
  {Hernquist}, {Weinberger}, \& {Genel}}]{2018MNRAS.477.1206N}
{Naiman} J.~P. {et~al.}, 2018, \mnras, 477, 1206

\bibitem[{{Nandra} {et~al}\mbox{.}(2013){Nandra}, {Barret}, {Barcons},
  {Fabian}, {den Herder}, {Piro}, {Watson}, {Adami}, {Aird}, {Afonso}, \&
  et~al.}]{2013arXiv1306.2307N}
{Nandra} K. {et~al.}, 2013, arXiv e-prints, arXiv:1306.2307

\bibitem[{{Nelson} {et~al}\mbox{.}(2015){Nelson}, {Pillepich}, {Genel},
  {Vogelsberger}, {Springel}, {Torrey}, {Rodriguez-Gomez}, {Sijacki}, {Snyder},
  {Griffen}, {Marinacci}, {Blecha}, {Sales}, {Xu}, \&
  {Hernquist}}]{2015A&C....13...12N}
{Nelson} D. {et~al.}, 2015, Astronomy and Computing, 13, 12

\bibitem[{{Nelson} {et~al}\mbox{.}(2019{\natexlab{a}}){Nelson}, {Pillepich},
  {Springel}, {Pakmor}, {Weinberger}, {Genel}, {Torrey}, {Vogelsberger},
  {Marinacci}, \& {Hernquist}}]{2019MNRAS.490.3234N}
{Nelson} D. {et~al.}, 2019{\natexlab{a}}, \mnras, 490, 3234

\bibitem[{{Nelson} {et~al}\mbox{.}(2018){Nelson}, {Pillepich}, {Springel},
  {Weinberger}, {Hernquist}, {Pakmor}, {Genel}, {Torrey}, {Vogelsberger},
  {Kauffmann}, {Marinacci}, \& {Naiman}}]{2018MNRAS.475..624N}
{Nelson} D. {et~al.}, 2018, \mnras, 475, 624

\bibitem[{{Nelson} {et~al}\mbox{.}(2019{\natexlab{b}}){Nelson}, {Springel},
  {Pillepich}, {Rodriguez-Gomez}, {Torrey}, {Genel}, {Vogelsberger}, {Pakmor},
  {Marinacci}, {Weinberger}, {Kelley}, {Lovell}, {Diemer}, \&
  {Hernquist}}]{2019ComAC...6....2N}
{Nelson} D. {et~al.}, 2019{\natexlab{b}}, Computational Astrophysics and
  Cosmology, 6, 2

\bibitem[{{Pillepich} {et~al}\mbox{.}(2018{\natexlab{a}}){Pillepich}, {Nelson},
  {Hernquist}, {Springel}, {Pakmor}, {Torrey}, {Weinberger}, {Genel}, {Naiman},
  {Marinacci}, \& {Vogelsberger}}]{2018MNRAS.475..648P}
{Pillepich} A. {et~al.}, 2018{\natexlab{a}}, \mnras, 475, 648

\bibitem[{{Pillepich} {et~al}\mbox{.}(2018{\natexlab{b}}){Pillepich},
  {Springel}, {Nelson}, {Genel}, {Naiman}, {Pakmor}, {Hernquist}, {Torrey},
  {Vogelsberger}, {Weinberger}, \& {Marinacci}}]{2018MNRAS.473.4077P}
{Pillepich} A. {et~al.}, 2018{\natexlab{b}}, \mnras, 473, 4077

\bibitem[{{Reines} {et~al}\mbox{.}(2020){Reines}, {Condon}, {Darling}, \&
  {Greene}}]{2020ApJ...888...36R}
{Reines} A.~E., {Condon} J.~J., {Darling} J., {Greene} J.~E., 2020, \apj, 888,
  36

\bibitem[{{Reines}, {Greene} \& {Geha}(2013){Reines}, {Greene}, \&
  {Geha}}]{2013ApJ...775..116R}
{Reines} A.~E., {Greene} J.~E., {Geha} M., 2013, \apj, 775, 116

\bibitem[{{Reines} \& {Volonteri}(2015)}]{2015arXiv150806274R}
{Reines} A.~E., {Volonteri} M., 2015, \apj, 813, 82

\bibitem[{{Rosario} {et~al}\mbox{.}(2012){Rosario}, {Santini}, {Lutz}, {Shao},
  {Maiolino}, {Alexander}, {Altieri}, {Andreani}, {Aussel}, {Bauer}, {Berta},
  {Bongiovanni}, {Brandt}, {Brusa}, {Cepa}, {Cimatti}, {Cox}, {Daddi}, {Elbaz},
  {Fontana}, {F{\"o}rster Schreiber}, {Genzel}, {Grazian}, {Le Floch},
  {Magnelli}, {Mainieri}, {Netzer}, {Nordon}, {P{\'e}rez Garcia}, {Poglitsch},
  {Popesso}, {Pozzi}, {Riguccini}, {Rodighiero}, {Salvato}, {Sanchez-Portal},
  {Sturm}, {Tacconi}, {Valtchanov}, \& {Wuyts}}]{2012A&A...545A..45R}
{Rosario} D.~J. {et~al.}, 2012, \aap, 545, A45

\bibitem[{{Rosas-Guevara} {et~al}\mbox{.}(2016){Rosas-Guevara}, {Bower},
  {Schaye}, {McAlpine}, {Dalla Vecchia}, {Frenk}, {Schaller}, \&
  {Theuns}}]{2016MNRAS.462..190R}
{Rosas-Guevara} Y., {Bower} R.~G., {Schaye} J., {McAlpine} S., {Dalla Vecchia}
  C., {Frenk} C.~S., {Schaller} M., {Theuns} T., 2016, Monthly Notices of the
  Royal Astronomical Society, 462, 190

\bibitem[{{Rosas-Guevara} {et~al}\mbox{.}(2015){Rosas-Guevara}, {Bower},
  {Schaye}, {Furlong}, {Frenk}, {Booth}, {Crain}, {Dalla Vecchia}, {Schaller},
  \& {Theuns}}]{2015MNRAS.454.1038R}
{Rosas-Guevara} Y.~M. {et~al.}, 2015, \mnras, 454, 1038

\bibitem[{{Schaye} {et~al}\mbox{.}(2015){Schaye}, {Crain}, {Bower}, {Furlong},
  {Schaller}, {Theuns}, {Dalla Vecchia}, {Frenk}, {McCarthy}, {Helly},
  {Jenkins}, {Rosas-Guevara}, {White}, {Baes}, {Booth}, {Camps}, {Navarro},
  {Qu}, {Rahmati}, {Sawala}, {Thomas}, \& {Trayford}}]{2015MNRAS.446..521S}
{Schaye} J. {et~al.}, 2015, \mnras, 446, 521

\bibitem[{{Sijacki} {et~al}\mbox{.}(2015){Sijacki}, {Vogelsberger}, {Genel},
  {Springel}, {Torrey}, {Snyder}, {Nelson}, \&
  {Hernquist}}]{2015MNRAS.452..575S}
{Sijacki} D., {Vogelsberger} M., {Genel} S., {Springel} V., {Torrey} P.,
  {Snyder} G.~F., {Nelson} D., {Hernquist} L., 2015, \mnras, 452, 575

\bibitem[{{Sparre} {et~al}\mbox{.}(2015){Sparre}, {Hayward}, {Springel},
  {Vogelsberger}, {Genel}, {Torrey}, {Nelson}, {Sijacki}, \&
  {Hernquist}}]{2015MNRAS.447.3548S}
{Sparre} M. {et~al.}, 2015, \mnras, 447, 3548

\bibitem[{{Springel}(2005)}]{2005MNRAS.364.1105S}
{Springel} V., 2005, \mnras, 364, 1105

\bibitem[{{Springel}(2010)}]{2010MNRAS.401..791S}
{Springel} V., 2010, \mnras, 401, 791

\bibitem[{{Springel} {et~al}\mbox{.}(2018){Springel}, {Pakmor}, {Pillepich},
  {Weinberger}, {Nelson}, {Hernquist}, {Vogelsberger}, {Genel}, {Torrey},
  {Marinacci}, \& {Naiman}}]{2018MNRAS.475..676S}
{Springel} V. {et~al.}, 2018, \mnras, 475, 676

\bibitem[{{Terrazas} {et~al}\mbox{.}(2017){Terrazas}, {Bell}, {Woo}, \&
  {Henriques}}]{2017ApJ...844..170T}
{Terrazas} B.~A., {Bell} E.~F., {Woo} J., {Henriques} B. M.~B., 2017, \apj,
  844, 170

\bibitem[{{Thomas} {et~al}\mbox{.}(2019){Thomas}, {Dav{\'e}},
  {Angl{\'e}s-Alc{\'a}zar}, \& {Jarvis}}]{2019MNRAS.487.5764T}
{Thomas} N., {Dav{\'e}} R., {Angl{\'e}s-Alc{\'a}zar} D., {Jarvis} M., 2019,
  \mnras, 487, 5764

\bibitem[{{Thomas} {et~al}\mbox{.}(2021){Thomas}, {Dav{\'e}}, {Jarvis}, \&
  {Angl{\'e}s-Alc{\'a}zar}}]{2021MNRAS.503.3492T}
{Thomas} N., {Dav{\'e}} R., {Jarvis} M.~J., {Angl{\'e}s-Alc{\'a}zar} D., 2021,
  \mnras, 503, 3492

\bibitem[{{Truong} {et~al}\mbox{.}(2020){Truong}, {Pillepich}, {Werner},
  {Nelson}, {Lakhchaura}, {Weinberger}, {Springel}, {Vogelsberger}, \&
  {Hernquist}}]{2020MNRAS.494..549T}
{Truong} N. {et~al.}, 2020, \mnras, 494, 549

\bibitem[{{Tzanavaris} {et~al}\mbox{.}(2016){Tzanavaris}, {Hornschemeier},
  {Gallagher}, {Lenki{\'c}}, {Desjardins}, {Walker}, {Johnson}, \&
  {Mulchaey}}]{2016ApJ...817...95T}
{Tzanavaris} P., {Hornschemeier} A.~E., {Gallagher} S.~C., {Lenki{\'c}} L.,
  {Desjardins} T.~D., {Walker} L.~M., {Johnson} K.~E., {Mulchaey} J.~S., 2016,
  \apj, 817, 95

\bibitem[{{Ueda} {et~al}\mbox{.}(2014){Ueda}, {Akiyama}, {Hasinger}, {Miyaji},
  \& {Watson}}]{2014ApJ...786..104U}
{Ueda} Y., {Akiyama} M., {Hasinger} G., {Miyaji} T., {Watson} M.~G., 2014,
  \apj, 786, 104

\bibitem[{{Vito} {et~al}\mbox{.}(2016){Vito}, {Gilli}, {Vignali}, {Brandt},
  {Comastri}, {Yang}, {Lehmer}, {Luo}, {Basu-Zych}, {Bauer}, {Cappelluti},
  {Koekemoer}, {Mainieri}, {Paolillo}, {Ranalli}, {Shemmer}, {Trump}, {Wang},
  \& {Xue}}]{2016MNRAS.463..348V}
{Vito} F. {et~al.}, 2016, \mnras, 463, 348

\bibitem[{{Vito} {et~al}\mbox{.}(2014){Vito}, {Gilli}, {Vignali}, {Comastri},
  {Brusa}, {Cappelluti}, \& {Iwasawa}}]{2014MNRAS.445.3557V}
{Vito} F., {Gilli} R., {Vignali} C., {Comastri} A., {Brusa} M., {Cappelluti}
  N., {Iwasawa} K., 2014, \mnras, 445, 3557

\bibitem[{{Vogelsberger} {et~al}\mbox{.}(2013){Vogelsberger}, {Genel},
  {Sijacki}, {Torrey}, {Springel}, \& {Hernquist}}]{2013MNRAS.436.3031V}
{Vogelsberger} M., {Genel} S., {Sijacki} D., {Torrey} P., {Springel} V.,
  {Hernquist} L., 2013, \mnras, 436, 3031

\bibitem[{{Vogelsberger} {et~al}\mbox{.}(2014{\natexlab{a}}){Vogelsberger},
  {Genel}, {Springel}, {Torrey}, {Sijacki}, {Xu}, {Snyder}, {Bird}, {Nelson},
  \& {Hernquist}}]{2014Natur.509..177V}
{Vogelsberger} M. {et~al.}, 2014{\natexlab{a}}, \nat, 509, 177

\bibitem[{{Vogelsberger} {et~al}\mbox{.}(2014{\natexlab{b}}){Vogelsberger},
  {Genel}, {Springel}, {Torrey}, {Sijacki}, {Xu}, {Snyder}, {Nelson}, \&
  {Hernquist}}]{2014MNRAS.444.1518V}
{Vogelsberger} M. {et~al.}, 2014{\natexlab{b}}, \mnras, 444, 1518

\bibitem[{{Volonteri} {et~al}\mbox{.}(2016){Volonteri}, {Dubois}, {Pichon}, \&
  {Devriendt}}]{2016MNRAS.460.2979V}
{Volonteri} M., {Dubois} Y., {Pichon} C., {Devriendt} J., 2016, \mnras, 460,
  2979

\bibitem[{{Weinberger} {et~al}\mbox{.}(2017){Weinberger}, {Springel},
  {Hernquist}, {Pillepich}, {Marinacci}, {Pakmor}, {Nelson}, {Genel},
  {Vogelsberger}, {Naiman}, \& {Torrey}}]{2017MNRAS.465.3291W}
{Weinberger} R. {et~al.}, 2017, \mnras, 465, 3291

\bibitem[{{Weinberger} {et~al}\mbox{.}(2018){Weinberger}, {Springel}, {Pakmor},
  {Nelson}, {Genel}, {Pillepich}, {Vogelsberger}, {Marinacci}, {Naiman},
  {Torrey}, \& {Hernquist}}]{2018MNRAS.479.4056W}
{Weinberger} R. {et~al.}, 2018, \mnras, 479, 4056

\bibitem[{{Zhang}, {Gilfanov} \& {Bogd{\'a}n}(2012){Zhang}, {Gilfanov}, \&
  {Bogd{\'a}n}}]{2012A&A...546A..36Z}
{Zhang} Z., {Gilfanov} M., {Bogd{\'a}n} {\'A}., 2012, \aap, 546, A36

\bibitem[{{Zinger} {et~al}\mbox{.}(2020){Zinger}, {Pillepich}, {Nelson},
  {Weinberger}, {Pakmor}, {Springel}, {Hernquist}, {Marinacci}, \&
  {Vogelsberger}}]{2020MNRAS.499..768Z}
{Zinger} E. {et~al.}, 2020, \mnras, 499, 768

\end{thebibliography}

\label{lastpage}
\end{document}